%% file: main.tex
\documentclass[12pt]{article}
\usepackage{amsmath}
\usepackage{enumerate}
\usepackage[semicolon]{natbib}
\setcitestyle{aysep={}} 
\usepackage{url} 
\usepackage{authblk}
\usepackage{amsthm}
\usepackage{amsfonts}
\usepackage{graphicx} 
\usepackage{subcaption}

\usepackage{physics}
\usepackage{multirow, makecell}
\usepackage{csvsimple}
\usepackage{bbm}
\usepackage{soul}
\usepackage{booktabs}
\usepackage{threeparttable}
\usepackage{eqparbox}
\usepackage{wrapfig}
\usepackage{bm}
\usepackage{mathtools}
\usepackage{pgf,tikz}
\usepackage{float}
\usepackage{tocloft}
\usepackage{parskip}
\usepackage[%
    colorlinks=true,
    pdfborder={0 0 0},
    linkcolor=blue,
    citecolor={blue}
]{hyperref}
\usepackage{multibib}
\newcites{supp}{Supp. References}

\newcommand{\blind}{1}

\input{cmd}
\begin{document}

\def\spacingset#1{\renewcommand{\baselinestretch}%
{#1}\small\normalsize} \spacingset{1}


\if1\blind
{
\title{An Experimental Design for Anytime-Valid Causal Inference on Multi-Armed Bandits}
\author[1]{Biyonka Liang}
\author[2]{Iavor Bojinov}
\affil[1]{Department of Statistics, Harvard University}
\affil[2]{Harvard Business School}
\date{\today}
  \maketitle
} \fi

\if0\blind
{
\title{An Experimental Design for Anytime-Valid Causal Inference on Multi-Armed Bandits}
\author[1]{}
  \maketitle
} \fi

\begin{abstract}
Experimentation is crucial for managers to rigorously quantify the value of a change and determine if it leads to a statistically significant improvement over the status quo. As companies increasingly mandate that all changes undergo experimentation before widespread release, two challenges arise: (1) minimizing the proportion of customers assigned to the inferior treatment and (2) increasing experimentation velocity by enabling data-dependent stopping. This paper addresses both challenges by introducing the Mixture Adaptive Design (MAD), a new experimental design for multi-armed bandit (MAB) algorithms that enables anytime-valid inference on the Average Treatment Effect (ATE) for \emph{any} MAB algorithm. Intuitively, MAD "mixes" any bandit algorithm with a Bernoulli design, where at each time step, the probability of assigning a unit via the Bernoulli design is determined by a user-specified deterministic sequence that can converge to zero. This sequence lets managers directly control the trade-off between regret minimization and inferential precision. Under mild conditions on the rate the sequence converges to zero, we provide a confidence sequence that is asymptotically anytime-valid and guaranteed to shrink around the true ATE. Hence, when the true ATE converges to a non-zero value, the MAD confidence sequence is guaranteed to exclude zero in finite time. Therefore, the MAD enables managers to stop experiments early while ensuring valid inference, enhancing both the efficiency and reliability of adaptive experiments. Empirically, we demonstrate that the MAD achieves finite-sample anytime-validity while accurately and precisely estimating the ATE, all without incurring significant losses in reward compared to standard bandit designs.
\end{abstract}

\noindent%
\newpage
\spacingset{1} 
\section{Introduction}\label{section:introduction}
Managers routinely use experimentation to quantify the value that a change brings to their business by randomly assigning customers to either the change (the treatment) or the status quo (the control). The resulting estimation of the \textit{average treatment effect} (ATE) from such experimentation can help managers augment their decision-making around deployment and investment opportunities \citep{thomke2020experimentation,kohavi2020trustworthy, koning2022experimentation, bojinovgupta}. As an additional benefit, experimentation de-risks innovation by reducing the proportion of people exposed to a potentially harmful change. Risk mitigation measures are particularly important for large organizations where only a small percentage of product changes produce positive effects \citep{kohavi2020trustworthy, bojinovgupta}, and one common risk mitigation approach is to harmful experiments as quickly as possible, as prolonging the experiment could cause substantial revenue loss and customer alienation, as exhibited in various case studies at large technology companies like Amazon, Netflix, and Microsoft \citep{thomke2020experimentation, bojinovgupta}. 

However, methods that are designed for experiments with a pre-specified sample size/time horizon can necessarily only provide inferential results \emph{after} that sample size/end time is attained. Using these approaches, managers could not continuously monitor their experiment to determine whether a statistically significant negative effect has occurred (motivating them to stop the experiment) without invalidating Type I error guarantees. For instance, even if they rely on heuristic thresholds (\emph{e.g.}, if $10{,}000$ customers drop out, as in the Netflix experiment described in \cite{ham}) to determine whether or not to stop the experiment, they would not be able to perform valid inference on the data collected using approaches designed for pre-specified time horizons, since a data-dependent stopping time was used. 

To address this limitation, researchers use sequential testing procedures to derive confidence sequences (i.e., sequences of confidence intervals with uniform type-I error rate control), which enables managers to continuously monitor the results of an experiment and generate inference that is ``anytime-valid''. Hence, managers can stop the experiment using any data-driven metric or threshold of choice, e.g., as soon as the ATE is statistically significantly different from zero, without invalidating error guarantees \citep{waudby2020estimating, howard, johari2022always, waudbysmith2022}. Various companies, such as Netflix \citep{lindon2022anytime}, Microsoft \citep{waudbysmith2022}, and Adobe \citep{waudbysmith_timeuniform} have recently begun exploring methodology in anytime-valid inference for their experimentation platforms. 

However, as customers often arrive sequentially, it is increasingly common for managers at companies such as those above to de-risk experimentation further by
using dynamic assignment mechanisms to alter the treatment assignment probabilities of new customers based on the observed data \citep{bojinovgupta, che2023adaptive, fiez2024best}. The most popular approach to dynamic experimentation is to use multi-armed bandit algorithms (MAB) because of their flexibility and ability to decrease the likelihood of new customers being assigned to subpar treatments. Unfortunately, this beneficial risk minimization property is somewhat at odds with the objective of estimating the ATE, especially in an anytime-valid manner. To minimize the estimation error for the ATE, the best practice statistically is to assign each customer's treatment independently and randomly, so that about an equal number of customers are allocated to the treatment and control \citep{wainwright2019high}. MAB algorithms are instead designed to minimize regret by assigning treatments dynamically, meaning that previous outcomes inform future assignment probabilities. Hence, statistical inference on MAB experiments is challenging due to this dependence across treatment assignments and the fact that fewer people tend to be assigned to subpar treatments over time, as observations from all arms are necessary to estimate the ATE with sufficient precision. Prior work has attempted to overcome this challenge by developing asymptotic estimators of the treatment effect in bandit experiments \citep{ luedtke2016statistical, deshpande2019online, hadad, zhang} or proposing various augmentations to the designs of MAB that ensure an adequate number of customers are assigned to the subpar treatment \cite[]{erraqabi2017trading, econometrica, simchi}.

 However, these approaches are designed to produce inferential results for MAB experiments given a pre-specified stopping time or sample size, and hence, do not naturally extend to continuous monitoring settings. Because continuous monitoring is often desirable for large-scale online experiments \citep{lindon2022anytime, ham}, the lack of inferential tools for the continuous monitoring of bandit experiments limits their utility in practice.
For example, if a manager chooses to stop a bandit experiment based on some heuristic threshold, there are essentially no existing approaches that would allow for valid statistical inference, thus limiting what could be learned from that data. While some existing works in sequential testing address adaptive experiments, they often focus on the pure exploration settings such as best arm identification \citep{jamieson2013lilucboptimal, nikolakakis2021quantile, howard2022sequential}, which aims to accurately identify the best arm given a fixed budget of pulls or desired confidence level, and hence, does not involve optimizing for cumulative regret \citep{audibert2010best, bubeck2010pureexplorationmultiarmedbandit, gabillon2012bestarm, jamieson2014best}. Therefore, such approaches are by construction not suited for risk mitigation in online experiments as they do not aim to minimize regret over time. Other works in the continuous monitoring of adaptive experiments which aim to balance exploration and exploitation place restrictive assumptions on the degree of exploitation allowed, such as requiring the bandit algorithm used to generate treatment assignment probabilities that are uniformly bounded away from zero and one \cite[]{howard, ham}, hence excluding nearly all standard bandit algorithms such as Thompson sampling \cite[]{thompson1933likelihood} and the Upper Confidence Bound algorithm \cite[]{auer}. 

\subsection{Our Contributions}
To enable anytime-valid inference for the MAB algorithms most commonly used in online experiments, we propose using a Mixture Adaptive Design (MAD): a new experimental design that at each time step $t$ ``mixes'' any MAB algorithm with a Bernoulli design through a deterministic sequence $\delta_t \in (0, 1]$, which controls the priority placed on the Bernoulli design as the sample size grows. For $\delta_t = \omega\left(\frac{1}{t^{1/4}}\right)$ \footnote{As is standard in asymptotic notation \citep{de1981asymptotic}, $\delta_t = \omega\left(\frac{1}{t^{1/4}}\right)$ implies that $\frac{1}{\delta_t} = o(t^{1/4})$. Intuitively, we can think of $\delta_t$ as a sequence approaching zero slower than $\frac{1}{t^{1/4}}$.}, we provide a confidence sequence that is asymptotically valid (a concept which we formalize in the following section) and guaranteed to shrink around the true ATE for any choice of bandit algorithm. To our knowledge, this confidence sequence is the first to guarantee ATE estimation in adaptively collected data settings that do not require the treatment assignment probabilities to be bounded away from zero and one. Thus, the MAD expands the utility of confidence sequences to nearly any bandit algorithm while providing guarantees on the validity and power of the ATE estimation. Practically, the MAD guarantees that, under an actual non-zero treatment effect, a manager wishing to stop the experiment once the confidence sequence does not cover zero is guaranteed to stop in a finite time. The condition that $\delta_t= \omega(\frac{1}{t^{1/4}})$ provides the manager with great flexibility to tune the experimental design based on the problem specifics. For instance, setting $\delta_t = 1$ recovers a Bernoulli design and setting $\delta_t= \omega(\frac{1}{t^{1/4}})$ prioritizes the MAB algorithm. We provide recommendations for setting $\delta_t$ based on the user's priorities; see Section~\ref{section:setting_delta}. Finally, we show empirically that the MAD generates confidence sequences that shrink quickly around the true ATE and achieves the correct coverage in finite samples without major losses in reward compared to a standard bandit design.

The paper is organized as follows. Section \ref{section:background} describes the related literature and the necessary background. Section \ref{section:problemstatement} formalizes our problem statement by defining the context and causal estimands. Section \ref{section:ourmethod} defines the MAD, and Section \ref{sec:theory} derives its theoretical properties. Section \ref{section:setting_delta} provides some practical recommendations on how to set $\delta_t$. Section \ref{section:simulations} describes our simulation study, and finally, Section \ref{section:conclusion} presents our conclusions. All theorem proofs are contained in the Appendix, which also provides additional simulations and extensions of our results.

\section{Background and Related Work}\label{section:background}
\subsection{Inference and Exploration in Bandit Experiments}
Simultaneously conducting inference on the ATE while minimizing regret via adaptive treatment assignment is particularly challenging; many statistical guarantees for common ATE estimators are designed for independent and identically distributed (i.i.d.) data, and such guarantees often no longer hold in MAB settings. For instance, the difference in sample averages between a treatment arm and a control arm is no longer unbiased for the true ATE between these arms \citep{ateunbiased,deshpande2019online} and may no longer be asymptotically normal \citep{dima, zhang}. While Inverse Propensity Weighting (IPW) estimators are often unbiased in the MAB setting \cite[]{horvitz1952generalization, hadad}, their variance can be highly unstable under adaptive assignment, thus making any subsequent hypothesis testing on the ATE essentially powerless. Specifically, the variance can rapidly increase as the probability that a bandit algorithm pulls a sub-optimal arm rapidly approaches zero \citep{tutorial_ts}. For instance, the well-known Thompson Sampling (TS) algorithm \cite[]{agrawal2012analysis} and variations of the Upper Confidence Bound (UCB) algorithm \cite[]{garivier2011kl, menard2017minimax, kaufmann2018bayesian} achieve a regret bound of $O(\log T)$ and are said to be asymptotically optimal as the regret of any algorithm must have at least a $\log(T)$ rate \cite[]{lai1985asymptotically}. Hence, we expect the number of draws from sub-optimal arms to be on the order of $\log(T)/T$ in the long run \cite[]{econometrica}. 

Hence, a growing body of work aims to develop statistical methods for performing inference on MABs. Many works focus on developing statistical tests or central limit theorems for MABs without significantly altering the bandit algorithm \citep{lai1982least, hadad, luedtke2016statistical, deshpande2019online, zhang, zhang_m_est, hadad, nair}, hence allowing for aggressive regret minimization. However, these approaches are not designed for continuous monitoring settings, and many of these works also require certain assumptions on the outcome distributions, such as moment bounds or i.i.d. outcome distributions within each treatment arm, to establish their theoretical guarantees; we provide a detailed discussion on why these assumptions may not be appropriate in the context of online experiments in Section~\ref{section:why_design_based}. 
In contrast, \textit{pure exploration} approaches prioritize information gathering to achieve specific inferential objectives, such as identifying the best arm or a subset of best arms \citep{ kim2006selecting, bubeck2011pure, chen2015ranking, russo2016simple, jamieson2013lilucboptimal, nikolakakis2021quantile, howard2022sequential} or estimating the value of all arms accurately \citep{antos2010active, carpentier2011upper}, \text{without} concern for minimizing regret.
Hence, the distinction between pure exploration approaches and the classic exploration-exploitation tradeoff in MAB problems lies in the emphasis on precision in identifying the best arm rather than optimizing long-term returns. While some work has explored continuous monitoring in best arm identification \citep{jamieson2013lilucboptimal, nikolakakis2021quantile, howard2022sequential}, these pure exploration approaches are, by construction, not designed for risk mitigation, which is the primary motivation for continuous monitoring in online experiments. Other works have focused on incorporating some degree of regret minimization in best arm identification tasks \citep{degenne2019bridging, zhang2024fastregretoptimalbest}, but are generally not focused on continuous monitoring settings.

In contrast to the above works, works in the experimental design of MABs aim to adapt MAB algorithms in a way that optimizes both inference and regret, often providing theoretical guarantees for both. Prior works in this area have primarily altered existing bandit algorithms to sample more from sub-optimal arms, making them more amenable to ATE estimation. For instance, \cite{econometrica} proposes a variation of Thompson sampling which re-weights the estimated mean reward of each arm to enforce additional exploration of sub-optimal arms. \cite{erraqabi2017trading} and \cite{simchi} formulates the trade-off between mean squared estimation error and regret as an 
optimization problem, then proposes a variation of existing bandit algorithms, \emph{i.e.}, UCB and EXP3 \cite[]{auer2002finite} respectively, 
which satisfies their characterization of optimality.
Intuitively, some degree of exploration must be imposed to produce inferential guarantees, as the precision of ATE inference essentially depends on the variance of the treatment arm which has the fewest number of observations. However, these works focus on adapting \textit{specific} bandit algorithms and thus, are difficult to generalize and apply broadly to other bandit algorithms. Additionally, these works propose designs that do not allow the user to tune the degree of exploration in the algorithm \cite[]{econometrica} or have limitations on when during the experiment this tuning can take place \cite[]{simchi}. 
\cite{hahn2011adaptive} proposes a two-stage adaptive design and proves an asymptotic normality result for ATE estimation, but requires the adaptive assignment algorithm to have assignment probabilities bounded away from zero and one, which is not satisfied by most common bandit algorithms such as UCB (a deterministic algorithm) and Thompson sampling.
\cite{onlinelearningex} introduces an adaptation of linear contextual bandit algorithms that re-weights the assignment probabilities using balancing methods from causal inference, but focus on establishing regret bounds rather than providing inferential guarantees for, e.g., the arm means or treatment effects. Notably, the above works are designed for the setting where the manager is conducting inference at the end of an experiment with a pre-specified time horizon, and hence, are not designed for continuous monitoring.

\subsection{Anytime-Valid Inference}
As discussed in Section~\ref{section:introduction}, it is often desirable in large-scale online experiments to continuously monitor results, often with the goal of allowing for a data-dependent stopping time in response to running inferential results for risk mitigation. 
Such a continuous monitoring setting requires statistical tests that uniformly control Type I error at every time point, \emph{i.e.}, are \textit{anytime-valid}. As classical tests do not satisfy this condition, prior works have proposed using \textit{confidence sequences} to enable valid inference \cite[]{darling1967confidence, lai1976boundary, lai1976confidence, johari2022always, waudbysmith_timeuniform}.

\begin{definition}[Confidence Sequence]\label{def:cs}
A confidence sequence (CS) is a set of confidence sets $\{C_t\}_{t=1}^\infty$ at level $\alpha \in (0, 1)$ such that, for a true (non-zero) treatment effect $\tau_t$, 
$$\mathbb{P}\left(\forall t, \tau_t \in C_t \right) \geq 1-\alpha.$$
\end{definition}

For example, \cite{johari2022always} provides a non-asymptotic confidence sequence for traditional A/B testing settings where the treatment allocations are not made adaptively. However, extending non-asymptotic confidence sequences to adaptively collected data often require specific assumptions on the data such as known bounds on the random variables and/or tail behavior assumptions such as sub-Gaussianity \cite[]{waudby2020estimating, howard, Howard_2022}. In particular, such assumptions prevent non-asymptotic CSs from being applicable or even possible to construct in many real-world settings.
To address this restriction, \cite{waudbysmith_timeuniform} first introduced the notion of an \textit{asymptotic confidence sequence} and derived a universal asymptotic CS that requires only CLT-like assumptions. 

\begin{definition}[Asymptotic Confidence Sequence]\label{def:asymp_cs}
$(\hat{\mu}_{t} \pm \hat{V}_t)$ is an asymptotic $1-\alpha$ confidence sequence for a target parameter $\mu_t$ if there exists some (unknown) non-asymptotic $1-\alpha$ confidence sequence $(\hat{\mu}_{t} \pm V^*_t)$ for $\mu_t$ such that
$$\frac{\hat{V}_t}{V^*_t} \stackrel{a.s.}{\to}1.$$
Furthermore, we say that $\hat{V}_t$ has approximation rate $R_t$ if $V^*_t = \hat{V}_t + o(R_t)$ almost surely, where $R_t$ is a deterministic (\emph{i.e.}, non-random) sequence. 
\end{definition}

 As the assumptions necessary for generating asymptotic CSs are comparatively much weaker, asymptotic CSs expand the utility of CSs to a wider array of settings, such as time series and panel experiments \citep{ham}. Hence, asymptotic CSs trade off small sample validity for flexibility and generalizability and has been shown in both simulation and practice to perform well, even for small and moderate-sized samples \cite{waudbysmith_timeuniform,ham}.
 However, existing work on continuous inference for MAB experiments is rather limited. While \cite{ham} and \cite{howard} discuss extensions to experiments with adaptive treatment assignment probabilities, their proposed CSs require \textit{probabilistic treatment assignment}, \emph{i.e.}, they assume that there exists $ 0 < p_{min} \leq 1/2$ s.t. $p_{t}(w) \in [p_{min}, 1-p_{min}]$ almost surely. This restriction excludes many commonly used bandit algorithms such as UCB, which has deterministic treatment assignments, and Thompson Sampling. Without this treatment assignment assumption, the asymptotic confidence sequences of \cite{ham} are \textit{not guaranteed to be asymptotically anytime-valid} \cite[]{waudbysmith_timeuniform}, and the non-asymptotic confidence sequences of \cite{howard} may be inapplicable since their validity results requires $p_{min}$ to exist and be known. Thus, applying such approaches to most commonly used MAB algorithms necessarily requires imposing lower bounds on the treatment assignment probabilities of the algorithm, such as clipping them to be within $[p_{min}, 1-p_{min}]$, restricting the adaptivity allowed and increasing regret. 

\section{Problem Setting}\label{section:problemstatement}
In this section, we formalize the multi-armed bandit problem setting, describe and justify our design-based approach, and define the causal estimands of interest. 

Generally, we assume that we observe a sequence of $t$ units  $\{W_i, Y_i\}_{i=1}^t$ where $W_i \in \{0, ..., K-1\}$ and $Y_i$ are the treatment assignment and outcome respectively for unit $i$. In the MAB literature, $\{W_i, Y_i\}_{i=1}^t$ are analogous to the sequence of actions and rewards commonly notated $\{A_i, R_i\}_{i=1}^t$. Although we will ultimately show that our method can be defined more broadly to any $K \geq 2$ number of treatments and the batched bandit setting, we first assume that we observe a single unit at each time and that \emph{we have binary treatment assignments}, \emph{i.e.}, $W_i \in \{0, 1\}$. Throughout, we assume that there is no population interference \citep{cox1958planning} and that each person receives the same version of treatment; typically, these two assumptions are combined into the popular Stable Unit Treatment Value Assumption (SUTVA) \citep{rubin1986comment}. Under these assumptions, in our binary treatment setting, each unit $i$ has a pair of potential outcomes $\{Y_i(1), Y_i(0)\}$.
The observed outcome $Y_i$ is a function of the random treatment assignment $W_i$ and the potential outcomes ${Y_i(1), Y_i(0)}$:
\begin{align*}
    Y_i = W_i Y_i(1) + (1-W_i)Y_i(0).
\end{align*}

Let $\tau_i = Y_i(1)-Y_i(0)$. Then, the Average Treatment Effect (ATE) at time $t$ is:
\begin{definition}[Average Treatment Effect (ATE) at time $t$]
    $$\bar{\tau}_t \coloneqq \frac{1}{t}\sum_{i=1}^t \tau_i.$$
\end{definition}
Our objective is to generate a confidence sequence for $\bar{\tau}_t$ in MAB experiments with anytime-valid Type I error control.

\subsection{Design-Based Causal Inference}\label{section:why_design_based}
Throughout, we adopt a design-based (also referred to as finite-population) perspective to analyzing bandit experiments. This perspective is often used in the analysis of randomized experiments \citep{neyman1935statistical,fisher1936design, ding2016randomization, aronow2017estimating, athey2022design, abadie2020sampling, bojinov2023design}, and is an alternative perspective to inference on the super-population perspective, which is more commonly assumed in the analysis of bandit experiments. Here, we provide an overview of the design-based perspective and discuss our reasoning for adopting such an approach for online experiments.

We first discuss the primary distinctions between the finite-population perspective and the super-population perspective. Under a super-population perspective, experimental units are assumed to be randomly drawn (often i.i.d.) from a hypothetical, infinite super-population, and hence, the outcomes for each treatment are assumed to be randomly drawn (often i.i.d.) from some underlying distribution, e.g., for all $i$, $w \in \{0, 1\}$, $Y_i(w) \stackrel{i.i.d.}{\sim} \mathcal{N}(\mu_{w}, 1)$.  Under this approach, estimands of interest are defined relative to this infinite super-population; the typical estimand is the average treatment effect across the population. In contrast, the finite population model does not presuppose the existence of such a super-population, with the causal estimand being specific to the actual sample gathered; here, the typical estimand is the average treatment effect across the sample (\cite{imbens2015causal}, Chapter 6). Because the population underlying the randomized experiment is finite under the design-based perspective, the set of all potential outcomes for the population can be treated as unknown but non-random due to the population being finite \citep{abadie2020sampling}. Hence, the randomness in the experiment is solely attributed to how treatments are allocated, i.e., in our notation above, $\{Y_i(1), Y_i(0)\}$ is non-random, so $Y_i$ is random only via the treatment assignments $W_i$.  Mathematically, this perspective can be represented as conditioning on the complete set of unobserved potential outcomes. Note that in contrast, the super-population perspective treats the potential outcomes as random variables, and hence, $Y_i$ is random via both the treatment assignments $W_i$ and $\{Y_i(1), Y_i(0)\}$.

Hence, the primary distinction between the super-population and finite-population methods concerns assumptions on the underlying population, which necessarily results in different target estimands and their interpretations. 
Importantly, in the context of online experiments, finite-population estimands have a meaningful interpretation. As one primary motivation for our work is to enable effective risk mitigation in bandit experiments, we are primarily interested in quantifying the risk incurred on the \textit{actual} users in the experiment, and not some hypothetical super-population of users. For instance, when determining a data-driven stopping time, managers want to stop the experiment when the cost to the users in the experiment is too high, not necessarily when the expected cost to some super-population of users is too high.

Additionally, though the super-population perspective is most commonly adopted in the analysis of randomized experiments, there exist various online experimentation settings where no natural super-population seems to exist, such as the common setting 
where the data is collected via a convenience sample \citep{abadie2020sampling}. 
 For instance, consider an experiment that is run on all users who visited a website on a particular day, which, as noted in \citep{abadie2020sampling}, cannot be justifiably considered a random sample from some underlying super-population. Hence, performing inference under a super-population perspective would require us to specify a reasonable super-population, determine the underlying sampling scheme generating this data, and quantify its uncertainty. As various case studies at companies such as Linkedin and Microsoft have shown that users within an experiment are often not reflective of the larger user base \citep{wang2019heavy, bojinovgupta, rajkumar2022causal}, accurately specifying this sampling scheme is necessary for generalizability. However, this task is effectively impossible without strong assumptions and/or highly informative analyses based on previous data. Under a design-based perspective, there is no need to specify this sampling scheme, as we treat the sample itself as a finite population, and aim to learn finite-population estimands rather than generalizing to some larger population. Even in settings where the sample could be justified as a random sample from a hypothetical super-population, assumptions on the outcome model are often necessary to establish theoretical results, for instance, that all treatment and control units have the same variance or commonly, that the outcomes for each treatment are drawn i.i.d. from some distribution \citep{johari2022always, waudbysmith_timeuniform}. However, such assumptions often do not hold in reality, as heterogenous treatment effects exist across different users \cite[]{onlinelearningex, gupta2019top, larsen2024statistical}. For example, various case studies in online experiments discuss how heavy users behave notably differently than the larger user base \citep{machmouchi2016principles, wang2019heavy, bojinovgupta}, and hence, their outcomes cannot reasonably be considered as coming from the same underlying outcome distribution as other users. However, violations of assumptions like those above can result in effective sample sizes that are much smaller than assumed, causing variance estimation to be highly conservative \cite[]{meng2018statistical}. In contrast, a design-based approach does not require us to specify the randomness underlying the potential outcomes, and hence, accommodates complex structure in the outcome model, such as heteroscedastic variance, distribution shift, and other forms of non-stationarity commonly present in practice \citep{bojinov_timeseries, ham}.

Hence we adopt a design-based perspective, as it enables us to account for complex structure in the outcome model known to be present in online experiments while providing estimands that have meaningful interpretations for risk mitigation in this problem area.


\subsubsection{Bandits under a Design-Based Framework}
Since a design-based approach allows us to handle non-stationarity in the outcome model, a natural question is whether it is sensible to use a bandit for treatment allocation in such settings. However, 
 bandits are often run in settings where non-stationarity, such as heterogenous treatment effects across different user groups and changes in user behavior across different times of day or days of the week, is very likely present \citep{nonstationaryrewards3, nonstationaryrewards2, nonstationaryrewards1,gupta2019top, fiez2024best, larsen2024statistical}. However, managers are required to assume a stationary outcome distribution by many existing (super-population) approaches in order to analyze the experiment. While it is valid to question when a bandit is appropriate in online experiments, our work addresses a different pressing question: given the use of bandits in real-world settings where some degree of non-stationary is often present, can we develop a robust method that allows for non-stationarity while maintaining error control? By providing anytime-valid inference on bandits that is valid regardless of non-stationarity in the outcome model, we allow managers to actually track non-stationary that is occurring in the ATE, facilitating a more accurate understanding of how their experiment is affecting their users and hence, enabling them to properly assess whether to stop the experiment based on running results.
Through various simulations, we find that the confidence sequence generated from the MAD tracks changes in non-stationarity over time, even in settings where there is a sign flip in the ATE and the regret can suffer greatly; see Appendix~\ref{appendix:nonstationary_mad}. 

However, in reality, highly adversarial (in terms of regret minimization) forms of non-stationary in the outcome model rarely occur, e.g. where the ATE oscillates from positive to negative rapidly over time. We generally expect the true ATE to have a smooth trend and maintain the same sign \citep{kohavi2020trustworthy} so the best, i.e., higher reward arm is consistent throughout the experiment even though the ATE may be shrinking. In such settings, the bandit generally consistently chooses the best arm with high probability. Hence, there is some justification for the use of bandits in non-stationary settings, and we refer interested readers to existing works that focus on this question \cite[]{nonstationaryrewards3, nonstationaryrewards2, nonstationaryrewards1}. 

\section{The Mixture Adaptive Design}\label{section:ourmethod}
We now formalize our proposed experimental design and state our main result. As in the MAB literature, we assume that the treatment assignment probabilities at a given time $t$ can depend on previously observed data up to time $t-1$. Formally, let $H_t \coloneqq \{W_i, Y_i\}_{i=1}^{t}$. For any arbitrary treatment assignment algorithm $\mathcal{A}$, even those that are adaptive and/or do not satisfy probabilistic treatment assignments, define the probability that the adaptive algorithm assigns treatment $w$ at time $i$ as
$p_{t}^{\mathcal{A}}(w) \coloneqq \mathbb{P}_{\mathcal{A}}\left(W_t = w \mid H_{t-1}\right)$, where $\mathbb{P}_{\mathcal{A}}$ denotes probability taken with respect to $\mathcal{A}$. For instance, if the manager wanted to set $\mathcal{A}$ as a Thompson sampling for their experiment, $p_{t}^{\mathcal{A}}(w)$ would be the assignment probability that Thompson sampling would assign treatment $w$ to unit $t$ given $H_{t-1}$.
 Recall, although we will ultimately show that our method (and all theoretical results) apply to any $K \geq 2$ number of treatments and the batched bandit setting, we will first present our design assuming we have binary treatment assignments, i.e., $W_i \in \{0, 1\}$. 

\begin{definition}[Mixture Adaptive Design (MAD)]\label{mad_def}
For any treatment assignment algorithm $\mathcal{A}$ and a real-valued sequence $\delta_i \in (0, 1]$, $w \in \{0, 1\}$, then the probability that the MAD will assign treatment $w$ at time $t$ is:
   \begin{align*}
      p^{\text{MAD}}_{t}(w) \coloneqq \mathbb{P}_{\text{MAD}}\left(W_t = w \mid H_{t-1}\right) = \delta_t \left(\frac{1}{2}\right)+ (1-\delta_t)p_{t}^{\mathcal{A}}(w).
\end{align*}
\end{definition}

Intuitively, the MAD ``mixes'' a (possibly adaptive and non-probabilistic) assignment algorithm $\mathcal{A}$ with a Bernoulli design. For instance, assuming $\mathcal{A}$ is a Thompson sampler, then the MAD will assign treatment $w$ according to a Bernoulli design with probability $\delta_t$, and according to the Thompson sampler with probability $1-\delta_t$, for all $t, w$. Note, $p^{\text{MAD}}_{t}(w) > 0$ even for deterministic assignment algorithms like UCB because $\delta_t > 0$ for all $t$, thus avoiding divisions by zero when computing IPW estimators. For instance, if the UCB algorithm has $p_{t}^{\mathcal{A}}(1) = 0$, then $p^{\text{MAD}}_{t}(1) = \delta_t \left(\frac{1}{2}\right) > 0$.
Definition~\ref{mad_def} can easily be extended to the $K \geq 2$ treatment settings with $w \in \{0, 1, ..., K-1\}$ by setting $ p^{\text{MAD}}_{t}(w) \coloneqq \mathbb{P}_{\text{MAD}}\left(W_t = w \mid H_{t-1}\right) = \delta_t \left(\frac{1}{K}\right)+ (1-\delta_t)p_{t}^{\mathcal{A}}(w)$.

The intuitive idea behind the MAD is that if we can balance the Bernoulli design and the adaptive design via $\delta_t$, we can gain the ATE precision while maintaining some of the regret minimization of the bandit algorithm. For simplicity of notation, we suppress $p^{\text{MAD}}_{t}(w)$'s dependence on $\mathcal{A}$. 

We now define our ATE estimators. Let $\mathcal{F}_{t}$ be the sigma-algebra that contains all pairs of potential outcomes $\{Y_i(1), Y_i(0)\}_{i=1}^t$ and $H_{t-1}$. 

Based on the estimator for $\tau_i$ proposed in \cite{bojinov_panel} for adaptive experiment settings, we set the estimator for $\tau_i$ as
\begin{align}\label{tau_hat_i}
    \hat{\tau}_{i} \coloneqq \frac{\mathbbm{1}\{W_i = 1\}Y_i}{p^{\text{MAD}}_{i}(1)} - \frac{\mathbbm{1}\{W_i = 0\}Y_i}{p^{\text{MAD}}_{i}(0)}, 
\end{align}
which is an unbiased estimator of $\tau_i$:
\begin{align*}
\mathbb{E}\left[\hat{\tau}_{i} \mid \mathcal{F}_{i} \right] = {\tau}_i,
\end{align*}
where, under our design-based approach, the expectation is taken with respect to the treatment assignment mechanism. Hence, 
\begin{align}\label{hatbar_tau}
    \hat{\bar{\tau}}_{t} \coloneqq \frac{1}{t}\sum_{i=1}^t  \hat{\tau}_{i},
\end{align}
is an unbiased estimator for $\bar{\tau}_t$:
\begin{align*}
 \frac{1}{t}\sum_{i=1}^t \mathbb{E}\left[\hat{\tau}_{i} \mid \mathcal{F}_{i} \right] = \bar{\tau}_t.
\end{align*}
Additionally, we have that, for all $i$, 
\begin{align}\label{eq:variance}  Var(\hat{\tau}_{i} \mid \mathcal{F}_{i}) \leq \sigma^2_{i}\text{, where } \sigma^2_{i}\coloneqq \frac{Y_i(1)^2}{p^{\text{MAD}}_{i}(1)} + \frac{Y_i(0)^2}{p^{\text{MAD}}_{i}(0)},
\end{align}
and thus, a natural unbiased estimator for $\sigma^2_{i}$ is 
\begin{align}\label{eq:variance_est}
    \hat{\sigma}^2_{i}\coloneqq \frac{Y_i(1)^2 \mathbbm{1}\{W_i = 1\}}{(p^{\text{MAD}}_{i }(1))^2} + \frac{Y_i(0)^2\mathbbm{1}\{W_i = 0\}}{(p^{\text{MAD}}_{i }(0))^2}.
\end{align}

Equations~\eqref{eq:variance} and~\eqref{eq:variance_est} follow from \cite{aronow2017estimating} and \cite{bojinov_timeseries}, which propose the estimator of Equation~\eqref{eq:variance_est} because the closed form does not admit a natural unbiased estimator. This is common in design-based inference where the variance includes the product of two potential outcomes, which cannot be observed simultaneously \citep{imbens2015causal}. 
Let $S_{t} \coloneqq \sum_{i=1}^t \sigma^2_{i}$ and $\hat{S}_{t} \coloneqq \sum_{i=1}^t \hat{\sigma}^2_{i}$.




\subsection{Extension to Batched Assignment Settings}\label{section:morearms}


Often, it is more practical to update the treatment assignment probabilities of an adaptive algorithm after observing a batch of units \citep{gao2019batched, zhang, nabi2022bayesian,che2023adaptive, gong2023bandits, ren2024dynamic}. 
 Assume we observe a sequence of batches, where for each batch $j$, we have a (non-random and finite) batch size of $B$. So, for each batch $j$, we observe $H^{\text{batched}}_{j} \coloneqq \left\{W^{(j)}_i, Y^{(j)}_i\right\}_{i=1}^{B}$, where $W^{(j)}_i \in \{0, 1\}$ and $Y^{(j)}_i$ are the treatment assignment and outcome for unit $i$ in batch $j$, respectively. We assume the treatment assignment probabilities are fixed within a batch, \emph{i.e.}, the treatment assignment probabilities are only (adaptively) updated after observing a batch of $B$ units.

Let $p_{j }^{\mathcal{A}_\text{batched}}(w) \coloneqq \mathbb{P}(W^{(j)}_i = w \mid H^{\text{batched}}_{j-1})$ be the assignment probability to treatment $w$ for unit $i$ in the $j$th batch for any user-provided (adaptive) batched algorithm $\mathcal{A}_\text{batched}$. 
\begin{definition}[Mixture Adaptive Design for Batched Assignment]\label{bmad}Given a batched (adaptive) assignment algorithm $\mathcal{A}_\text{batched}$ and a real-valued sequence $\delta_j \in (0, 1]$,
for $w \in \{0, 1, ..., K-1\}$, the assignment probability to treatment $w$ in batch $j$ is:
   \begin{align*}
   p^{\text{MAD}_\text{batched}}_{j }\left(w\right) =  \delta_j\left(\frac{1}{K}\right) + (1-\delta_j)p_{j }^{\mathcal{A}_\text{batched}}(w) 
\end{align*}
\end{definition}

In batched settings, treatment assignment probabilities update after each batch of $B$ units rather than after every individual observation, so treatments within a batch are assigned independently (that is, for each unit $i$ in batch $j$, we assign treatment by rolling a $K$-sided dice where each treatment $w \in \mathcal{W}$ has probability $p^{\text{MAD}_\text{batched}}_{j}\left(w \right)$ of appearing, and $p^{\text{MAD}_\text{batched}}_{j}\left(w \right)$ is fixed within the batch $j$).  In fact, we can think of Definition~\ref{bmad} as a more general version of Definition~\ref{mad_def}, where Definition~\ref{mad_def} is just Definition~\ref{bmad} with the batch size $B = 1$ and $K=2$ treatments. 

\section{Theoretical Guarantees of the MAD}\label{sec:theory}
We now provide both inferential and regret guarantees for the MAD. We will present results for the MAD with binary treatments in Definition~\ref{mad_def}, and fully formalize and prove the analogous results for $K\geq2$ treatments and batched assignment settings in Appendix~\ref{sec:k_morethan2} and~\ref{sec:appendix_batched}.

\subsection{Inferential Guarantees}
We first state the assumptions necessary to establish our inferential results.

\begin{assump}[Bounded (Realized) Potential Outcomes]\label{a_bound}
   There exists $M \in \mathbb{R}$ such that $$\limsup_{t \to \infty}|Y_t(w)| \leq M < \infty$$ for all $w \in \mathcal{W}$.
\end{assump}

This assumption is used in existing work on CSs for the ATE \cite[]{howard, ham} and commonly assumed in design-based causal inference settings \cite[]{bojinov_timeseries, bojinov_panel, lei2021regression}. Note, this is an assumption on the \textit{realized} potential outcomes. Although this assumption may seem limiting, the realized outcomes in any real-world experiment are almost always bounded, as the limitations of computing precision guarantee that any realized outcome collected using existing computing resources will be bounded by, \emph{e.g.}, the highest floating point number via IEEE-754 standards. Hence, even if $Y_t(w)$ were drawn from a Gaussian distribution using existing computing resources, the \textit{realized} potential outcomes will never exceed this upper limit on floating point precision. Therefore, we believe that this assumption is satisfied in almost all settings. Additionally, the confidence sequence we provide in Theorem~\ref{thrm1} does not take in $M$ as input, and hence, the value of $M$ need not be known by the manager and does not affect the width of the confidence sequence.

We also require that the average conditional variance of our estimator $\frac{1}{t}\sum_{i=1}^t Var(\hat{{\tau}}_{i} \mid \mathcal{F}_{i} )$ is not vanishing. Specifically, we require that the cumulative conditional variances $\sum_{i=1}^t Var(\hat{{\tau}}_{i} \mid \mathcal{F}_{i} )$ grow at least linearly in $t$.
Define $\Omega(t)$ such that if $f(t) = \Omega(t)$, there exists $k>0$ and $t_0 > 0$ such that for all $t \geq t_0$, $f(t) \geq kt$. 

\begin{assump}[At Least Linear Rate of Cumulative Conditional Variances]\label{a_var2}
$$\sum_{i=1}^t Var(\hat{{\tau}}_{i} \mid \mathcal{F}_{i} ) = \Omega(t).$$
\end{assump}

Assumption~\ref{a_var2} states that the sums of the conditional variances go to infinity at least as fast as a constant rate. Existing works such as \cite{ham} and \cite{waudbysmith_timeuniform} require that $\sum_{i=1}^t Var(\hat{{\tau}}_{i} \mid \mathcal{F}_{i} ) \to \infty$, but do not assume a specific rate. For instance, they allow for the average conditional variance to vanish superlinearly, which would violate Assumption~\ref{a_var2}. Although Assumption~\ref{a_var2} is slightly stronger, this assumption should be satisfied in most realistic experimental settings 
so long as we do not have an adversarial sequence $(Y_i(1), Y_i(0))_{i=1}^\infty$ which would ``cancels out'' the rate of the MAD assignment probabilities. For instance, if 
 there existed some time $\tilde{t}$ such that beyond $\tilde{t}$, all $\{Y_i(1), Y_i(0)\}_{i \geq \tilde{t}}$ are constantly $0$, this assumption would not hold. Of course, such scenarios are unusual in practice and may indicate practical issues with the experiment.
Importantly, this assumption is the only condition we impose on the user-chosen bandit algorithm (the $p^{\mathcal{A}}_{i}(w)$ in $p_{i}^{\text{MAD}}(w)$), and therefore, we expect our result to be valid for \textit{almost any bandit algorithm}.

As is standard in asymptotic notation \citep{de1981asymptotic}, let $\delta_t = \omega(1/t^a)$ denote that $\delta_t t^a \to \infty $, \emph{i.e.}, that $1/\delta_t = o(t^a)$.  Intuitively, we can think of $\delta_t$ as a sequence approaching zero slower than $1/t^a$.

\begin{customthm}{1}\label{thrm1}  Let $(\hat{\tau}_t)_{t=1}^\infty$ be the sequence of random variables where $W_t=w$ with probability $p_{t}^{\text{MAD}}(w)$, as in Definition~\ref{mad_def}, with respect to some treatment assignment algorithm $\mathcal{A}$. Let
$$\hat{V}_t =  \sqrt{\frac{2(\hat{S}_{t}\eta^2 + 1)}{t^2\eta^2}\log\left(\frac{\sqrt{\hat{S}_{t}\eta^2 + 1}}{\alpha} \right)}.$$ Then, under Assumptions~\ref{a_bound} and~\ref{a_var2} and setting $\delta_t = \omega\left(\frac{1}{t^{1/4}}\right)$, $(\hat{\bar{{\tau}}}_{t} \pm \hat{V}_t)$ is a valid $(1-\alpha)$ asymptotic CS for $\bar{\tau}_t$ and $\hat{V}_t \stackrel{a.s.}{\to} 0$.
\end{customthm}

The proof is in Appendix~\ref{sec:theorem1proof}.
Intuitively, Theorem~\ref{thrm1} holds because a Bernoulli design provides this guarantee (set $\delta_t = 1$ above), so a design that stochastically injects a Bernoulli design into a MAB experiment will maintain the same guarantee as long as the experiment does not deviate from the Bernoulli design too quickly, \emph{i.e.}, the rate at which $\delta_t$ approaches $0$ is controlled. 

At a high level, proving the validity of Theorem~\ref{thrm1} proceeds by showing that $\delta_t= \omega\left(\frac{1}{t^{1/4}}\right)$ along with Assumptions~\ref{a_bound} and~\ref{a_var2} ensure that the ATE estimator of Equation~\eqref{tau_hat_i} satisfies a Lindeberg-type uniform integrability (see Lemma~\ref{lemma_a1} in Appendix~\ref{sec:theorem1proof}), thus allowing us to apply the universal asymptotic CS of \cite{waudbysmith_timeuniform} in this setting. The shrinking variance result proceeds by showing that $\delta_t = \omega\left(\frac{1}{t^{1/4}}\right)$ and Assumption~\ref{a_bound} guarantees that $\hat{S}_t \log \hat{S}_t = o(t^2)$ almost surely, and hence, $\hat{V}_t$ is shrinking towards $0$ almost surely. Note, this CS is not a function of $M$ in Assumption~\ref{a_bound}; $M$ is only used to ensure the Lindeberg-type uniform integrability result and to control the asymptotic rate of $\hat{S}_t$. Therefore, $M$ can be set arbitrarily large to satisfy Assumption~\ref{a_bound} without affecting the width of our CS. 

Practically, Theorem~\ref{thrm1} enables managers to perform dynamic experiments using the MAD and, as long as Assumptions~\ref{a_bound} and~\ref{a_var2} hold, be guaranteed both valid and powerful inference. While \cite{ham} and \cite{howard} propose confidence sequences for the ATE that allows for adaptive assignment, their CS requires a probabilistic assignment assumption for their validity guarantees. Both \cite{ham} and \cite{howard} also require Assumption~\ref{a_bound}, so Theorem~\ref{thrm1} removes the probabilistic assignment assumption entirely while only requiring a slightly stronger assumption on the cumulative conditional variances (which should be easily satisfied in most practical settings, as discussed above). Additionally, we prove the novel result that the width of our asymptotic CS is shrinking by deriving the rate of $\hat{S}_t$ for the MAD. While \cite{ham} discusses how their asymptotic CS scales with the variance of the estimator and state that they expect it to shrink in many adaptive settings, they do not prove that $\hat{S}_t$ has the correct rate to ensure a shrinking CS, even with probabilistic assignments.
 
As discussed in Section~\ref{section:background}, managers may desire to define a stopping rule for the experiment, such as when zero is outside of the CS \cite{ham}. The MAD design guarantees that as long as $\bar{\tau}_t$ is truly non-zero in the long-term, such a stopping rule will occur in finite time.

Let $T_{MAD} \coloneqq \inf_{t}\left\{t: 0 \notin (\hat{\bar{\tau}}_t \pm \hat{V}_t) \right\}$, that is, this is the first time $0$ is not within the confidence sequence specified in Theorem~\ref{thrm1}.

\begin{customthm}{2}\label{thrm2} Under the conditions of Theorem \ref{thrm1}, if $\bar{\tau}_t \to c$ for some $|c| > 0$ as $t \to \infty$, then
$$\mathbb{P}\left(T_{MAD} < \infty\right) = 1.$$
\end{customthm}
The proof is provided in Appendix~\ref{sec:theorem2proof}. The assumption that $\bar{\tau}_t \to c$ is satisfied immediately in stationary outcome setting; for example, if we assume $\tau_i = c$ for all $i$ or if the joint potential outcomes are i.i.d. draws for some distribution $\mathcal{D}$ with finite expectation. However, this assumption still allows for non-stationary as long as the long-term behavior of $\bar{\tau}_t$ stabilizes towards some value. For instance, settings where the non-stationarity is limited to the beginning of the experiment then stabilizes as more samples are observed would satisfy this assumption. However, the presence of non-stationarity may invalidate the use of certain stopping rules, like when zero is outside the CS, as indicative of a  ``true'' or long-term positive ATE. Though we provide this result to establish theoretical guarantees in settings when it is appropriate to use $T_{MAD}$ as a stopping rule, in practice, managers should use their best judgment to determine when it is appropriate to stop their specific experiment. For instance, to determine whether $T_{MAD}$ is an appropriate stopping rule, they may wait until a sufficient number of samples has been observed to assess whether the ATE is stabilizing towards some value or exhibiting non-stationarity (in which case, they may want to use a different stopping rule altogether). For example, in our simulations (most of which have $\bar{\tau}_t \to c, |c|>0$), more than 1000 observations generally produces a CS that a tight enough around the true ATE to be reflective of the true behavior.

Importantly, as part of the proof for Theorem~\ref{thrm2}, 
we prove that the CS of Theorem~\ref{thrm1} shrinks around the true ATE.
 Hence, even if there is long-term non-stationarity and $\bar{\tau}_t$ does not stabilize around some constant value, the CS will still shrink around the true ATE with enough samples; we explore this tracking behavior under non-stationarity in the simulations of Appendix~\ref{appendix:nonstationary_mad}. Hence, managers can observe non-stationarity in their experiment and from there, determine a stopping rule based on their specific experiment.


\subsubsection{Inference for $K \geq 2$ Treatment and Batched Assignment Settings}

For any pair of treatments $w, w'$ where a non-zero treatment effect exists, analogous results of Theorem~\ref{thrm1} and Theorem~\ref{thrm2} hold for the MAD with $\geq 2$ treatments for both the standard and batched assignment (Defintion~\ref{bmad}) setting; full statement and proof are in Appendix~\ref{sec:k_morethan2} and~\ref{sec:appendix_batched}.

In particular, settings where $K>2$ showcases the advantages of our anytime-valid guarantees, as the manager can iteratively exclude ``unpromising'' treatments from consideration after observing sufficient inferential evidence, thus reducing harm to the experimental units. For example, assume the manager has $K-1$ total treatments, $w_1, ..., w_{K-1}$ under consideration and one control, $w_0$. At each time step $t$, the manager can compute the CS of Theorem~\ref{thrm1} for each of the $K-1$ treatment-control pairs $(w_j, w_0)$, $j = 1, ..,. K-1$. Then, if the manager observes sufficient evidence of strong treatment effects between certain treatments and the control (\emph{e.g.}, the confidence sequence excludes $0$ and/or seems to be centered around a non-zero value) while others seem to have weak or possibly no treatment effects, the manager can decide to remove these ``weak'' treatments from consideration and continue the experiment with only the most promising treatments.

\subsubsection{Connections to Exact Anytime-Valid Inference}
Since $T_{MAD}$ is a stopping time based on an \emph{asymptotic} CS, a reasonable concern is that the CS at any finite time point $t$ will not have finite-sample error control, and hence, any inferential conclusion drawn based on the level-$\alpha$ asymptotic CS at $T_{MAD}$ may not be statistically valid. However, recalling the definition of an asymptotic CS in Definition~\ref{def:asymp_cs}, there is some underlying exact (\emph{i.e.}, non-asymptotic) CS that the asymptotic CS of Theorem~\ref{thrm1} is converging towards almost surely. We show that, under the same assumptions of Theorem~\ref{thrm2}, this unknown, exact CS is also \emph{guaranteed to have a finite stopping time}. This results implies that, if we have access to the true underlying exact CS, we would make the \emph{same inferential conclusion} as the asymptotic CS, just at a later time point in the experiment.


Formally, define $T^*_{MAD} \coloneqq \inf_{t}\left\{t: 0 \notin (\hat{\bar{\tau}}_t \pm V^*_t) \right\}$ as the analogous stopping time to $T_{MAD}$ for the \emph{exact} anytime-valid CS underlying the asymptotic CS of Theorem~\ref{thrm1}.


\begin{customthm}{3}\label{thrm3}
Under the conditions of Theorem \ref{thrm2}, 
$\mathbb{P}\left(T^*_{MAD} < \infty\right) = 1$. 
\end{customthm}

 The theorem shows that even though the asymptotic CS cannot guarantee exact error control at the stopping time, it will make the same inferential conclusion as its underlying exact counterpart. In particular, we show in the proof of Theorem~\ref{thrm3} that $V^*_t = \hat{V}_t + o(1)$ almost surely for $\delta_t = \omega(1/t^{1/4})$, and hence, the asymptotic CS approaches its exact counterpart at a \emph{constant} rate which does not depend on $t$.  In practice, we recommend running the experiment for a few additional samples after $T_{MAD}$ is observed to check whether the CS for $t>T_{MAD}$ still excludes zero. 

We can also use the MAD to perform exact anytime-valid inference. We prove the same stopping time guarantee for the MAD using the exact anytime-valid CS for ATE estimation provided in Corollary 2 of \cite{howard}. 

Define $\tilde{T}_{MAD} \coloneqq \inf_{t}\left\{t: 0 \notin (\hat{\bar{\tau}}_t \pm u(V_t)/t \right\}$ where $u(V_t)$ is defined as on Corollary 2 of \cite{howard} 
and $V_t = \sum_{i=1}^t (\hat{\tau}_i - \hat{\bar{\tau}}_i)^2$.

\begin{customlemma}{3.1}\label{lemma3.1}
Under the conditions of Theorem \ref{thrm2}, assume that there exists $p_{min} \in (0,1/2)$ such that $p_{t}^{\text{MAD}}(w) \geq \min\{p_{min}, 1-p_{min}\}$, then $\mathbb{P}\left(\tilde{T}_{MAD} < \infty\right) = 1$.

\end{customlemma}

However, using the exact CS has a number of practical limitations. First, to guarantee exact anytime validity and shrinking width, this exact CS requires the MAD to have probabilistic treatment assignments, \emph{i.e.}, that all treatment assignment probabilities are bounded within some $[p_{min}, 1-p_{min}]$. This probabilistic treatment assignment condition greatly restricts the user's choices of $\delta_t$ and the assignment algorithm $\mathcal{A}$, since the treatment assignment probabilities can no longer approach $0$ or $1$ as more samples are observed. The only way to ensure  $p_{t}^{\text{MAD}}(w) \geq \min\{p_{min}, 1-p_{min}\}$ is to either choose a $\delta_t$ which can approach $0$ but an $\mathcal{A}$ which has assignment probabilities bounded away from $0$ and $1$, which would exclude most common adaptive algorithms like Thompson sampling and UCB, or to use any $\mathcal{A}$ but restrict $\delta_t \geq \min\{p_{min}, 1-p_{min}\}$, so the MAD is forced to assign treatment via a Bernoulli design at a greater than $p_{min}$ rate for all time $t$. 

Second, as discussed in previous works \citep{ham, waudbysmith_timeuniform}, exact CSs such as those of \cite{howard} not only require restrictive distributional assumptions on the data, such as known bounds on the estimator and/or its moments, to establish validity but can also generate far less precise inference than asymptotic CSs. In fact, we show in Appendix~\ref{simulation_results_appendix} that the CS of \cite{howard} shrinks towards the true ATE far slower than our asymptotic CS of Theorem~\ref{thrm1} and, thus, requires far more samples to produce precise inference. Importantly, our asymptotic CS tends to maintain the desired coverage and error control even in moderate sample sizes. In fact, we find in Sections~\ref{section:simulations} and~\ref{simulation_results_appendix} that the coverage and error control of our asymptotic CS is \emph{nearly indistinguishable} from that of a Bernoulli design across a wide range of experimental settings.

Finally, the statistics literature has a long history of using asymptotic tests such as the Wald test in finite-sample settings \cite[]{vdv}, where the primary and largely accepted justification used is that such asymptotic tests can well approximate finite-sample settings once a sufficient sample size has been reached, and thus, will provide close-to-exact error control. We show through our empirical results that our asymptotic CS provides essentially exact error control even for small sample sizes. Based on these results, and the inferential congruence guarantee of Theorem~\ref{thrm3}, we believe it is statistically justifiable to use of the asymptotic CS of Theorem~\ref{thrm1} in practice, especially given the scale of most online experiments.

\subsection{Regret Guarantees}\label{sec:regret}
In many applications, managers run adaptive algorithms with the goal of simultaneously minimizing harm and conducting inference, usually with some prioritization between these two goals. Because the MAD incorporates a Bernoulli design into an adaptive assignment algorithm to gain inferential guarantees, a natural next question is: what is lost in terms of regret to ensure valid inference? We show that the regret of the MAD at any time point $t$ is a weighted sum between the regret of the underlying adaptive algorithm $\mathcal{A}$ and the regret of a Bernoulli design, where the weights are given by $\delta_t$. 

Assume binary treatments with the control arm having mean outcome $\mu^0_t$ and the treatment arm having mean outcome $\mu^1_t$. 
Let $Y_t^{\text{obs}}$ be the observed outcome/reward at time $t$. Let $\mathbb{E}_{\mathcal{A}}[Y_t^{\text{obs}}]$, $\mathbb{E}_{\text{Bern}}[Y_t^{\text{obs}}]$, and $\mathbb{E}_{\text{MAD}}[Y_t^{\text{obs}}]$ denote the expected outcome (often referred to as the reward in the reinforcement learning literature) taken with respect to the an adaptive assignment algorithm $\mathcal{A}$, a Bernoulli design, and the MAD using $\mathcal{A}$ as the adaptive algorithm, respectively\footnote{Often, when considering regret, researchers take the expectation with respect to the outcome distribution; this is a straightforward extension of the discussion in this section.}. Let $\mu^*_t =  \max(\mu^1_t, \mu^0_t)$ and define the regret of a design $\pi$ (often referred to as the policy in the reinforcement learning literature) at time $t$ as $R_t(\pi) =  - \mathbb{E}_{\pi}[Y_t^{\text{obs}}]$, where $\pi \in \{\mathcal{A}, \text{Bern}, \text{MAD}\}$.
Hence, the regret of a design $\pi$ is the expected difference in outcome between the outcome-maximizing design (which assigns treatment with probability $1$) and $\pi$. Then,
\begin{align*}
R_t(\text{MAD}) & = \mu^*_t  - \mathbb{E}_{\text{MAD}}[Y_t^{\text{obs}}] ,\\
\\& = \mu^*_t  - \left(\delta_t \mathbb{E}_{\text{Bern}}[Y_t^{\text{obs}}] + (1-\delta_t) \mathbb{E}_{\mathcal{A}}[Y_t^{\text{obs}}]\right),\\
    &= \mu^*_t  - \left(\delta_t \left(\frac{\mu^0_t + \mu^1_t}{2}\right) + (1-\delta_t) \mathbb{E}_\mathcal{A}[Y_{t}^{\text{obs}}]\right),
\end{align*}
and hence,
\begin{align*}  R_t(\text{MAD})- R_t(\mathcal{A}) & =  \mu^*_t  - \left(\delta_t \left(\frac{\mu^0_t + \mu^1_t}{2}\right) + (1-\delta_t)  \mathbb{E}_{\mathcal{A}}\left[Y_t^{\text{obs}}\right]\right) - \left(\mu^*_t -\mathbb{E}_{\mathcal{A}}\left[Y_t^{\text{obs}}\right]\right) \\
    & = \delta_t\left(\mathbb{E}_{\mathcal{A}}\left[Y_t^{\text{obs}}\right]-\frac{\mu^0_t + \mu^1_t}{2} \right)
\end{align*}
At each time point, the difference in regret between the MAD and the adaptive algorithm $\mathcal{A}$ shrinks towards $0$ at a $\delta_t$ rate. Hence, the expected reward of the MAD and the bandit at each time point will asymptotically be the same, i.e., the two algorithms will asymptotically make the same decision at every time point. So, if the bandit achieves low regret, so too will the MAD. But under dependence or non-stationary settings which are adversarial to bandit algorithms, naturally, the bandit will not achieve low regret and MAD will not either. For instance, settings where an optimal arm exists at least asymptotically (i.e., $\mu^*_t \to c $ for some fixed $c > \mu^0_t$ as $t \to \infty$) are more amenable to the bandit correctly identifying the best arm asymptotically, and a bandit algorithm could still be used for effective regret minimization.

However, even in adversarial non-stationary settings which causes the adaptive algorithm to accrue high regret (i.e., the ATE is positive before some $t$, and negative after), the CS generated from the MAD still accurately tracks the true ATE over time; see Appendix~\ref{appendix:nonstationary_mad}. This ``tracking'' highlights the advantages of anytime-valid inference under a design-based perspective, as the MAD enables the experimenter to observe non-stationary behavior in the ATE while maintaining sharp and valid inference on the ATE, even in settings where the bandit algorithm can have poor regret minimization.

In contrast with experimental design works such as \cite{econometrica} which do not allow the user to tune the degree of regret minimization in the algorithm and \cite{simchi}, where this tuning is a function of various hyperparameters which may tradeoff with one another, the regret of the MAD is explicitly controlled by a single user-chosen parameter, $\delta_t$. \footnote{Recall also that \cite{econometrica} and \cite{simchi} each present a specific design for a specific adaptive algorithm, and thus, cannot be used generally for any adaptive algorithm.} Additionally, both these works only characterize the asymptotic behavior of their design's regret while we provide a finite-sample characterization. 


\section{Recommendations for Setting $\delta_t$}\label{section:setting_delta}

Theorems~\ref{thrm1} and~\ref{thrm2} hold for any $\delta_t= \omega(\frac{1}{t^{1/4}})$, therefore, the MAD provides flexibility to select $\delta_t$ based on the specifics of the problem. In this section, we provide a few recommendations of $\delta_t$ based on the manager's priorities.

\textit{Example 1: Prioritizing regret minimization.} Setting $\delta_t = 1/t^{0.24}$  prioritizes the bandit design, and hence regret minimization, while still maintaining validity and the shrinking width of the CS guarantees. 

\textit{Example 2: Prioritizing learning.} Setting $\delta_t = c$, for some $c \in (0, 1]$, maintains a constant exploration rate (\emph{i.e.}, the Bernoulli design), which is valuable when the manager wants to learn the causal effect and stop the experiment as soon as possible. Setting $\delta_i=c<1$ still allows for some limited exploitation, whereas $\delta_i=1$ reduces to a Bernoulli design.  

\textit{Example 3: First prioritize regret minimization, then learning.} Setting  $\delta_t = \max\{\tilde{\delta}_t, c\}$ for some $c \in (0, 1]$, with $\tilde{\delta}_t$ a sequence in $(0, 1]$ prioritizes the bandit design up to a point, then maintain a constant rate of exploration in the long run. For instance, since managers often have a desired stopping point $N$, say after two weeks, they could set $\tilde{\delta}_t =t^{0.24}$ and $ c = 1/N^{0.24}$. Such a $\delta_i$ can be useful in settings where the manager primarily prioritizes regret minimization and, hence, initially chooses an aggressive $\tilde{\delta}_t$ to see if they can achieve statistically significant results before the $N$ samples/time steps. If such results are not achieved by $N$ samples, then the manager will maintain a constant rate of Bernoulli assignment from that point onward to sharpen the inference further.

\textit{Example 4: First prioritize learning, then regret minimization.} Setting $\delta_t = \min\{\tilde{\delta}_t, c\}$ for some $c \in (0, 1]$, with $\tilde{\delta}_t$ a sequence in $(0, 1]$ maintains a constant rate of exploration up to a point, then gradually places more priority on regret minimization from that point onward. In many tech experiments, effect sizes can be small such that early adaptation can be ineffective , as the learning algorithm does not have a sufficiently large sample size to resolve the difference between the two treatments \citep{bojinovgupta}. In such settings, it may be more sample efficient to prioritize inference early, then introduce adaptivity after a certain sample size has been achieved.

Generally, managers can flexibly determine a $\delta_t$ that balances the trade-off between regret minimization and statistical power throughout the experiment. Note also, setting $\delta_t$ as in Examples 2 and 3 is similar to clipping in which the assignment probabilities of a stochastic adaptive assignment algorithm like TS are clipped to never fall below a certain value $p_{\text{min}}>0$ \cite[]{jin2020mots, dima}. For instance, if we set $\delta_t = c$ for some $c>0$, then for a two-armed MAB, TS would choose, at each timepoint $t$, arm $1$ with probability $(1-c)p^{\mathcal{A}}_{t}(1) + c/2$ and arm $0$ with probability $1$ with probability $(1-c)p^{\mathcal{A}}_{t}(0) + c/2$, and hence, the assignment probability is forced to never drop below $c/2$, which is similar to clipping setting $p_{min} = c/2$. Similarly, setting $\delta_t = \max\{\tilde{\delta}_t, c\}$ would be like clipping with a $p_{min}$ that varies with $t$ but is still lower bounded by $c/2$.

\section{Simulation Study}\label{section:simulations}
We now explore the empirical properties of our proposed design through an extensive simulation study. The key takeaway is that the CS of Theorem~\ref{thrm1} using the MAD with an appropriately set $\delta_i$ maintains finite-sample anytime validity while powerfully detecting non-zero ATEs. In this section, we present the results for binary outcomes; Appendix~\ref{appendix:nonstationary_mad} explores non-stationary ATE settings, Appendix~\ref{appendix:stopping_mad} provides further exploration on the regret of the MAD vs. the Bernoulli when using a stopping rule, and Appendix~\ref{simulation_results_appendix} contains the analogous results for Normal, Cauchy, and t-distributed outcome models.

Throughout, we assume $Y_t(1) \stackrel{i.i.d.}{\sim} \text{Bernoulli}(p_1)$ and $Y_t(0)  \stackrel{i.i.d.}{\sim} \text{Bernoulli}(p_0)$ for all $t$, i.e, $\bar{\tau}_t = \bar{\tau} = p_1-p_0$ constantly. To understand how the treatment effect affects the empirical performance of the MAD, we explore three possible outcome distributions: $(p_0, p_1) = \{(0.5, 0.5), (0.6, 0.8), (0.2, 0.8)\}$, so the ATE = $0, 0.2, 0.6$, respectively. 

For the adaptive part of the MAD, we consider the popular Beta-Bernoulli Thompson sampler (TS) initialized with uniform priors on both arms; in the appendix, we also provide results for the UCB algorithm. For $\delta_t$, we evaluate \emph{$\delta_t = \frac{1}{t^{0.24}}$}, which we call the \emph{unclipped MAD}, and \emph{$\delta_t = \max(\frac{1}{t^{0.24}}, 0.2)$},  which we call the \emph{clipped MAD}. We also define two baseline designs: a \textit{Bernoulli design (i.e., $\delta_t = 1$)} and a \textit{Standard Bandit design (i.e., $\delta_t=0$)}. 

For each of the four designs, the CS of Theorem~\ref{thrm1} is computed over $T=10,000$ time steps and $N=100$ independent replicates. Note that since we do not adjust the assignment probabilities for the Standard Bandit Design, there are no validity guarantees for this design. We set $\alpha = 0.05$ for computing the CS throughout. For the $\eta$ parameter of the CS of Theorem~\ref{thrm1}, we set $\eta = \sqrt{\frac{-2\log(\alpha) +\log(-2\log(\alpha)+1)}{t^*}}\approx 0.028$ with $t^* = 10{,}000$; we choose this value following the recommendations in Appendix B.2 of \cite{waudbysmith_timeuniform} on setting $\eta$ to optimize the width of the CS for a specific time $t^*$, where we take $t^*$ as the end of the time horizon we set for this experiment, $T$. 


\subsection{Results}

\begin{figure}
    \centering
\includegraphics[width=\linewidth]{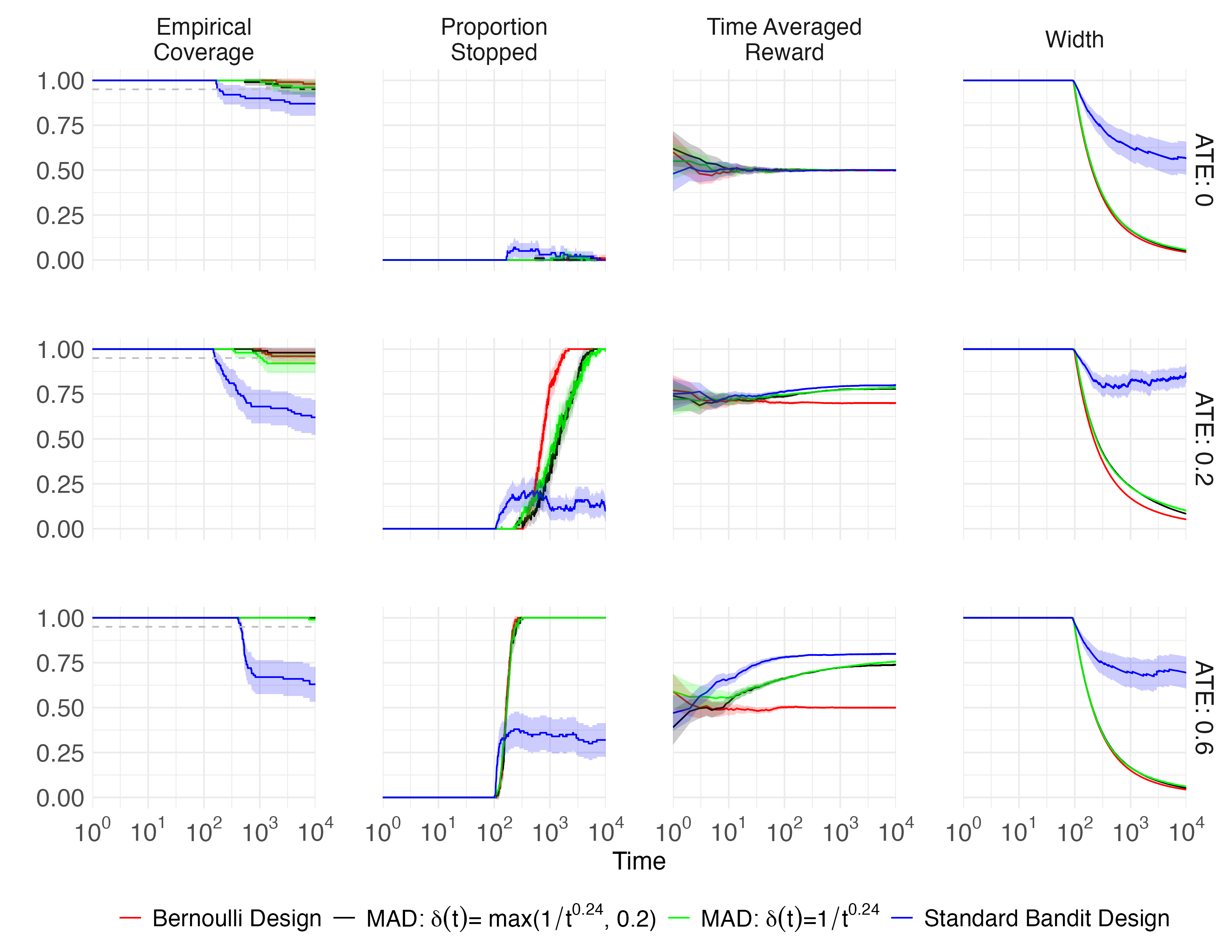}
    \caption{
Empirical coverage, proportion stopped, time averaged reward, and width of the CS of Theorem~\ref{thrm1} across $N=100$ random seeds for different experimental designs under a two-armed bandit setting with a Bernoulli outcome model using TS as the bandit algorithm; see Section~\ref{section:simulations} for full description of the experimental setting and each metric. The dashed grey line represents $1-\alpha$. Error bands depict $\pm 2$ SEs.}
    \label{fig:all_metrics_bern}
\end{figure}

Figure~\ref{fig:all_metrics_bern} depicts various metrics of the designs across the different ATE values for the Bernoulli outcome experiments. The first column of Figure~\ref{fig:all_metrics_bern} shows the empirical coverage at each time point, which we calculate as the proportion of CSs up to and including $t$ for which the true ATE is inside the CS at each time point, \emph{i.e.}, we compute $\frac{1}{t}\sum_{i=1}^t \mathbbm{1}\left\{ \bar{\tau} \in \left(\hat{\bar{\tau}}_i \pm \hat{V}_i\right)\right\}$. We use this empirical coverage to estimate our anytime-valid error control as given in Definition~\ref{def:cs}. The clipped and unclipped MADs have $1-\alpha$ error control across all time as desired, even when $t$ is small, suggesting that the CS of Theorem~\ref{thrm1} can maintain proper finite-sample anytime validity even though it is an asymptotic CS. Notably, the Standard Bandit design does not maintain the proper coverage when the ATE is non-zero, exhibiting how naively using a Standard Bandit design cannot ensure anytime validity. When the ATE is zero, the Standard Bandit design maintains the proper coverage intuitively because the bandit algorithm never strongly favors one arm over the other since the two arms are indistinguishable.

The second column of Figure~\ref{fig:all_metrics_bern} explores the performance of the most common stopping rule: stopping the first time $0$ is outside the CS. The results shows that the proportion of replicates that do not contain $0$ at time $t$. The MAD excludes $0$ with much fewer samples than the Standard Bandit design for both choices of $\delta_t$, essentially matching that of the Bernoulli design when the ATE is large. Notably, when the ATE is non-zero, a significant proportion of the CSs for the Standard Bandit design never exclude $0$ by the end of the time horizon.

The third column of Figure~\ref{fig:all_metrics_bern} shows the time-average reward of each of the designs, which we compute as $\frac{1}{t} \sum_{i=1}^t Y_i^{\text{obs}}$ for each $t$. As expected, based on the results of Section~\ref{sec:regret}, the reward of the MAD quickly approaches that of the Standard Bandit Design. Our results show that the MAD maintains anytime validity even in finite samples and has a CS width and stopping time behavior similar to the Bernoulli design while achieving a reward closer to the Standard Bandit design. Note also that the clipped MAD can have a slightly smaller width over time but also a somewhat smaller reward than the unclipped MAD because the clipped MAD enforces greater exploration rather than exploitation. We expect that with a more aggressive clipping rate, this difference will be starker.

Finally, the last column of Figure~\ref{fig:all_metrics_bern} shows that the width of the CSs under both the clipped and unclipped MADs shrinks much faster than the Standard Bandit Design, nearly matching that of the Bernoulli design. Note, the Standard Bandit design tends to produce wider CSs than the MAD but still lacks the correct coverage when the ATE is non-zero. Intuitively, this result occurs because as time increases, the assignment probability for arm $0$ becomes negligibly small so that the algorithm essentially only draws from arm $1$. Hence, the ATE estimator $\hat{\bar{\tau}}_t$ as in Equation~\eqref{hatbar_tau} is computed mostly only using observations from arm $1$. This skewing is worse when the ATE is larger because the assignment probability for arm $0$ more quickly approaches $0$.

Using the clipped MAD, we also perform a subset of the Bernoulli outcome experiments using the non-asymptotic CS of \cite{howard}; see Appendix~\ref{simulation_results_appendix}. As expected, the CS of \cite{howard} is generally much wider than our asymptotic CS and tends to overcover. As our asymptotic CS maintains finite-sample anytime validity across a wide range of outcomes models and sample sizes and has a smaller width, we recommend using our asymptotic CS in practice.

Notably, the MAD mitigates risk both by enabling anytime-valid inference while allowing for adaptive treatment assignments. However, the Bernoulli design provides the best inferential precision, and hence, will generally satisfy a stopping rule (like zero being outside the CS) faster than the MAD. A natural question is whether experiments run with the MAD will still achieve smaller regret than if it had been run with a Bernoulli design that may have stopped sooner. We explore this question further in the simulations of Appendix~\ref{appendix:stopping_mad}, and find that the MAD will generally achieve higher reward than the Bernoulli design, even though the Bernoulli design tends to stop sooner, where the improvement in reward varies with the size of the ATE.
\if1\blind
{
The script used to reproduce all simulation results is provided at \url{https://github.com/biyonka/mixture_adaptive_design}.
} \fi

\if0\blind
{
The script used to reproduce all simulation results is provided at [Url removed for anonymity].
}\fi
\section{Conclusion}\label{section:conclusion}
We propose the Mixture Adaptive Design (MAD), an experimental design for valid and
powerful anytime-valid, design-based causal inference in adaptive experiments. By controlling the rate of a bandit algorithm's adaptive assignment probabilities via ``mixing'' with a Bernoulli design, the MAD provides a confidence sequence for the ATE that is asymptotically valid with width guaranteed to shrink to zero for nearly any choice of bandit algorithm. As existing approaches do not guarantee validity in general bandit settings, the MAD provides a practical tool for online and anytime-validcausal inference in both standard and batched bandit settings.

\if1\blind
{
\section{Acknowledgements}
We thank Nathan Cheng, P. Aronow, Panos Toulis, and the entire Data Science and AI Operations $D^3$ Lab at HBS for their helpful discussions and feedback.


} \fi
\if0\blind
{} \fi
\bibliographystyle{agsm}
\spacingset{1}
\bibliography{bibtex}
\bigskip

\clearpage
\begin{center}
{\large\bf SUPPLEMENTARY MATERIAL}
\end{center}
\appendix
\spacingset{1}
\section{Proof of Theorem 1}\label{sec:theorem1proof}
To prove Theorem 1, we must first show that $(\hat{\tau}_{i})_{i=1}^\infty$ satisfies a Lindeberg-type uniform integrability condition (Condition L-2 of \cite{waudbysmith_timeuniform}).


\begin{customlemma}{A.1}\label{lemma_a1}
Let $(\hat{\tau}_t)_{t=1}^\infty$ be a sequence of random variables where 
 $W_t=w$ with probability $p_{t}^{\text{MAD}}(w)$ for $w \in \{0, 1\}$ and $\delta_t \in (0, 1]$ such that $\delta_t = \omega\left(\frac{1}{t^{1/4}}\right)$. Assume Assumptions~\ref{a_bound} and~\ref{a_var2} hold. Then, $(\hat{\tau}_t)_{i=1}^\infty$ satisfies a Lindeberg-type uniform integrability condition, i.e., there exists $\kappa \in (0, 1)$ such that
$$\sum_{t=1}^\infty \frac{\mathbb{E}\left[(\hat{{\tau}}_{t}-{\tau}_{t})^2\mathbbm{1}\{(\hat{{\tau}}_{t}-{\tau}_{t})^2 > (B_{t})^\kappa \}\right]}{ (B_{t})^\kappa} < \infty \text{ a.s.}$$
where $B_{t} = \sum_{i=1}^t Var(\hat{{\tau}}_{i} \mid \mathcal{F}_{i} )$.
\end{customlemma}

\begin{proof}

By Assumption~\ref{a_bound},
\begin{align*}
   \left|\hat{{\tau}}_{t} \right|  = \left| \frac{Y_t(1)\mathbbm{1}\{ W_t = 1\}}{p^{\text{MAD}}_{t}(1)} - \frac{Y_t(0)\mathbbm{1}\{ W_t= 0\}}{p^{\text{MAD}}_{t}(0)}\right| \leq \frac{2M}{\min(p^{\text{MAD}}_{t}(0), p^{\text{MAD}}_{t}(1))}
\end{align*}
and $|{\tau}_t| \leq 2M$ for all $t$.

Hence,
\begin{align*}
(\hat{{\tau}}_{t}-{\tau}_{t})^2 &\leq
\left(\frac{2M}{\min(p^{\text{MAD}}_{t}(0), p^{\text{MAD}}_{t}(1))}\right)^2 + (2M)^2 + 2\left(\frac{2M}{\min(p^{\text{MAD}}_{t}(0), p^{\text{MAD}}_{t}(1))}\right)2M \\
& \leq \frac{24M^2}{(\min(p^{\text{MAD}}_{t}(0), p^{\text{MAD}}_{t}(1)))^2}
\end{align*}
 First, note that for all $w$, $p^{\text{MAD}}_{t}(w) \geq \delta_t (1/2)$ and so $\frac{1}{p^{\text{MAD}}_{t}(w)} = o(t^{1/4})$. 
  So, $(\hat{{\tau}}_{t}-{\tau}_{t})^2 = o(t^{1/2}$) almost surely.


By Assumption~\ref{a_var2}, $B_{t} = \Omega(t)$. Therefore, $B_{t}^\kappa = \Omega(t^{\kappa})$. Set $\kappa > 1/2$. Then, there exists some $\tilde{t}$ such that for all $t \geq \tilde{t}$,
$B_{t}^\kappa > (\hat{{\tau}}_{t}-{\tau}_{t})^2$ a.s..
Hence, for all $t \geq \tilde{t}$, $\mathbbm{1}\{(\hat{{\tau}}_{t}-{\tau}_{t})^2 > (B_t)^\kappa \} = 0$, and so,
$$\sum_{i=1}^t \frac{\mathbb{E}\left[(\hat{{\tau}}_{t}-{\tau}_{t})^2\mathbbm{1}\{(\hat{{\tau}}_{t}-{\tau}_{t})^2 > (B_t)^\kappa \}\right]}{ (B_t)^\kappa} < \infty \text{ a.s.}.$$

\end{proof}

\begin{customthm}{1}
 Let $(\hat{\tau}_t)_{t=1}^\infty$ be the sequence of random variables where $W_t=w$ with probability $p_{t}^{\text{MAD}}(w)$, as in Definition~\ref{mad_def}, with respect to some treatment assignment algorithm $\mathcal{A}$. Let
$$\hat{V}_t =  \sqrt{\frac{2(\hat{S}_{t}\eta^2 + 1)}{t^2\eta^2}\log\left(\frac{\sqrt{\hat{S}_{t}\eta^2 + 1}}{\alpha} \right)}.$$ Then, under Assumptions~\ref{a_bound} and~\ref{a_var2} and setting $\delta_t = \omega\left(\frac{1}{t^{1/4}}\right)$, $(\hat{\bar{{\tau}}}_{t} \pm \hat{V}_t)$ is a valid $(1-\alpha)$ asymptotic CS for $\bar{\tau}_t$ and $\hat{V}_t \stackrel{a.s.}{\to} 0$.
\end{customthm}

\begin{proof}

By Assumptions~\ref{a_bound} and~\ref{a_var2} imply that Lemma~\ref{lemma_a1} holds.
Together, Lemma~\ref{lemma_a1} and Assumption~\ref{a_var2} satisfy Conditions L-1 and L-2 of Theorem 2.5 in \cite{waudbysmith_timeuniform}. Let $B_t = \sum_{i=1}^t Var(\hat{{\tau}}_{i} \mid \mathcal{F}_{i} )$ and
$$V^*_t =  \sqrt{\frac{2(B_{t}\eta^2 + 1)}{t^2\eta^2}\log\left(\frac{\sqrt{B_{t}\eta^2 + 1}}{\alpha} \right)}$$

By Steps 1 and 2 of the proof of Theorem 2.5 in \cite{waudbysmith_timeuniform}, $(\hat{\bar{{\tau}}}_{t} \pm V^*_t)$ is a valid $(1-\alpha)$asymptotic CS.

 As noted in Equation~\eqref{eq:variance}, 
\begin{align} Var(\hat{\tau}_{i} \mid \mathcal{F}_{i}) \leq \sigma^2_{i}\text{, where } \sigma^2_{i}\coloneqq \frac{Y_i(1)^2}{p_{i}(1)} + \frac{Y_i(0)^2}{p_{i}(0)}.
\end{align}

Hence, 
$(\hat{\bar{{\tau}}}_{t} \pm \tilde{V}_t)$ where 
$$\tilde{V}_t =  \sqrt{\frac{2(S_{t}\eta^2 + 1)}{t^2\eta^2}\log\left(\frac{\sqrt{S_{t}\eta^2 + 1}}{\alpha} \right)}$$
 is still a valid $(1-\alpha)$asymptotic CS, where $S_t = \sum_{i=1}^t \sigma_i^2$. This holds because replacing $B_t$ with $S_{t}$ only makes the CS wider since $S_t \geq B_t$.

As noted in Equation~\eqref{eq:variance_est}, an unbiased estimator for $\sigma^2_{i}$ is 
\begin{align}
    \hat{\sigma}^2_{i}\coloneqq \frac{Y_i(1)^2 \mathbbm{1}\{W_i = 1\}}{p^2_{i}(1)} + \frac{Y_i(0)^2\mathbbm{1}\{W_i = 0\}}{p^2_{i}(0)}.
\end{align}
and we define $\hat{S}_t = \sum_{i=1}^t \hat{\sigma}_i^2$.

To establish the validity of the CS in Theorem 1, we must show that $\frac{1}{t}\hat{S}_t - \frac{1}{t}S_t \stackrel{a.s.}{\to} 0$. Then, Condition L-3 of Theorem 2.5 of \cite{waudbysmith_timeuniform} is satisfied and we can conclude that 
$(\hat{\bar{{\tau}}}_{t} \pm \hat{V}_t)$ where 
$$\hat{V}_t =  \sqrt{\frac{2(\hat{S}_{t}\eta^2 + 1)}{t^2\eta^2}\log\left(\frac{\sqrt{\hat{S}_{t}\eta^2 + 1}}{\alpha} \right)}$$
 is still a valid $(1-\alpha)$ asymptotic CS.

First, note that 
\begin{align*}
    (\hat{\sigma}_i^2)^2 &\leq \left(\frac{M^2}{p_{i}^{MAD}(1)^2}+\frac{M^2}{p_{i}^{MAD}(0)^2}\right)^2\\
    & = \frac{M^4}{p_{i}^{MAD}(1)^4} + \frac{M^4}{p_{i}^{MAD}(0)^4} + 2\frac{M^2}{p_{i}^{MAD}(1)^2p_{i}^{MAD}(0)^2}\\
    & \leq \frac{M^4}{(\delta_i/2)^4} + \frac{M^4}{(\delta_i/2)^4} + 2\frac{M^2}{(\delta_i/2)^4}\\
    & = M^4\left(2^4 \frac{1}{\delta_i^4} + 2^4 \frac{1}{\delta_i^4} + 2^5 \frac{1}{\delta_i^4} \right)\\
    & = o(i)
\end{align*}
where the last line follows because $\frac{1}{\delta_i} = o\left(i^{1/4}\right)$.
Define $X_i = \hat{\sigma}_i^2 - \sigma_i^2$ and note that $X_i$ is a martingale difference sequence. Hence,
\begin{align*}
    \mathbb{E}[X_i] &= E[(\hat{\sigma}_i^2)^2]- (\sigma_i^2)^2\\
    & \leq E[(\hat{\sigma}_i^2)^2]\\
    & = o(i)
\end{align*}

So, $\frac{\mathbb{E}[X_i^2]}{i^2} = \frac{o(i)}{i^2}$ and hence, $\sum_{i=1}^\infty \frac{\mathbb{E}[X_i^2]}{i^2} < \infty$. For instance, if $\delta_i = \frac{1}{i^a}$ for some $0 \leq a < 1/4$, then there exists a constant $C < \infty$ and $i_0>0$ such that for all $i > i_0$,
$\frac{\mathbb{E}[X_i^2]}{i^2} \leq Ci^{4a-2}$, and since $4a-2 < -1$, $\sum_{i=1}^\infty \frac{\mathbb{E}[X_i^2]}{i^2} \leq C \sum_{i=1}^\infty i^{4a-2} < \infty$ by the p-series test.

Then, by the SLLN for martingale difference sequences (Theorem 1 of \cite[]{csorgHo1968strong}), we can conclude that
$$\frac{1}{t} \sum_{i=1}^t X_i \stackrel{a.s.}{\to} 0.$$

Hence, Condition L-3 of \cite{waudbysmith_timeuniform} is satisfied and by Step 3 of the proof for Theorem 2.5 of \cite{waudbysmith_timeuniform}, we can conclude that 
$(\hat{\bar{{\tau}}}_{t} \pm \hat{V}_t)$ where 
$$\hat{V}_t =  \sqrt{\frac{2(\hat{S}_{t}\eta^2 + 1)}{t^2\eta^2}\log\left(\frac{\sqrt{\hat{S}_{t}\eta^2 + 1}}{\alpha} \right)}$$
 is still a valid $(1-\alpha)$ asymptotic CS.

To show $\hat{V}_t \stackrel{a.s.}{\to} 0$, note that 
\begin{align*}
    \hat{\sigma}_i^2 &\leq \frac{M^2}{p_{i}^{MAD}(1)^2}+\frac{M^2}{p_{i}^{MAD}(0)^2}\\
    & \leq  \frac{M^2}{(\delta_i/2)^2}+\frac{M^2}{(\delta_i/2)^2}\\
        & = \frac{4M^2}{\delta_i^2}+\frac{4M^2}{\delta_i^2}\\
    & = 8M^2 \frac{1}{\delta_i^2}
\end{align*}
Hence, $\hat{S}_t \leq 8M^2 \sum_{i=1}^t 1/\delta_i^2 = 8M^2 \sum_{i=1}^t o(i^{1/2}) $ a.s..

 Therefore, there exists positive real numbers $N$ and $x_0$ such that for all $t \geq x_0$, $\hat{S}_{t} \leq N \sum_{i=1}^t i^{1/2} < N \sum_{i=1}^t t^{1/2} = N t^{3/2}$, allowing us to conclude that 
$\hat{S}_t = O(t^{\frac{3}{2}})$.
 
 Next, we show that $\log\left(\hat{S}_t\right) = o(t^{1/2})$ a.s.. For $t \geq x_0$,
\begin{align*}
    \frac{\log(\hat{S}_{t})}{t^{1/2}} \leq     \frac{\log(Nt^{3/2})}{t^{1/2}} =    \frac{\log(N) + (3/2)\log(t))}{t^{1/2}} 
\end{align*}
and $\frac{\log(N) + (3/2)\log(t))}{t^{1/2}}  \to 0$ as $t \to \infty$.  Therefore, $\log(\hat{S}_{t}) = o(t^{1/2})$ a.s. and we conclude that
\begin{align}\label{thrm5_eqref1}
    \hat{S}_t \log(\hat{S}_t) = o(t^2) \hspace{0.5em} \text{ a.s.}
\end{align}
Hence, $
\hat{V}_t = \frac{2(\hat{S}_{t}\eta^2 + 1)}{t^2\eta^2}\log\left(\frac{\sqrt{\hat{S}_{t}\eta^2 + 1}}{\alpha} \right) = o(1)$ a.s..

\end{proof}
Note, if $1/\delta_i$ was increasing at a rate faster than $i^{1/4}$ asymptotically, we are not guaranteed that  $\hat{V}_t = o(1)$ a.s..

\section{Proof of Theorem 2}\label{sec:theorem2proof}
\begin{customthm}{2}  Under the conditions of Theorem 1,
assume $\bar{\tau}_t \to c$ for some $|c| > 0$ as $i \to \infty$.
 Then,
$$\mathbb{P}\left(T_{MAD} < \infty\right) = 1.$$
\end{customthm}

\begin{proof}

    We will first show that $\frac{1}{t}\sum_{i=1}^t \hat{\tau}_{i} \stackrel{a.s.}{\to}  c $.  Let $u_i \coloneqq \hat{\tau}_{i} -\tau_{i}$.

    Note that 
    \begin{align*}
        \mathbb{E}[u_i^2] &=  \mathbb{E}[\hat{\tau}_i^2]-\tau_i^2\\
        &\leq  \mathbb{E}[\hat{\tau}_i^2]
    \end{align*}
    where
    \begin{align*}
         \hat{\tau}_i^2 &= \left(\frac{Y_i(1)\mathbbm{1}\{W_i = 1\}}{p_{i}^{MAD}(1)} - \frac{Y_i(0)\mathbbm{1}\{W_i = 0\}}{p_{i}^{MAD}(0)}\right)^2\\
        & \leq \frac{Y_i(1)^2}{(\delta_i/2)^2} + \frac{Y_i(0)^2}{(\delta_i/2)^2}\\
        & = 8M^2 \frac{1}{\delta_i^2}\\
        & = 8M^2 o(i^{1/2})
    \end{align*}

    Hence,
\begin{align*}
    \sum_{i=1}^ t \frac{\mathbb{E}[u_i^2]}{i^2} \leq \sum_{i=1}^ t \frac{8M^2 o(i^{1/2})}{i^2} < \infty
\end{align*}
   by the p-series test.

Hence, by the Strong Law of Large Numbers for martingale difference sequences (Theorem 1 of \cite[]{csorgHo1968strong}), we can conclude that
\begin{align}\label{thrm6_eqref1}
    \frac{1}{t} \sum_{i=1}^t u_i \stackrel{a.s.}{\to} 0.
\end{align}

By Theorem~\ref{thrm1}, we have that $\hat{V}_t \stackrel{a.s.}{\to} 0$.

Therefore, applying Slutsky's Theorem, we conclude that
\begin{align}\label{thrm5_eqref2}
    \hat{\bar{{\tau}}}_{t}   +\hat{V}_t \stackrel{a.s.}{\to} c,
\end{align}
and
\begin{align}\label{thrm5_eqref3}
    \hat{\bar{{\tau}}}_{t}   - \hat{V}_t \stackrel{a.s.}{\to} c.
\end{align}
So,
\begin{align*}
&\mathbb{P}\left(0 \notin \left(\hat{\bar{{\tau}}}_{t }   \pm \hat{V}_t\right) \right)\\
&= \mathbb{P}\left(0 <   \hat{\bar{{\tau}}}_{t }  - \hat{V}_t\right) + \mathbb{P}\left(0 > \hat{\bar{{\tau}}}_{t }    +\hat{V}_t \right)\\
& \to 1,
\end{align*}
where the last line follows because $\mathbb{P}\left(0 > \hat{\bar{{\tau}}}_{t}  +\hat{V}_t\right) \to 0$ and $\mathbb{P}\left(0 <  \hat{\bar{{\tau}}}_{t}   -\hat{V}_t\right) \to 1$.

Let $p_t = \mathbb{P}\left(0 \in (\hat{\bar{{\tau}}}_{t}  \pm \hat{V}_t)\right)$. Since $p_t \to 0$ as $t \to \infty$, there exists a subsequence $t_k$ such that $p_{t_k} \leq \frac{1}{k^2}$, for all $k \geq 1$. Let $A_{t_k}$ be the event $\{0 \in (\frac{1}{t_k} \sum_{i=1}^{t_k} \hat{\tau}_{i} \pm \hat{V}_{t_k})\}$. Then, $\sum_{k=1}^\infty \mathbb{P}\left(A_{t_k}\right) < \infty$, and by the Borel-Cantelli lemma,
\begin{align}\label{thrm5_eqref4}
\mathbb{P}\left(\limsup_{k \to \infty} A_{t_k}\right) = 0.
\end{align}

Hence,
\begin{align}\label{thrm5_eqref5}
    \mathbb{P}\left(\title T_{MAD} = \infty\right) \leq \mathbb{P}\left(\limsup_{k \to \infty} A_{t_k}\right) = 0.
\end{align}
\end{proof}

\section{The MAD for $K\geq2$ Treatments}\label{sec:k_morethan2}
We first formalize the problem setting for $K\geq 2$ treatments, which we call the \emph{Generalized Mixture Adaptive Design}, and prove more general versions of Theorems~\ref{thrm1} and Theorem~\ref{thrm2} for $K \geq 2$ treatments. Assume $W_t \in \mathcal{W} \coloneqq \{0, ..., K-1\}$. 

Let $\mathcal{F}_{t, n}$ be the sigma-algebra that contains all  potential outcomes $\{(Y_i(w))_{w \in \mathcal{W}}\}_{i=1}^t$ and all observed data $\{W_i, Y_i\}_{i=1}^n$ where $n \leq t$. 

As before, let the assignment probabilities for any user-chosen adaptive algorithm be denoted as $p_{i}^{\mathcal{A}}(w) = \mathbb{P}(W_t = w \mid H_{t-1})$ where $H_{t-1} = \{W_i, Y_i\}_{i=1}^{t-1}$. Hence, the  Generalized Mixture Adaptive Design has assignment probabilities $p^{\text{MAD}}_{i}(w) = \delta_i \frac{1}{K} + (1-\delta_i)p_{i}^{\mathcal{A}}(w)$, for all $w \in \mathcal{W}$. 

For any pair of treatments $w, w' \in \mathcal{W}$, let $\tau_i(w, w') = Y_i(w)-Y_i(w')$ and define the Average Treatment Effect (ATE) between $w$ and $w'$ up to $t$ as $\bar{\tau}_t(w, w') \coloneqq \frac{1}{t}\sum_{i=1}^t \tau_i(w, w')$. So, using the Generalized Mixture Adaptive Design, our corresponding estimator for the ATE is:
\begin{align*}
    \hat{\bar{\tau}}_t(w, w') = \frac{1}{t}\sum_{i=1}^t \hat{\tau_i}(w, w'),
\end{align*}
where $   \hat{\tau_i}(w, w') \coloneqq \frac{\mathbbm{1}\{W_i = w\}Y_i(w)}{p^{\text{MAD}}_{i}(w)} - \frac{\mathbbm{1}\{W_i = w'\}Y_i(w')}{p^{MAD}_{i}(w')}$. We also have the corresponding upper bound on the variance:
\begin{align*}
Var(\hat{\tau}_{i}(w, w') \mid \mathcal{F}_{i}) \leq \sigma^2_{i}(w, w') \text{, where } \sigma^2_{i}(w, w') \coloneqq \frac{Y_i(w)^2}{p^{\text{MAD}}_{i}(w)} + \frac{Y_i(w')^2}{p^{MAD}_{i}(w')},
\end{align*}
and the analogous unbiased estimator of $\sigma^2_{i}(w, w')$:
\begin{align*}
    \hat{\sigma}^2_{i}(w, w') \coloneqq \frac{Y_i(w)^2 \mathbbm{1}\{W_i = w\}}{(p^{\text{MAD}}_{i}(w))^2} + \frac{Y_i(w')^2\mathbbm{1}\{W_i = w'\}}{(p^{MAD}_{i}(w'))^2}.
\end{align*}

Finally, let $S_{t}(w, w') \coloneqq \sum_{i=1}^t \sigma_{i}^2(w, w')$ and $\hat{S}_{t}(w, w') \coloneqq \sum_{i=1}^t \hat{\sigma}_{i}^2(w, w')$.

Define $T_{MAD}(w, w') \coloneqq \inf_{t}\left\{t: 0 \notin (\hat{\bar{\tau}}_t(w, w') \pm \hat{V}_t(w, w')) \right\}$ where the treatment assignments are generated via the Generalized Mixture Adaptive Design.

Though an analogous result of Theorem 1 for this setting follows almost directly from the fact that we still have $1/\delta_i = o(i^{1/4})$, we provide a full statement of the result and its proof here for completeness.

For precision, we need to state our cumulative conditional variance condition for the $K \geq 2$ setting. 
\begin{assump}[At Least Linear Rate of Cumulative Conditional Variances for All Pairs of Treatments]\label{a_var3}
For all $w, w' \in \mathcal{W}$,
$\sum_{i=1}^t Var(\hat{{\tau}}_{i}(w, w') \mid \mathcal{F}_{i} ) = \Omega(t).$
\end{assump}


We also establish the following lemma, which we will use in the proof of Theorem~\ref{thrm1star}.

\begin{customlemma}{A.2}\label{lemma_a2} Let $w, w' \in \{0, 1, ..., K-1\}$.
Let $\{\hat{\tau}_t(w, w')\}_{t=1}^\infty$ be a sequence of random variables where 
 $W_t=w$ with probability $p^{\text{MAD}}_{t}(w) = \delta_t \frac{1}{K} + (1-\delta_t)p_{t}^{\mathcal{A}}(w)$, for all $w \in \mathcal{W}$ and $\delta_t \in (0, 1]$ such that $\delta_t = \omega\left(\frac{1}{t^{1/4}}\right)$. Assume Assumptions~\ref{a_bound} and~\ref{a_var3} hold. Then, $\{\hat{\tau}_t(w, w')\}_{i=1}^\infty$ satisfies the Lindeberg-type uniform integrability condition of \cite{waudbysmith_timeuniform}, \emph{i.e.}, there exists $\kappa \in (0, 1)$ such that
$$\sum_{t=1}^\infty \frac{\mathbb{E}\left[(\hat{{\tau}}_{t}(w, w')-{\tau}_{t}(w, w'))^2\mathbbm{1}\{(\hat{{\tau}}_{t}(w, w')-{\tau}_{t}(w, w'))^2 > (B_{t}(w, w'))^\kappa \}\right]}{ (B_{t}(w, w'))^\kappa} < \infty \text{ a.s.}$$
where $B_{t}(w, w') = \sum_{i=1}^t Var(\hat{{\tau}}_{i}(w, w') \mid \mathcal{F}_{i} )$.
\end{customlemma}

\begin{proof}

By Assumption~\ref{a_bound}, for all $t, w, w'$,
\begin{align*}
   \left|\hat{{\tau}}_{t}(w, w') \right|  = \left| \frac{Y_t(w)\mathbbm{1}\{ W_t = w\}}{p^{\text{MAD}}_{t}(w)} - \frac{Y_t(w')\mathbbm{1}\{ W_t= w'\}}{p^{\text{MAD}}_{t}(w')}\right| \leq \frac{2M}{\min(p^{\text{MAD}}_{t}(w'), p^{\text{MAD}}_{t}(w))}
\end{align*}
and $|{\tau}_t(w, w')| \leq 2M$.

Hence,
\begin{align*}
&(\hat{{\tau}}_{t}(w, w')-{\tau}_{t}(w, w'))^2 \\
&\leq
\left(\frac{2M}{\min(p^{\text{MAD}}_{t}(w), p^{\text{MAD}}_{t}(w'))}\right)^2 + (2M)^2 + 2\left(\frac{2M}{\min(p^{\text{MAD}}_{t}(w), p^{\text{MAD}}_{t}(w'))}\right)2M \\
& \leq \frac{24M^2}{(\min(p^{\text{MAD}}_{t}(w), p^{\text{MAD}}_{t}(w')))^2}.
\end{align*}
 First, note that for all $w \in \mathcal{W}$, $p^{\text{MAD}}_{t}(w) \geq \delta_t (1/K)$ and so $\frac{1}{p^{\text{MAD}}_{t}(w)} = o(t^{1/4})$. 
  So, $(\hat{{\tau}}_{t}(w, w')-{\tau}_{t}(w, w'))^2= o(t^{1/2})$ almost surely.


By Assumption~\ref{a_var2}, $B_{t}(w, w') = \Omega(t)$. Therefore, $B_{t}^\kappa(w, w') = \Omega(t^{\kappa})$. Set $\kappa > 1/2$. Then, there exists some $\tilde{t}$ such that for all $t \geq \tilde{t}$,
$B_{t}^\kappa(w, w') > (\hat{{\tau}}_{t}(w, w')-{\tau}_{t}(w, w'))^2$ a.s..
Hence, for all $t \geq \tilde{t}$, $\mathbbm{1}\{(\hat{{\tau}}_{t}(w, w')-{\tau}_{t}(w, w'))^2 > (B_t)^\kappa(w, w') \} = 0$, and so,
$$\sum_{t=1}^\infty \frac{\mathbb{E}\left[(\hat{{\tau}}_{t}(w, w')-{\tau}_{t}(w, w'))^2\mathbbm{1}\{(\hat{{\tau}}_{t}(w, w')-{\tau}_{t}(w, w'))^2 > (B_{t}(w, w'))^\kappa \}\right]}{ (B_{t}(w, w'))^\kappa} < \infty \text{ a.s.}$$
\end{proof}

\begin{customthm}{1$^*$}\label{thrm1star} 
For $w, w' \in \{0, 1, ..., K-1\}$, let $(\hat{\tau}_t(w, w'))_{t=1}^\infty$ be the sequence of random variables where $W_t=w$ with probability $p_{t}^{\text{MAD}}(w) = \frac{1}{K}\delta_t + (1-\delta_t)p^{\mathcal{A}}_{t}(w)$, and $\delta_t \in (0, 1]$ such that  $\delta_t = \omega\left(\frac{1}{t^{1/4}}\right)$.
Assume Assumptions~\ref{a_bound} and~\ref{a_var3} hold. Then $(\hat{\bar{{\tau}}}_{t}(w, w') \pm \hat{V}_t(w, w'))$ where 
$$\hat{V}_t(w, w') =  \sqrt{\frac{2(\hat{S}_{t}(w, w')\eta^2 + 1)}{t^2\eta^2}\log\left(\frac{\sqrt{\hat{S}_{t}(w, w')\eta^2 + 1}}{\alpha} \right)}$$
 is a valid $(1-\alpha)$ asymptotic CS for $\bar{\tau}_t(w, w')$ and $\hat{V}_t(w, w') \stackrel{a.s.}{\to} 0$.
\end{customthm}

\begin{proof}
By Assumptions~\ref{a_bound} and~\ref{a_var3} imply that Lemma~\ref{lemma_a2} holds.
Lemma~\ref{lemma_a2} and Assumption~\ref{a_var3} satisfy Conditions L-1 and L-2 of Theorem 2.5 in \cite{waudbysmith_timeuniform}, so, by Steps 1 and 2 of the proof of Theorem 2.5 in \cite{waudbysmith_timeuniform},

$(\hat{\bar{{\tau}}}_{t}(w, w') \pm V^*_t(w, w'))$ where 
$$V^*_t(w, w') =  \sqrt{\frac{2(B_{t}(w, w')\eta^2 + 1)}{t^2\eta^2}\log\left(\frac{\sqrt{B_{t}(w, w')\eta^2 + 1}}{\alpha} \right)}$$
 is a valid $(1-\alpha)$asymptotic CS and $B_t(w, w') = \sum_{i=1}^t Var(\hat{{\tau}}_{i}(w, w') \mid \mathcal{F}_{i} )$.

 As noted in Equation~\eqref{eq:variance}, 
\begin{align} Var(\hat{\tau}_{i}(w, w') \mid \mathcal{F}_{i}) \leq \sigma^2_{i}\text{, where } \sigma^2_{i}\coloneqq \frac{Y_i(w)^2}{p^{\text{MAD}}_{i}(w)} + \frac{Y_i(w')^2}{p^{MAD}_{i}(w')}.
\end{align}

Hence, 
$(\hat{\bar{{\tau}}}_{t}(w, w') \pm \tilde{V}_t(w, w'))$ where 
$$\tilde{V}_t(w, w')=  \sqrt{\frac{2(S_{t}(w, w')\eta^2 + 1)}{t^2\eta^2}\log\left(\frac{\sqrt{S_{t}(w, w')\eta^2 + 1}}{\alpha} \right)}$$
 is still a valid $(1-\alpha)$asymptotic CS, where $S_t(w, w') = \sum_{i=1}^t \sigma_i^2(w, w')$. This holds because replacing $B_t(w, w')$ with $S_{t}(w, w')$ only makes the CS wider since $S_t(w, w') \geq B_t(w, w')$.

As noted in Equation~\eqref{eq:variance_est}, an unbiased estimator for $\sigma^2_{i}(w, w')$ is 
\begin{align}
    \hat{\sigma}^2_{i}(w, w')\coloneqq \frac{Y_i(w)^2 \mathbbm{1}\{W_i = w\}}{(p^{\text{MAD}}_{i}(w))^2} + \frac{Y_i(w')^2\mathbbm{1}\{W_i = w'\}}{(p^{MAD}_{i}(w'))^2}.
\end{align}
and we define $\hat{S}_t(w, w') = \sum_{i=1}^t \hat{\sigma}_i^2(w, w')$.

To establish the validity of the CS in Theorem 1, we must show that $\frac{1}{t}\hat{S}_t(w, w') - \frac{1}{t}S_t(w, w') \stackrel{a.s.}{\to} 0$. Then, Condition L-3 of Theorem 2.5 of \cite{waudbysmith_timeuniform} is satisfied and we can conclude that 
$(\hat{\bar{{\tau}}}_{t}(w, w') \pm \hat{V}_t(w, w'))$ where 
$$\hat{V}_t(w, w') =  \sqrt{\frac{2(\hat{S}_{t}(w, w')\eta^2 + 1)}{t^2\eta^2}\log\left(\frac{\sqrt{\hat{S}_{t}(w, w')\eta^2 + 1}}{\alpha} \right)}$$
 is still a valid $(1-\alpha)$ asymptotic CS.

First, note that 
\begin{align*}
    (\hat{\sigma}_i^2(w, w'))^2 &\leq \left(\frac{M^2}{p_{i}^{MAD}(w)^2}+\frac{M^2}{p_{i}^{MAD}(w')^2}\right)^2\\
    & = \frac{M^4}{p_{i}^{MAD}(w)^4} + \frac{M^4}{p_{i}^{MAD}(w')^4} + 2\frac{M^2}{p_{i}^{MAD}(w)^2p_{i}^{MAD}(w')^2}\\
    & \leq \frac{M^4}{(\delta_i/K)^4} + \frac{M^4}{(\delta_i/K)^4} + 2\frac{M^2}{(\delta_i/K)^4}\\
    & = M^4\left(K^4 \frac{1}{\delta_i^4} + K^4 \frac{1}{\delta_i^4} + K^5 \frac{1}{\delta_i^4} \right)\\
    & = o(i)
\end{align*}
where the last line follows because $\frac{1}{\delta_i} = o\left(i^{1/4}\right)$ as shown in Lemma~\ref{lemma_a1}.
Define $X_i(w, w') = \hat{\sigma}_i^2(w, w') - \sigma_i^2(w, w')$ and note that $X_i(w, w')$ is a martingale difference sequence. Hence,
\begin{align*}
    \mathbb{E}[X_i(w, w')] &= E[(\hat{\sigma}_i(w, w')^2]- (\sigma_i(w, w')^2\\
    & \leq E[(\hat{\sigma}_i(w, w')^2]\\
    & = o(i)
\end{align*}

So, $\frac{\mathbb{E}[X_i^2(w, w')]}{i^2} = \frac{o(i)}{i^2}$ and hence, $\sum_{i=1}^\infty \frac{\mathbb{E}[X_i^2(w, w')]}{i^2} < \infty$. For instance, if $\delta_i = \frac{1}{i^a}$ for some $0 \leq a < 1/4$, then
$\frac{\mathbb{E}[X_i^2(w, w')]}{i^2} \leq Ci^{4a-2}$ for some constant $C < \infty$, and since $4a-2 < -1$, $\sum_{i=1}^\infty \frac{\mathbb{E}[X_i^2(w, w')]}{i^2} \leq C \sum_{i=1}^\infty i^{4a-2} < \infty$ by the p-series test.

Then, by the SLLN for martingale difference sequences (Theorem 1 of \cite[]{csorgHo1968strong}), we can conclude that
$$\frac{1}{t} \sum_{i=1}^t X_i(w, w') \stackrel{a.s.}{\to} 0.$$

Hence, Condition L-3 of \cite{waudbysmith_timeuniform} is satisfied and by Step 3 of the proof for Theorem 2.5 of \cite{waudbysmith_timeuniform}, we can conclude that 
$(\hat{\bar{{\tau}}}_{t}(w, w') \pm \hat{V}_t(w, w'))$ where 
$$\hat{V}_t(w, w') =  \sqrt{\frac{2(\hat{S}_{t}(w, w')\eta^2 + 1)}{t^2\eta^2}\log\left(\frac{\sqrt{\hat{S}_{t}(w, w')\eta^2 + 1}}{\alpha} \right)}$$
 is still a valid $(1-\alpha)$ asymptotic CS.

To show $\hat{V}_t(w, w') \stackrel{a.s.}{\to} 0$, note that 
\begin{align*}
    \hat{\sigma}_i^2 &\leq \frac{M^2}{p_{i}^{MAD}(w)^2}+\frac{M^2}{p_{i}^{MAD}(w')^2}\\
    & \leq  \frac{M^2}{(\delta_i/K)^2}+\frac{M^2}{(\delta_i/K)^2}\\
        & = \frac{K^2M^2}{\delta_i^2}+\frac{K^2M^2}{\delta_i^2}\\
    & = 2K^2M^2 \frac{1}{\delta_i^2}\\
    &  =   2K^2M^2 o(i^{1/2})
\end{align*}
Therefore, $\hat{S}_t(w, w') \leq 2K^2M^2 \sum_{i=1}^t o(i^{1/2}) $ a.s..

 Hence, there exists positive real numbers $N$ and $x_0$ such that for all $t \geq x_0$, $\hat{S}_{t}(w, w') \leq N \sum_{i=1}^t i^{1/2} < N \sum_{i=1}^t t^{1/2} = N t^{1+1/2}$, allowing us to conclude that 
$\hat{S}_t(w, w') = O(t^{1+\frac{1}{2}})$.
 
 We can show that $\log\left(\hat{S}_t(w, w')\right) = o(t^{1/2})$ a.s..
For $t \geq x_0$,
\begin{align*}
    \frac{\log(\hat{S}_{t}(w, w'))}{t^{1/2}} \leq     \frac{\log(Nt^{3/2})}{t^{1/2}} =    \frac{\log(N) + (3/2)\log(t))}{t^{1/2}} 
\end{align*}
and $\frac{\log(N) + (3/2)\log(t))}{t^{1/2}}  \to 0$ as $t \to \infty$.  Therefore, $\log(\hat{S}_{t}(w, w')) = o(t^{1/2})$ a.s. and $\hat{S}_t(w, w') \log(\hat{S}_t(w, w')) = o(t^2)$ a.s.. Hence, $
\hat{V}_t(w, w') = \frac{2(\hat{S}_{t}(w, w')\eta^2 + 1)}{t^2\eta^2}\log\left(\frac{\sqrt{\hat{S}_{t}(w, w')\eta^2 + 1}}{\alpha} \right) = o(1)$ a.s..

Note, if $1/\delta_i$ was increasing at a rate faster than $i^{1/4}$ asymptotically, \emph{e.g.}, $\delta_i = O(i^{1/2})$,  we are not guaranteed that  $\hat{V}_t(w, w') = o(1)$ a.s..
\end{proof}

\begin{customthm}{2$^*$}\label{thrm2star}  For $w, w' \in \{0, 1, ..., K-1\}$, let $(\hat{\tau}_i(w, w'))_{i=1}^\infty$ be the sequence of random variables where 
 $W_i=w$ with probability $p_{i}^{\text{MAD}}(w) = \delta_i \left(\frac{1}{2}\right) + (1-\delta_i) p^{\mathcal{A}}_i(w)$ for $w \in \mathcal{W}$ and $\delta_i = \omega\left(\frac{1}{i^{1/4}}\right)$.
Assume Assumptions~\ref{a_bound} and~\ref{a_var3} hold.
Assume $\bar{\tau_i}(w, w') \to c$ for some $|c| > 0$ as $i \to \infty$.
 Then,
$$\mathbb{P}\left( T_{MAD}(w, w') < \infty\right) = 1.$$
\end{customthm}

\begin{proof}

    We will first show that $\frac{1}{t}\sum_{i=1}^t \hat{\tau}_{i}(w, w') \stackrel{a.s.}{\to}  c $.  Let $u_i(w, w') \coloneqq \hat{\tau}_{i}(w, w')  -\tau_{i}(w, w') $.

    Note that 
    \begin{align*}
        \mathbb{E}[u_i(w, w')^2] &=  \mathbb{E}[\hat{\tau}_i(w, w')^2]-\tau_i(w, w')^2\\
        &\leq  \mathbb{E}[\hat{\tau}_i(w, w')^2]
    \end{align*}
    where
    \begin{align*}
         \hat{\tau}_i(w, w')^2 &= \left(\frac{Y_i(w)\mathbbm{1}\{W_i = w\}}{p_{i}^{MAD}(w)} - \frac{Y_i(w')\mathbbm{1}\{W_i = w'\}}{p_{i}^{MAD}(w')}\right)^2\\
        & \leq \frac{Y_i(w)^2}{(\delta_i/K)^2} + \frac{Y_i(w')^2}{(\delta_i/K)^2}\\
        & = 2K^2M^2 \frac{1}{\delta_i^2}\\
        & = 2K^2M^2 o(i^{1/2})
    \end{align*}

    Hence,
\begin{align*}
    \sum_{i=1}^ t \frac{\mathbb{E}[u_i(w, w')^2]}{i^2} \leq \sum_{i=1}^ t \frac{8M^2 o(i^{1/2})}{i^2} < \infty
\end{align*}
   by the p-series test.
Hence, by the Strong Law of Large Numbers for martingale difference sequences (Theorem 1 of \cite[]{csorgHo1968strong}), we can conclude that
$$\frac{1}{t} \sum_{i=1}^t u_i(w, w') \stackrel{a.s.}{\to} 0.$$

By Theorem~\ref{thrm1star}, we have that $\hat{V}_t(w, w') \stackrel{a.s.}{\to} 0$.

Therefore, applying Slutsky's Theorem, we conclude that
\begin{align*}
    \hat{\bar{{\tau}}}_{t}(w, w')   +\hat{V}_t(w, w') \stackrel{a.s.}{\to} c,
\end{align*}
and
\begin{align*}
    \hat{\bar{{\tau}}}_{t}(w, w')   - \hat{V}_t(w, w') \stackrel{a.s.}{\to} c.
\end{align*}
So,
\begin{align*}
&\mathbb{P}\left(0 \notin \left(\hat{\bar{{\tau}}}_{t }(w, w')   \pm \hat{V}_t(w, w')\right) \right)\\
&= \mathbb{P}\left(0 <   \hat{\bar{{\tau}}}_{t }(w, w')  - \hat{V}_t(w, w')\right) + \mathbb{P}\left(0 > \hat{\bar{{\tau}}}_{t }(w, w')    +\hat{V}_t(w, w') \right)\\
& \to 1,
\end{align*}
where the last line follows because $\mathbb{P}\left(0 > \hat{\bar{{\tau}}}_{t}(w, w')  +\hat{V}_t(w, w')\right) \to 0$ and 

$\mathbb{P}\left(0 <  \hat{\bar{{\tau}}}_{t}(w, w')   -\hat{V}_t(w, w')\right) \to 1$.

Let $p_t = \mathbb{P}\left(0 \in (\hat{\bar{{\tau}}}_{t}(w, w')  \pm \hat{V}_t(w, w'))\right)$. Since $p_t \to 0$ as $t \to \infty$, there exists a subsequence $t_k$ such that $p_{t_k} \leq \frac{1}{k^2}$, for all $k \geq 1$. Let $A_{t_k}$ be the event $\{0 \in (\frac{1}{t_k} \sum_{i=1}^{t_k} \hat{\tau}_{i}(w, w') \pm \hat{V}_{t_k}(w, w'))\}$. Then, $\sum_{k=1}^\infty \mathbb{P}\left(A_{t_k}\right) < \infty$, and by the Borel-Cantelli lemma,
$$\mathbb{P}\left(\limsup_{k \to \infty} A_{t_k}\right) = 0.$$

Hence,
$$\mathbb{P}\left(\title \tilde{T}_{MAD}(w, w') = \infty\right) \leq \mathbb{P}\left(\limsup_{k \to \infty} A_{t_k}\right) = 0.$$
\end{proof}

\section{The MAD for Batched Assignment Algorithms}\label{sec:appendix_batched}

We now define the causal estimators and estimands for the batched setting, where the algorithm can only be updated at pre-defined points. 
Let $\{(Y_i^{(j)}(w))_{w \in \mathcal{W}}\}$ be the set of all potential outcome for unit $i$ in batch $j$.
Given a pair of treatments $w, w' \in \mathcal{W} \coloneqq \{0, 1, ..., K-1\}$ for $K \geq 2$, we can define the Average Treatment Effect \textit{within} a batch $j$ as:
\begin{align*}
    \tau^{\text{batched}}_j(w, w') = \frac{1}{B}\sum_{i=1}^{B} Y_i^{(j)}(w) - Y_i^{(j)}(w'),
\end{align*}
and the corresponding unbiased estimator as:
\begin{align*}
    \hat{\tau}^{\text{batched}}_j(w, w') = \frac{1}{B}\sum_{i=1}^{B} \hat{\tau}^{(j)}_i(w, w').
\end{align*}
where $$ \hat{\tau}^{(j)}_i(w, w') = \frac{Y_i^{(j)}(w)\mathbbm{1}\{W^{(j)}_i = w\}}{p^{\text{MAD}_\text{batched}}_{j}(w)} - \frac{Y^{(j)}_i(w')\mathbbm{1}\{W^{(j)}_i = w'\}}{p^{\text{MAD}_\text{batched}}_{j}(w')}.$$
Thus, our target estimand, the \textit{Batched-Average Treatment Effect} up to batch $b$, is:
\begin{align*}
    \bar{\tau}^{\text{batched}}_b(w, w') = \frac{1}{b}\sum_{j=1}^{b}  \tau^{\text{batched}}_j(w, w').
\end{align*}
and the corresponding unbiased estimator is:
\begin{align*}
    \hat{\bar{\tau}}^{\text{batched}}_b(w, w') = \frac{1}{b}\sum_{j=1}^{b} \hat{\tau}^{\text{batched}}_j(w, w').
\end{align*}
Hence, we define $\mathcal{F}^{\text{batched}}_{b}$ as the filtration generated by all potential outcomes for all units up to and including the $b$th batch $\{\{(Y_i^{(j)}(w))_{w \in \mathcal{W}}\}_{i=1}^{B_h}\}_{j=1}^{b}$ and all observed history up to batch $b-1$, $\{H^{\text{batched}}_j\}_{j=1}^{b-1}$.

Then, for each $i= 1, ..., B$, $Var\left(\hat{\tau}^{(j)}_i(w, w') \mid \mathcal{F}^{\text{batched}}_{j} \right) \leq \sigma_i^{(j)2}(w, w')$
where $$\sigma_i^{(j)2}(w, w') = \frac{Y_i(w)^2}{p^{\text{MAD}_\text{batched}}_{j}(w)} + \frac{Y_i(w')^2}{p^{\text{MAD}_\text{batched}}_{j}(w')}.$$

So, 
$ \hat{\sigma}_i^{(j)2}(w, w') = \frac{Y_i(w)^2\mathbbm{1}\{W_i = w\}}{(p^{\text{MAD}_\text{batched}}_{j}(w))^2} + \frac{Y_i(w')^2\mathbbm{1}\{W_i = w'\}}{(p^{\text{MAD}_\text{batched}}_{j}(w'))^2}.$

Hence, 
$$Var\left(\hat{\tau}^{\text{batched}}_j(w, w') \mid \mathcal{F}^{\text{batched}}_{j} \right) = \frac{1}{B^2} \sum_{i=1}^{B} Var\left(\hat{\tau}^{(j)}_i(w, w') \mid \mathcal{F}^{\text{batched}}_{j} \right) \leq \frac{1}{B^2} \sum_{i=1}^{B} \sigma_i^{(j)2}(w, w').$$

Hence, we define

$S^{\text{batched}}_{b}(w, w') \coloneqq \sum_{j=1}^b \frac{1}{B^2} \sum_{i=1}^{B} \sigma_i^{(j)2}(w, w')$ and
 $\hat{S}^{\text{batched}}_{b}(w, w') \coloneqq \sum_{j=1}^b \frac{1}{B^2} \sum_{i=1}^{H_j} \hat{\sigma}_i^{(j)2}(w, w')$.

Finally, we state the analogous assumptions as Assumptions~\ref{a_bound} and \ref{a_var2} for the batched assignment setting. 
\begin{assump}[Bounded Potential Outcomes for Batched Bandits]\label{a_bound_batched}
   There exists $M \in \mathbb{R}$ such that $$|Y^{(j)}_i(w)| \leq M < \infty$$ for all $j, i \in \mathbb{N}^+, w \in \mathcal{W}$.
\end{assump}

\begin{assump}[At Least Linear Rate of Cumulative Conditional Variances for Batched Bandits]\label{a_var_batched}
For all $w, w' \in \mathcal{W}$,
$$\sum_{j=1}^{b}Var\left(\hat{\tau}^{\text{batched}}_j(w, w') \mid \mathcal{F}^{\text{batched}}_{j} \right)  = \Omega(b).$$
\end{assump}

\begin{customthm}{$1^\dagger$}\label{thrm1_dag} For $w, w' \in \mathcal{W}$, let $(\hat{\tau}^{\text{batched}}_j(w, w'))_{j=1}^{\infty}$ be the sequence of random variables where 
 $W^{(j)}_i=w$ with probability $p_{j}^{\text{MAD}_\text{batched}}(w)$ as in Definition~\ref{bmad}, $w \in \mathcal{W}$, and $\delta_j \in (0, 1]$ such that  $\delta_j = \omega\left(\frac{1}{j^{1/4}}\right)$.
Assume Assumptions~\ref{a_bound_batched} and~\ref{a_var_batched} hold. Then $(\hat{\bar{{\tau}}}^{\text{batched}}_{b}(w, w') \pm \hat{V}^{\text{batched}}_b(w, w'))$ where 
$$\hat{V}^{\text{batched}}_b(w, w') \coloneqq  \sqrt{\frac{2(\hat{S}^{\text{batched}}_{b}\eta^2 + 1)}{t^2\eta^2}\log\left(\frac{\sqrt{\hat{S}^{\text{batched}}_{b}\eta^2 + 1}}{\alpha} \right)}$$ 
 is a valid $(1-\alpha)$ asymptotic CS for $\bar{\tau}^{\text{batched}}_b$ and $\hat{V}^{\text{batched}}_b(w, w') \stackrel{a.s.}{\to} 0$.
\end{customthm}

\begin{proof}
    Note, the above result is equivalent to Theorem~\ref{thrm1star} except we replace $\hat{\tau}_i(w, w')$ with $\hat{\tau}_j^{\text{batched}}(w, w')$, $B_t(w, w')$ with
$B^{\text{batched}}_b(w, w') = \sum_{j=1}^b Var\left(\hat{\tau}^{\text{batched}}_j(w, w') \mid \mathcal{F}^{\text{batched}}_{j, j-1} \right)$, and $\hat{S}_t(w, w')$ and  $S_t(w, w')$ with $S^{\text{batched}}_{b}(w, w') \coloneqq \sum_{j=1}^b \frac{1}{B^2} \sum_{i=1}^{B} \sigma_i^{(j)2(w, w')}$ and $\hat{S}^{\text{batched}}_{b}(w, w') \coloneqq \sum_{j=1}^b \frac{1}{B^2} \sum_{i=1}^{H_j} \hat{\sigma}_i^{(j)2(w, w')}$ respectively. 
Assumption~\ref{a_var_batched} ensures $\sum_{j=1}^b Var\left(\hat{\tau}^{\text{batched}}_j(w, w') \mid \mathcal{F}^{\text{batched}}_{j, j-1} \right) = \Omega(b)$ and, because $p_{j}^{\text{MAD}_{\text{batched}}}(w) \geq \frac{1}{K}\delta_j$, we have that $(\hat{\tau}_j^{\text{batched}}(w, w')-{\tau}_j^{\text{batched}}(w, w'))^2 = o(j^{1/2})$ and following the steps of the proof of Lemma~\ref{lemma_a2}, we establish that 
 $\{\hat{\tau}_j^{\text{batched}}(w, w')\}_{j=1}^\infty$ satisfies the Lindeberg-type uniform integrability condition of \cite{waudbysmith_timeuniform}, \emph{i.e.}, there exists $\kappa \in (0, 1)$ such that
$$\sum_{b=1}^\infty \frac{\mathbb{E}\left[(\hat{\tau}_b^{\text{batched}}(w, w')-{\tau}_b^{\text{batched}}(w, w'))^2\mathbbm{1}\{(\hat{\tau}_b^{\text{batched}}(w, w')-{\tau}_b^{\text{batched}}(w, w'))^2 > (B^{\text{batched}}_b(w, w'))^\kappa \}\right]}{ (B^{\text{batched}}_b(w, w'))^\kappa} < \infty $$
almost surely. 
Hence, the remainder of the proof follows directly from Theorem~\ref{thrm1star}, replacing the corresponding terms for the batched bandit setting, because the fact that $p_{j}^{\text{MAD}_{\text{batched}}}(w) \geq \frac{1}{K}\delta_j$ ensures that all analogous terms have the same rates as in Theorem~\ref{thrm1star}.
\end{proof}

For $w, w' \in \mathcal{W}$, define $T_{\text{MAD}_{\text{batched}}}(w, w') \coloneqq \inf_{t}\left\{b: 0 \notin (\hat{\bar{\tau}}^{\text{batched}}_b(w, w') \pm V^{\text{batched}}_b(w, w')) \right\}$. We now state and prove an analogous result to Theorem~\ref{thrm2} in the batched assignment setting.

\begin{customthm}{$2^\dagger$}\label{thrm4} For $w, w' \in \mathcal{W}$, let $(\hat{\tau}^{\text{batched}}_j(w, w'))_{j=1}^{\infty}$ be the sequence of random variables where
 $W^{(j)}_i=w$ with probability $p_{j}^{\text{MAD}_{\text{batched}}}(w)$ as in Defintion~\ref{bmad}, and $\delta_j \in (0, 1]$ such that  $\delta_j = \omega\left(\frac{1}{j^{1/4}}\right)$.
Assume Assumptions~\ref{a_bound_batched} and~\ref{a_var_batched} hold. Then, if $ \hat{\bar{\tau}}^{\text{batched}}_b(w, w')  \to c$ as $b \to \infty$ for some $|c| > 0$,
$$\mathbb{P}\left( T_{\text{MAD}_{\text{batched}}}(w, w') < \infty\right) = 1.$$
\end{customthm}

\begin{proof}
    Having established Theorem~\ref{thrm1_dag}, the remainder of the proof follows from Theorem~\ref{thrm2star}, replacing the corresponding terms for the batched bandit setting, because the fact that $p_{j}^{\text{MAD}_{\text{batched}}}(w) \geq \frac{1}{K}\delta_j$ ensures that all analogous terms have the same rates as in Theorem~\ref{thrm2star}.
\end{proof}

\section{Proof of Theorem~\ref{thrm3}}

Recall, the definition of asymptotic confidence sequence for $\hat{\bar{\tau}}_{t}$ states that
$(\hat{\bar{\tau}}_{t} \pm \hat{V}_t)$ is a two-sided $(1-\alpha)$ asymptotic CS for $\hat{\bar{\tau}}_{t}$ if there exists an (unknown) non-asymptotic confidence sequence $(\hat{\bar{\tau}}_{t} \pm V^*_t)$ for $\tau_{t}$ such that $\frac{V^*_t}{\hat{V}_t}\stackrel{a.s.}{\to} 1$. Furthermore, we say that $\hat{V}_t$ has approximation rate $R_t$ if $V^*_t = \hat{V}_t + o(R_t)$ almost surely, where $R_t$ is a deterministic (\emph{i.e.}, non-random) sequence. 

Define $T^*_{MAD} \coloneqq \inf_{t}\left\{t: 0 \notin (\hat{\bar{\tau}}_t \pm V^*_t) \right\}$, i.e., this is the analogous stopping time to $T_{MAD}$ for the \emph{non-asymptotic} anytime-validCS.

We will show that, under the same setting as Theorem~\ref{thrm2}, $T^*_{MAD} < \infty$ with probability $1$.

\begin{customthm}{3}
Let $(\hat{\tau}_t)_{t=1}^\infty$ be the sequence of random variables where 
 $W_t=w$ with probability $p_{t}^{\text{MAD}}(w) = \frac{1}{2}\delta_t + (1-\delta_t)p^{\mathcal{A}}_{t}(w)$, $w \in \{0, 1\}$, and $\delta_t= \omega\left(\frac{1}{t^{1/4}}\right)$.
Assume Assumptions~\ref{a_bound} and~\ref{a_var2} hold. Assume $\bar{\tau}_t \to c$ for some $|c| > 0$ as $i \to \infty$. Then,  $\mathbb{P}\left(T^*_{MAD} < \infty\right) = 1$. 
\end{customthm}

At a high level, the proof proceeds by showing that $\delta_t$ guarantees that $V^*_t = \hat{V}_t + o(1)$ and follows almost immediately from results already established in the proofs of Theorem~\ref{thrm1} and~\ref{thrm2}. We prove this result for the binary treatment setting, though it can be naturally extended to more than two treatments, as in Theorem~\ref{thrm2star}, and the batched bandits setting, as in Theorem~\ref{thrm4}.


\proof{
As established Theorem 2.5 of \cite{waudbysmith_timeuniform}, the asymptotic CS of Theorem~\ref{thrm1} has approximation rate $R_t = \sqrt{S_{t}\log(S_{t})}/t$. For $\delta_t = \omega(1/t^{1/4})$, we have that
$$\sqrt{S_{t}\log(S_{t})}/t \leq \sqrt{\hat{S}_{t}\log(\hat{S}_{t})}/t = o(1)$$
where the final equality follows from Equation~\eqref{thrm5_eqref1} in the proof of Theorem~\ref{thrm1}.
Hence, for the MAD with $\delta_i = \omega(1/i^{1/4})$ for all $i=1, ..., t$, $V^*_t = \hat{V}_t + o(1)$ almost surely.

By Equations~\eqref{thrm5_eqref2} and~\eqref{thrm5_eqref3} in the proof of Theorem~\ref{thrm2},
\begin{align}
   \hat{\bar{\tau}}_{t}+ V^*_t &\stackrel{a.s.}{=} \hat{\bar{\tau}}_{t} + \hat{V}_t + o(1) \stackrel{a.s.}{\to} c\\
    \hat{\bar{\tau}}_{t}-V^*_t &\stackrel{a.s.}{=} \hat{\bar{\tau}}_{t} - \hat{V}_t - o(1) \stackrel{a.s.}{\to} c.
\end{align}
So, we can conclude that $\mathbb{P}\left(0 \in (\hat{\bar{\tau}}_{t}-V^*_t) \right) \to 1$ and the fact that $\mathbb{P}(T^*_{MAD} < \infty) = 1$ follows immediately from Equations~\eqref{thrm5_eqref4} and~\eqref{thrm5_eqref5} of the proof of Theorem~\ref{thrm2}.

}

\section{Proof of Lemma 3.1}

Recall, we define $\tilde{T}_{MAD} \coloneqq \inf_{t}\left\{t: 0 \notin (\hat{\bar{\tau}}_t \pm u(W_t)/t \right\}$ where $u(W_t)$ is defined as on Corollary 2 of \cite{howard} 
and $W_t = \sum_{i=1}^t (\hat{\tau}_i - \hat{\bar{\tau}}_i)^2$.

\begin{customthm}{3}
Let $(\hat{\tau}_t)_{t=1}^\infty$ be the sequence of random variables where 
 $W_t=w$ with probability $p_{t}^{\text{MAD}}(w) = \frac{1}{2}\delta_t + (1-\delta_t)p^{\mathcal{A}}_{t}(w)$, $w \in \{0, 1\}$, where there exists some $0 < p_{min} < 1$ such that $\delta_t \geq \min\{p_{min}, 1-p_{min}\}$ for all $i \in [1, \infty)$.
Assume Assumptions~\ref{a_bound} and~\ref{a_var2} hold. Assume $\bar{\tau}_t \to c$ for some $|c| > 0$ as $i \to \infty$. Then, $\mathbb{P}\left(\tilde{T}_{MAD} < \infty\right) = 1$. 
\end{customthm}

\proof{
Because $\delta_i \geq \min\{p_{min}, 1-p_{min}\}$ for all $i \in [1, \infty)$, $p_{i}^{\text{MAD}}(w) \geq \frac{p_{min}}{2}$.
By Theorem 1 of \cite{howard}, $u(W_t)/t$ is guaranteed to shrink to zero as $t$ grows. By Equation~\eqref{thrm6_eqref1} of the proof of Theorem~\ref{thrm2}, we have that $\hat{\bar{\tau}}_t \stackrel{a.s.}{\to}  c$. 

Therefore, applying Slutsky's Theorem, we conclude that
\begin{align}
    \hat{\bar{{\tau}}}_{t}   +u(W_t)/t \stackrel{a.s.}{\to} c,
\end{align}
and
\begin{align}
    \hat{\bar{{\tau}}}_{t}   - u(W_t)/t \stackrel{a.s.}{\to} c.
\end{align}
and the stopping time result follows immediately from the Borel Cantelli argument of Equations~\eqref{thrm5_eqref4} and~\eqref{thrm5_eqref5} of the proof of Theorem~\ref{thrm2}.
}

Hence, choosing $\delta_i$ as we did for the results of Section~\ref{section:simulations} enables the manager to have a guaranteed stopping time result using the MAD with a non-asymptotic CS. However, we find empirically that MAD with the asymptotic CS of Theorem~\ref{thrm1} often achieves the correct coverage in finite samples while shrinking significantly faster than the non-asymptotic CS of \cite{howard}, and hence, we recommend the use of the asymptotic CS of Theorem~\ref{thrm1} in practice.

\section{Additional Simulation Results}

\subsection{MAD under a Non-Stationary ATE}\label{appendix:nonstationary_mad}
One feature of our design-based approach is MAD's ability to handle non-stationary outcome models. Since the reward of a bandit algorithm can suffer as a result of non-stationarity, one valid question is whether inference stays sharp even as reward suffers. 
To exhibit our proof of concept, we run the Bernoulli outcome experiment of Section~\ref{section:simulations}, except now the Bernoulli parameters vary with time. Specifically, we assume $Y_t(1) \stackrel{i.i.d.}{\sim} \text{Bernoulli}(p_{1,t})$,  $Y_t(0) \stackrel{i.i.d.}{\sim} \text{Bernoulli}(p_{0,t})$ with $(p_{0,t}, p_{1,t}) = (0.2, 0.8)$ when $t \leq 500$, and we explore two different settings for  when $t > 500$: $(p_{0,t}, p_{1,t}) =(0.2, 0.4)$ and $(p_{0,t}, p_{1,t}) = (0.2, 0.1)$. Hence, in the second parameterization, the ATE flips sign after $500$ observations. As discussed in \cite{kohavi2020trustworthy}, we generally expect shifts in the ATE to maintain the same sign and have a smooth trend, so such step-wise changes in the ATE should be especially adversarial for the bandit algorithm. We run this experiment using the Bernoulli design and the MAD (using UCB, $\delta_t = \frac{1}{t^{0.24}}$) across $t=10{,}000$ total time steps.  

\begin{figure}[htb!]
\centering
\begin{subfigure}{0.5\textwidth}
  \centering  \includegraphics[width=\linewidth]{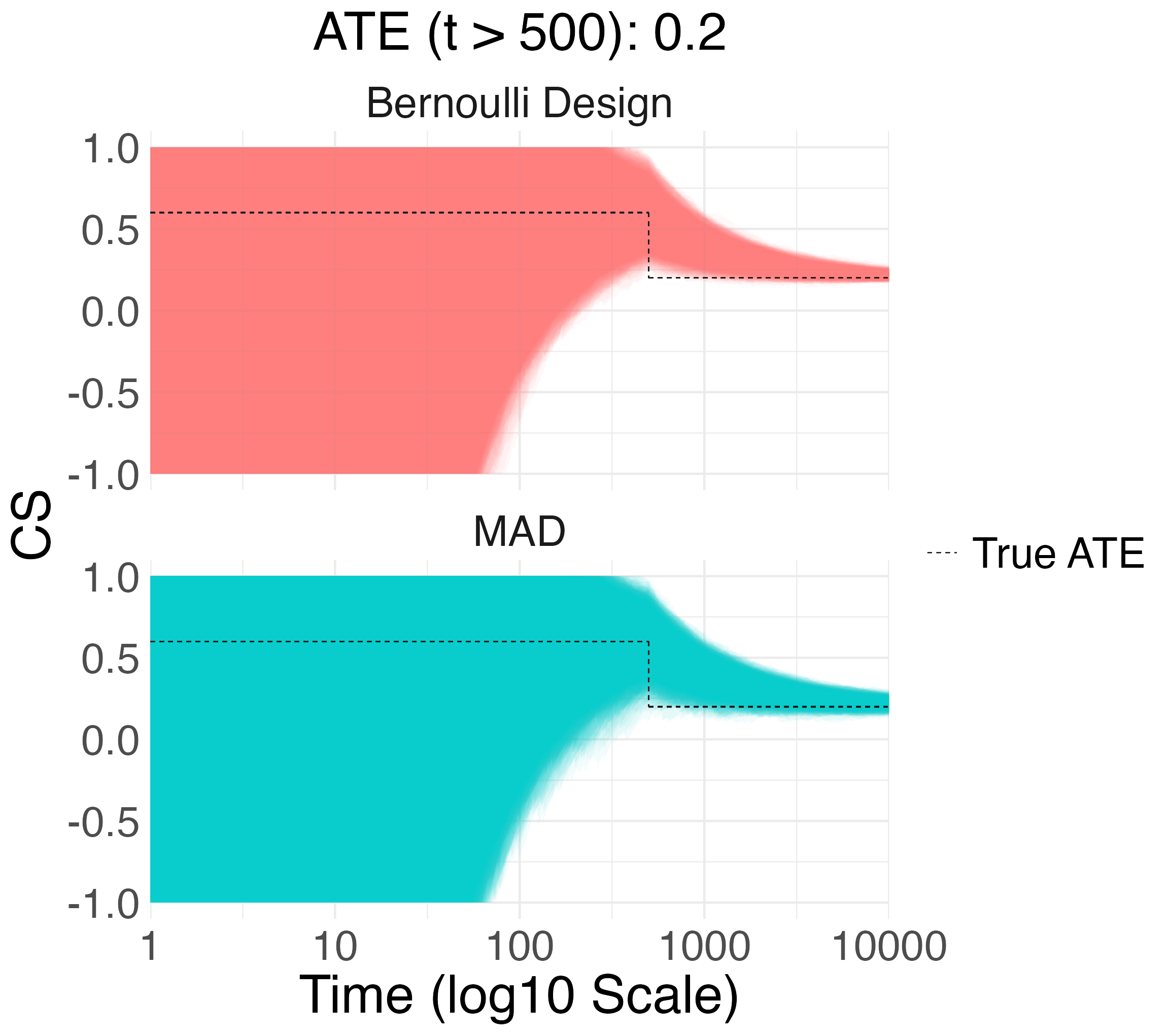}
  \caption{}
  \label{fig:no_flip}
\end{subfigure}%
\begin{subfigure}{0.5\textwidth}
  \centering
  \includegraphics[width=\linewidth]{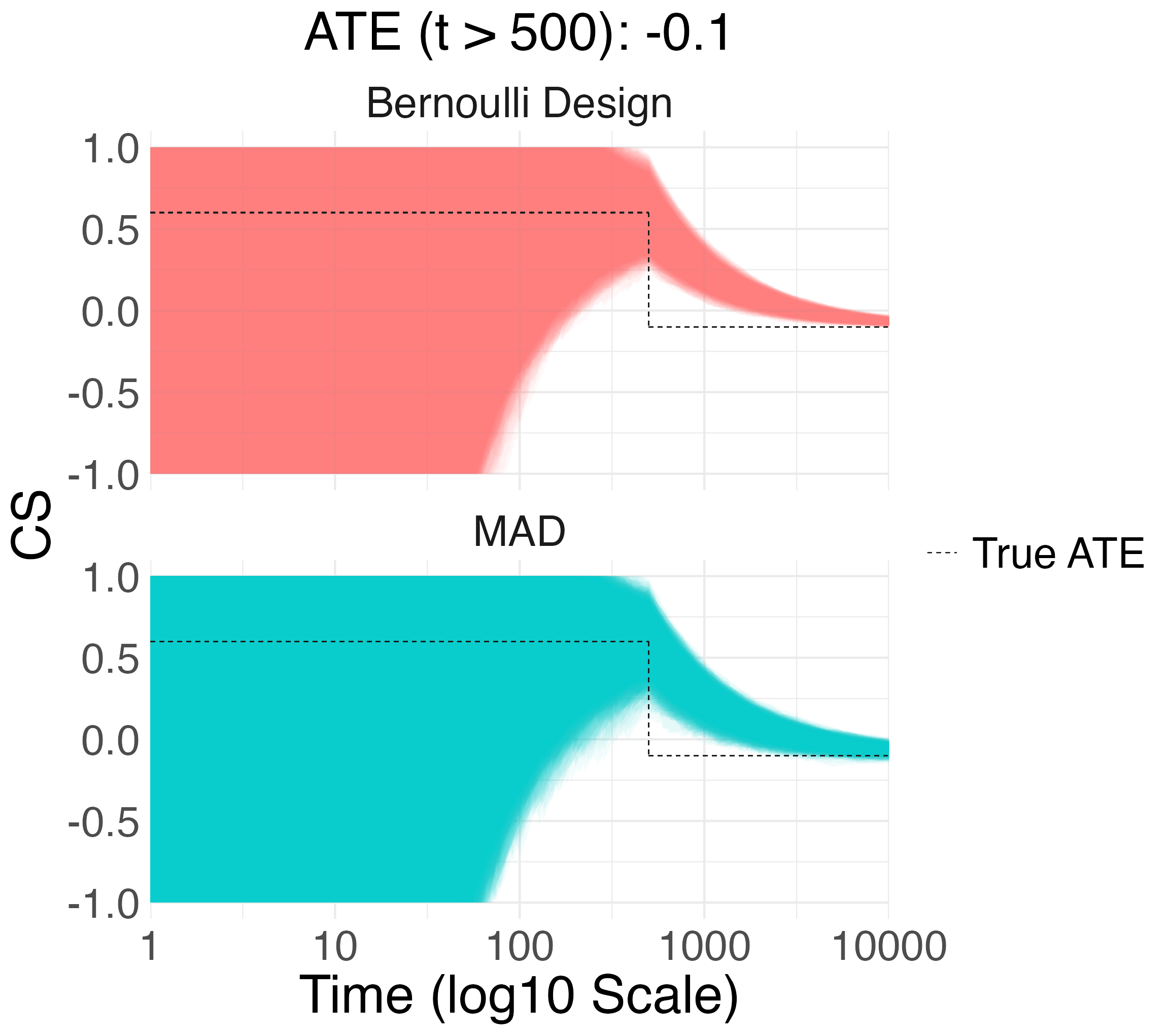}
  \caption{}
  \label{fig:flip}
\end{subfigure}
\caption{All CSs generated from the experiment of Appendix~\ref{appendix:nonstationary_mad} across $100$ random seeds (with transparency to show the overlaying of the CSs) with $(p_{0,t}, p_{1,t}) = (0.2, 0.8)$ when $t \leq 500$ and when $t > 500$, (a): $(p_{0,t}, p_{1,t}) =(0.2, 0.4)$ (ATE goes from $0.6$ to $0.1$) and (b): $(p_{0,t}, p_{1,t}) = (0.2, 0.1)$ (ATE goes from $0.6$ to $-0.1$).}
\label{fig:nonstationary_css}
\end{figure}

\begin{figure}[htb!]
\centering
\begin{subfigure}{0.5\textwidth}
  \centering  \includegraphics[width=\linewidth]{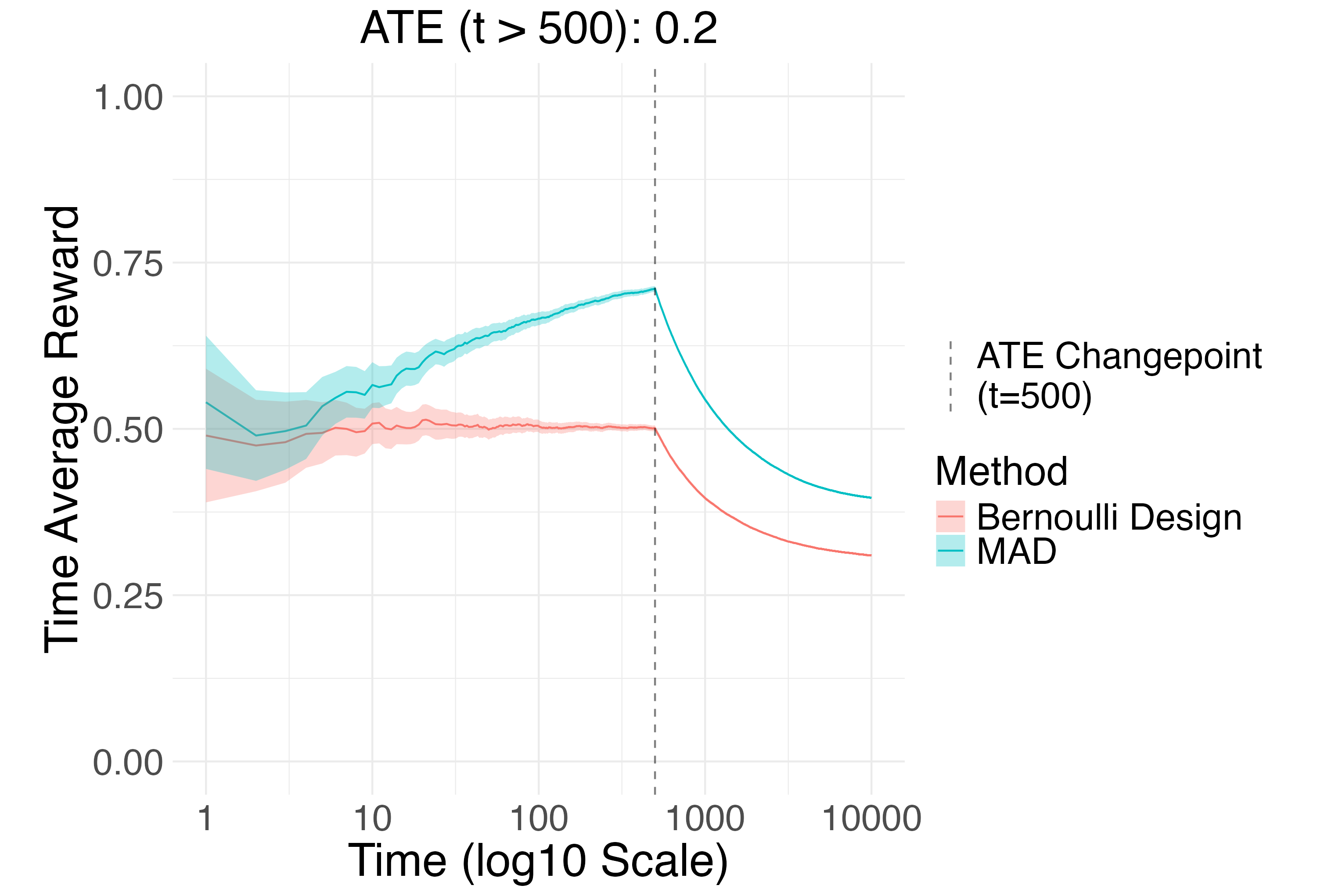}
  \caption{}
  \label{fig:no_flip}
\end{subfigure}%
\begin{subfigure}{0.5\textwidth}
  \centering
  \includegraphics[width=\linewidth]{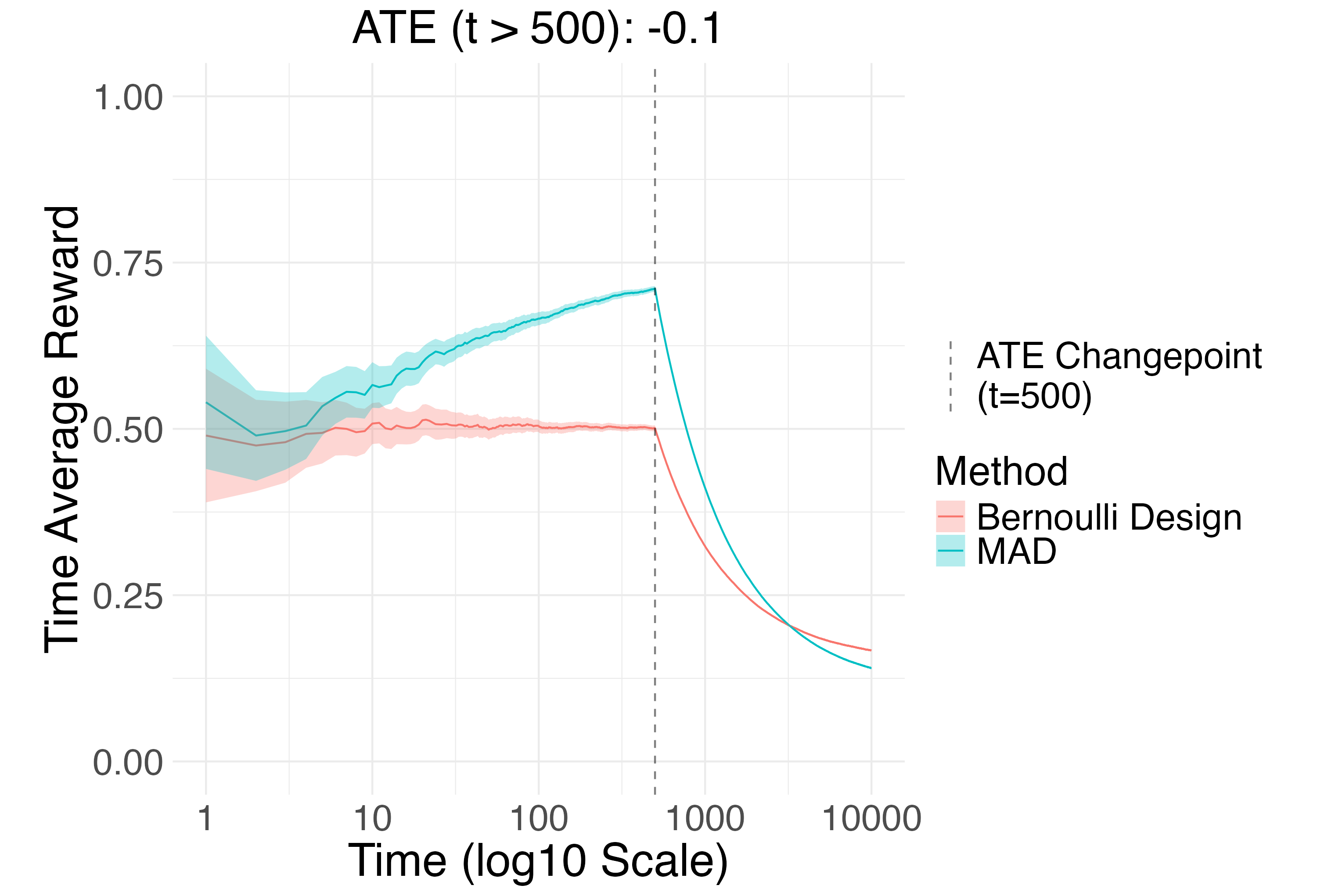}
  \caption{}
  \label{fig:flip}
\end{subfigure}
\caption{Time-averaged reward of the experiment of Appendix~\ref{appendix:nonstationary_mad} across $100$ random seeds (with transparency to show the overlaying of the CSs) with $(p_{0,t}, p_{1,t}) = (0.2, 0.8)$ when $t \leq 500$ and when $t > 500$, (a): $(p_{0,t}, p_{1,t}) =(0.2, 0.4)$ (ATE goes from $0.6$ to $0.1$) and (b): $(p_{0,t}, p_{1,t}) = (0.2, 0.1)$ (ATE goes from $0.6$ to $-0.1$). Error bars depict $\pm2$ standard errors.}
\label{fig:nonstationary_rewards}
\end{figure}

We repeat the above experiment for $100$ random seeds. Figure~\ref{fig:nonstationary_css} plots 
the CSs across the $100$ random seeds with transparency to show the overlaying of the CSs, 
and Figure~\ref{fig:nonstationary_rewards} cumulative reward of each of the algorithms across the $100$ random seeds. Like the Bernoulli design, the MAD CS tracks the shift in ATE with sufficient samples, even though the reward associated with the bandit can suffer dramatically due to the ATE shift. Note, with the ATE maintains the same sign, the MAD still outperforms the Bernoulli design in terms of reward, and hence, it can be argued that it is still reasonable to use an adaptive algorithm in this case.
Though a rather large number of samples is needed to capture this shift, we emphasize that most super-population approaches would not be valid in non-stationarity settings as they often rely on stationarity assumptions to establish their guarantees, and other approaches which can handle non-stationarity like \cite{howard} and \cite{ham} would only be valid when the assignment probabilities are bounded away from $0$ and $1$.

\subsection{MAD vs. Bernoulli Regret under Stopping Rules}\label{appendix:stopping_mad}
The MAD mitigates risk both by enabling anytime-valid inference while allowing for adaptive treatment assignments. However, the Bernoulli design provides the best inferential precision, and hence, will generally satisfy a stopping rule (like zero being outside the CS) faster than the MAD. A natural question is whether experiments run with the MAD will still achieve smaller regret than if it had been run with a Bernoulli design that may have stopped sooner.

To explore this question, we running our Bernoulli outcome experiment of Section~\ref{section:simulations} using both a high signal ($(p_{0}, p_{1}) = (0.2, 0.8)$) and low signal ($(p_{0}, p_{1}) = (0.2, 0.3)$) setting across $100$ random seeds. For each setting and seed, we first generate the full set of all potential outcomes across $t=10{,}000$, then run an experiment with a Bernoulli design and the MAD (using UCB, with $\delta_t = \frac{1}{t^{0.24}}$). For both designs, we use the first time in which zero is outside the CS as a stopping rule.\footnote{Because we generate a stationary ATE setting for the sake of this example, it is reasonable to use this stopping rule as indicative of a true non-zero ATE. In practice, managers should exercise their best judgment when determining a stopping rule.} We stop running both experiments when the MAD experiment satisfies the stopping rule.
If the Bernoulli design satisfies the stopping rule before the MAD, we allow the Bernoulli experiment to draw its identified best arm until the MAD satisfies the stopping rule (at which time both experiments stop), hence allowing the Bernoulli experiment to accrue reward until the MAD experiment is able to achieve sharp enough inference to stop. In the instances where the MAD stops before the Bernoulli, we stop running both experiments at the MAD stopping time.

\begin{figure}[htb!]
\centering
\begin{subfigure}{0.5\textwidth}
  \centering  \includegraphics[width=\linewidth]{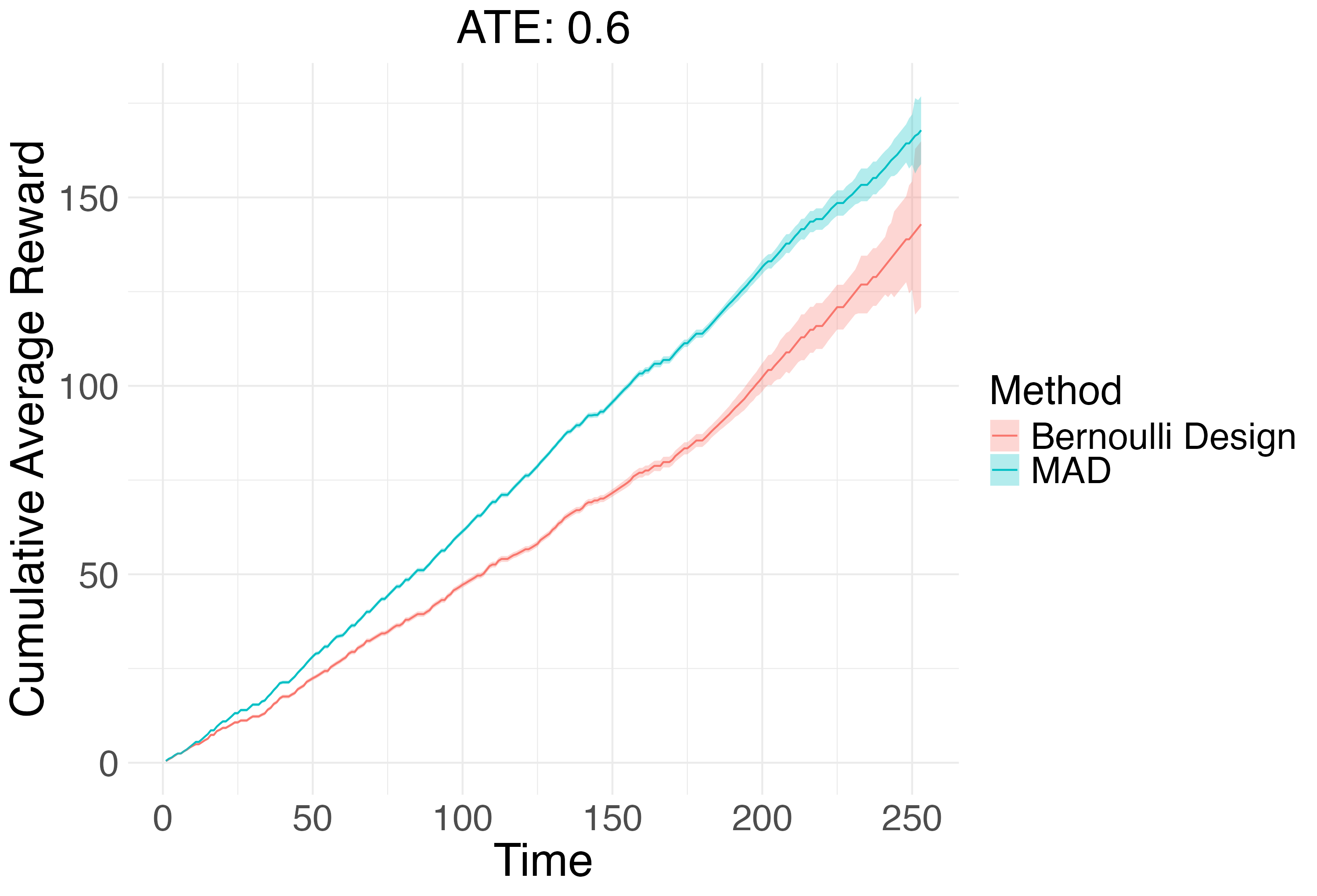}
  \caption{}
  \label{fig:bigsig}
\end{subfigure}%
\begin{subfigure}{0.5\textwidth}
  \centering
  \includegraphics[width=\linewidth]{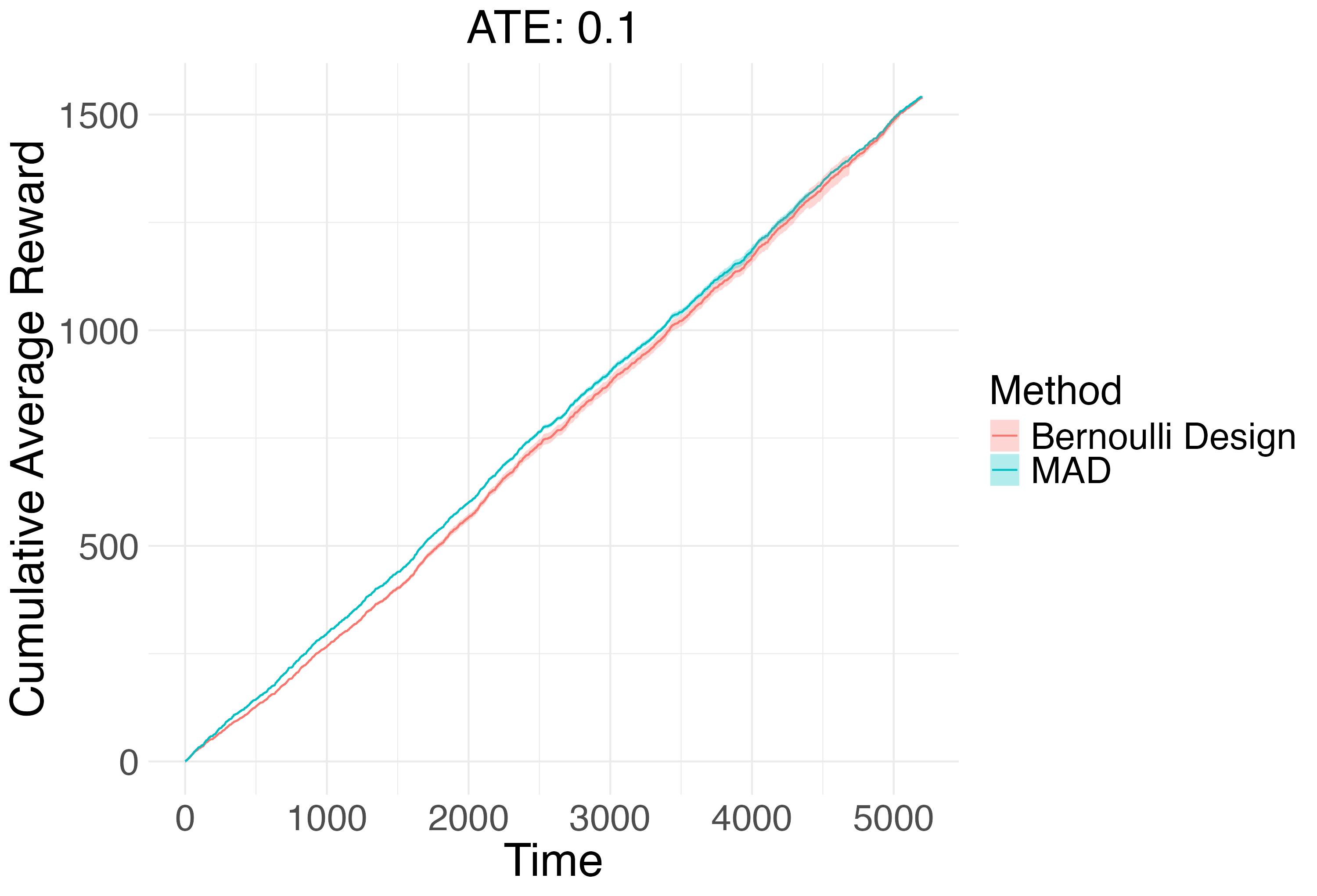}
  \caption{}
  \label{fig:smallsig}
\end{subfigure}
\caption{Cumulative average reward generated from the experiment of Appendix~\ref{appendix:stopping_mad} across $100$ random seeds with  (a): $(p_{0}, p_{1}) =(0.2, 0.8)$ and (b): $(p_{0}, p_{1}) = (0.2, 0.3)$. Error bars depict $\pm 2$ standard errors. Note, as each experiment was run for a different time based on when the MAD for that random seed stopped, the standard errors for larger $t$ can be larger since there are fewer data points where the MAD still has not stopped at that $t$.}
\label{fig:stopping_rewards}
\end{figure}

\begin{figure}[htb!]
\centering
\begin{subfigure}{0.5\textwidth}
  \centering  \includegraphics[width=\linewidth]{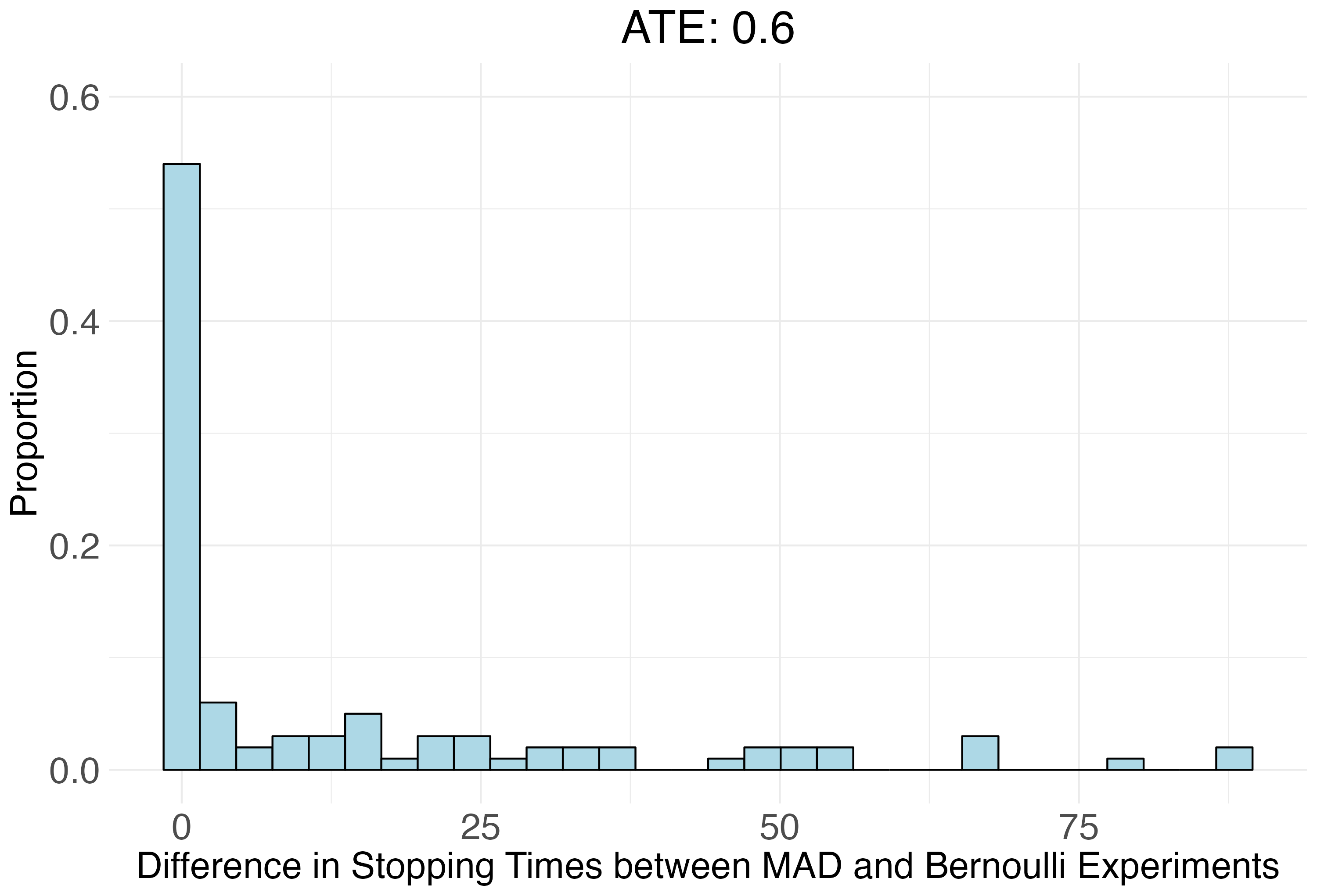}
  \caption{}
  \label{fig:big_sig}
\end{subfigure}%
\begin{subfigure}{0.5\textwidth}
  \centering
  \includegraphics[width=\linewidth]{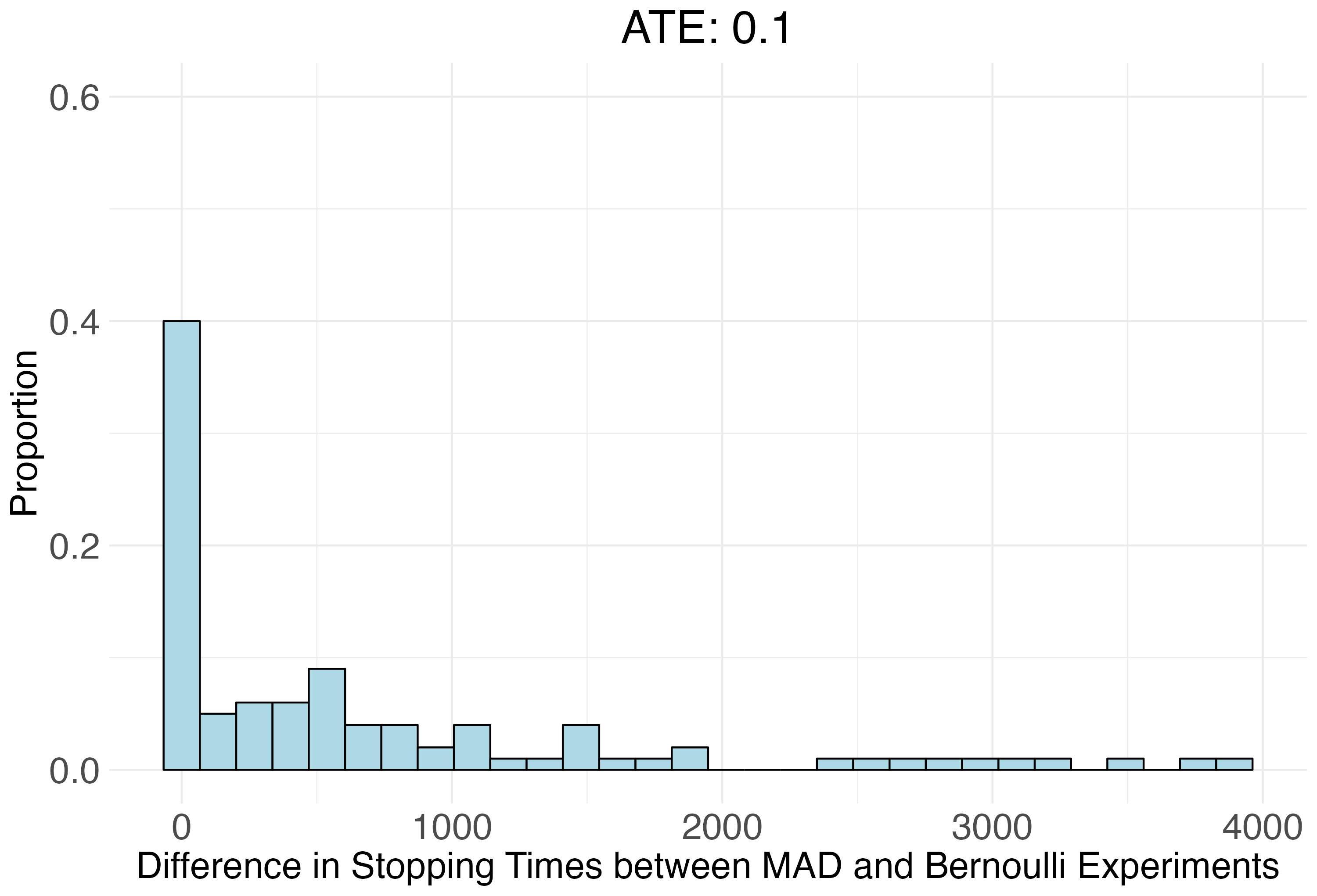}
  \caption{}
  \label{fig:small_sig}
\end{subfigure}
\caption{Histogram of the differences between the MAD and Bernoulli stopping times for the the experiment of Appendix~\ref{appendix:stopping_mad} across $100$ random seeds with  with (a): $(p_{0}, p_{1}) =(0.2, 0.8)$ and (b): $(p_{0}, p_{1}) = (0.2, 0.3)$. }
\label{fig:stopping_times}
\end{figure}

Figure~\ref{fig:stopping_rewards} shows the cumulative average reward accrued for the Bernoulli and MAD experiments across the $100$ random seeds and Figure~\ref{fig:stopping_times} shows a histogram of the difference in stopping times between the MAD and the Bernoulli experiments for both signal settings. Across the $100$ random seeds, the Bernoulli experiment stopped at maximum only $88$ time steps sooner than the MAD for the high signal example, and $3895$ time steps sooner in the low signal example. The median difference in stopping time was $0$ in the high signal example (the MAD often stops at the same time or sooner than the Bernoulli), $275$ in the low signal example. The MAD still achieves higher reward than the Bernoulli experiment over time in both the high and low signal settings, though for the experiments that ran for longer in the low signal setting, the Bernoulli and MAD experiments effectively achieve the same reward. Hence, these simulations showcase that the MAD's risk mitigation through both anytime-valid inference and adaptive treatment assignment can be more effective than solely targeting inferential precision and stopping sooner via a Bernoulli design, especially in high signal settings, and that the MAD can provide at least equivalent risk mitigation in relatively low signal settings.

\clearpage
 
\subsection{Exploring Different Outcome Models and Adaptive Algorithms}\label{simulation_results_appendix}
\begin{figure}
    \centering
\includegraphics[width=\linewidth]{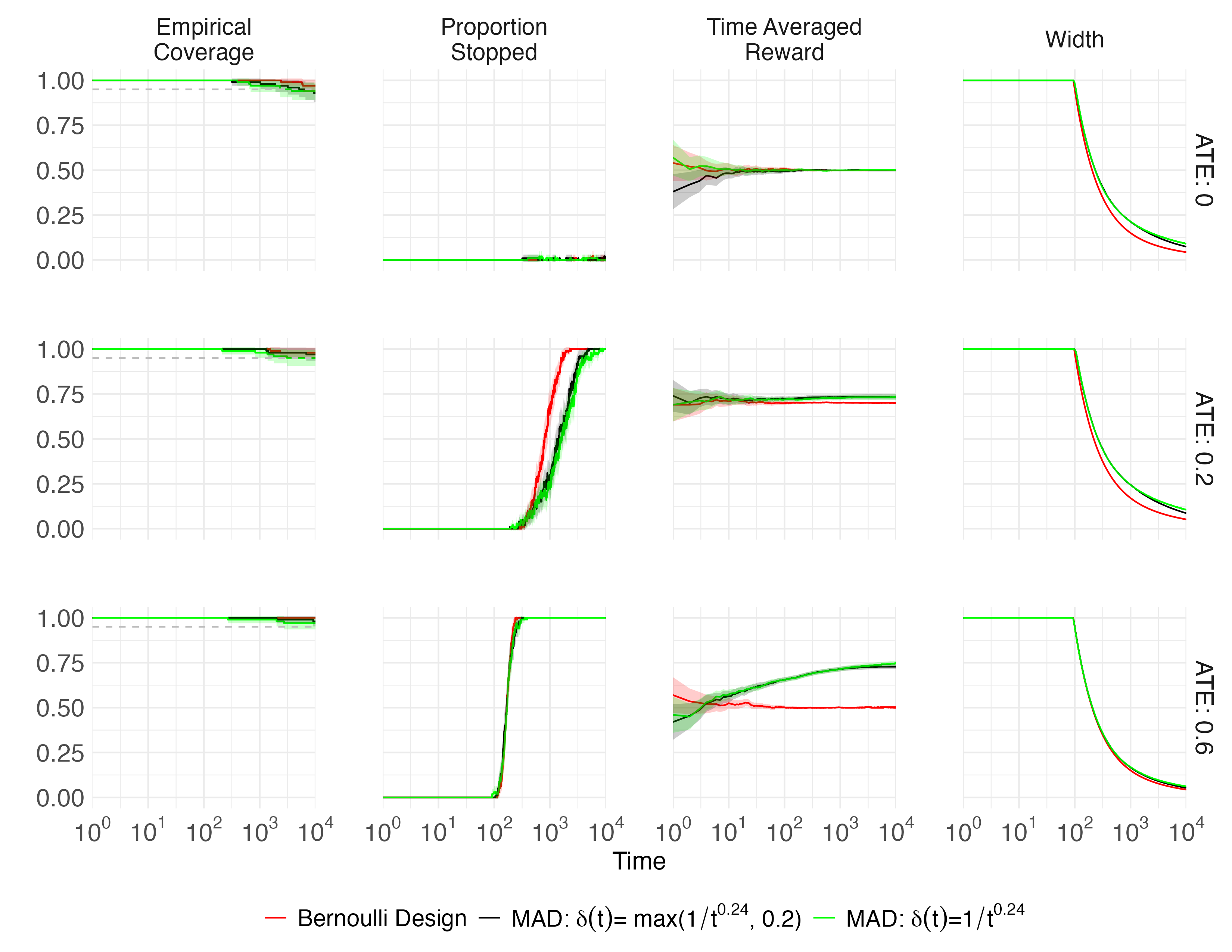}
    \caption{Empirical coverage, proportion stopped, time averaged reward, and width of the CS of Theorem~\ref{thrm1} across $N=100$ random seeds for different experimental designs under a two-armed bandit setting with a Bernoulli outcome model using UCB for the bandit algorithm; see Section~\ref{section:simulations} for full description of the experimental setting and each metric. $1-\alpha$ is depicted by the dotted grey line. Error bands depict $\pm 2$ SEs.}
    \label{fig:all_metrics_bern_ucb}
\end{figure}

We repeat versions of the Bernoulli outcome experiments in Section~\ref{section:simulations} using UCB as the adaptive algorithm. Figure~\ref{fig:all_metrics_bern_ucb} show the empirical coverage, stopping behavior, time averaged reward, and average width of this experiment, respectively. As in the TS experiments in Section~\ref{section:simulations}, we see that the MAD maintains finite-sample anytime validity, has a stopping time and width similar to the Bernoulli design while having significantly higher time averaged reward.

We also performed analogous experiments with different outcome models. As in the Bernoulli experiments, we consider both a Thompson sampler and UCB and run all experiments across $N=100$ random seeds. As baselines, we use the Standard Bandit design (for the experiments that use TS) and the Bernoulli design. MAD is implemented with $\delta(t) = \frac{1}{t^{0.24}}$. First, we consider a two-arm bandit with a Normal outcome model, i.e., $Y_i(1) \stackrel{i.i.d.}{\sim} \mathcal{N}(\mu_0, 1)$ and $Y_i(0)  \stackrel{i.i.d.}{\sim} \mathcal{N}(\mu_1, 1)$, so $\bar{\tau}_t = \mu_1-\mu_0$ for all $t$. The Thompson sampler is implemented with independent standard Normal priors on both arms. We consider $(\mu_0, \mu_1) = \{(1, 1), (1, 2), (1, 4)\}$ so the ATE = $0, 1, 3$, respectively. We set $T=1000$.

\begin{figure}
    \centering
\includegraphics[width=\linewidth]{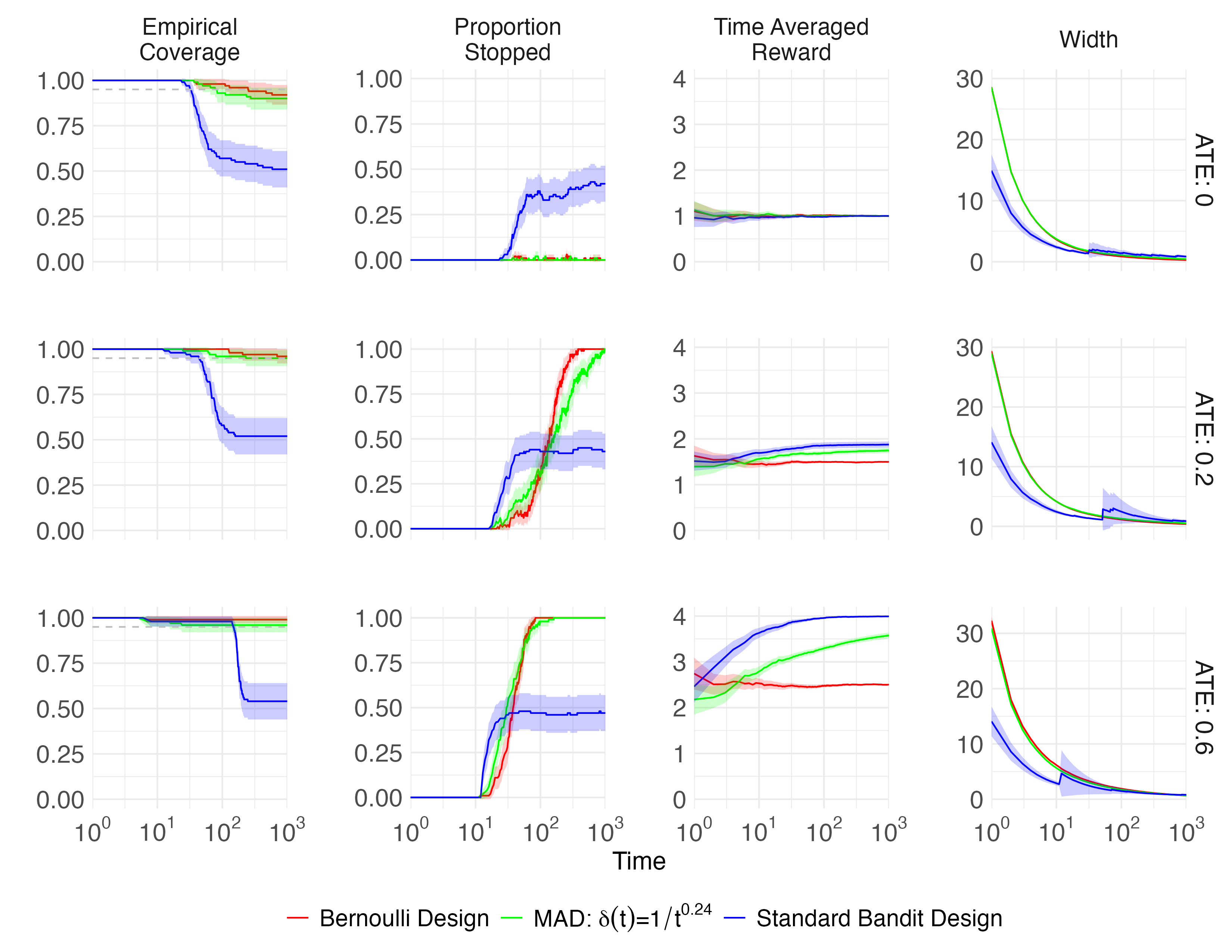}
    \caption{Empirical coverage, proportion stopped, time averaged reward, and width of the CS of Theorem~\ref{thrm1} across $N=100$ random seeds for different experimental designs under a two-armed bandit setting with a Normal outcome model using TS as the bandit algorithm, as described in Appendix~\ref{simulation_results_appendix}. The dashed grey line represents $1-\alpha$. Error bands depict $\pm 2$ SEs.}
    \label{fig:all_metrics_norm}
\end{figure}

\begin{figure}
    \centering
\includegraphics[width=\linewidth]{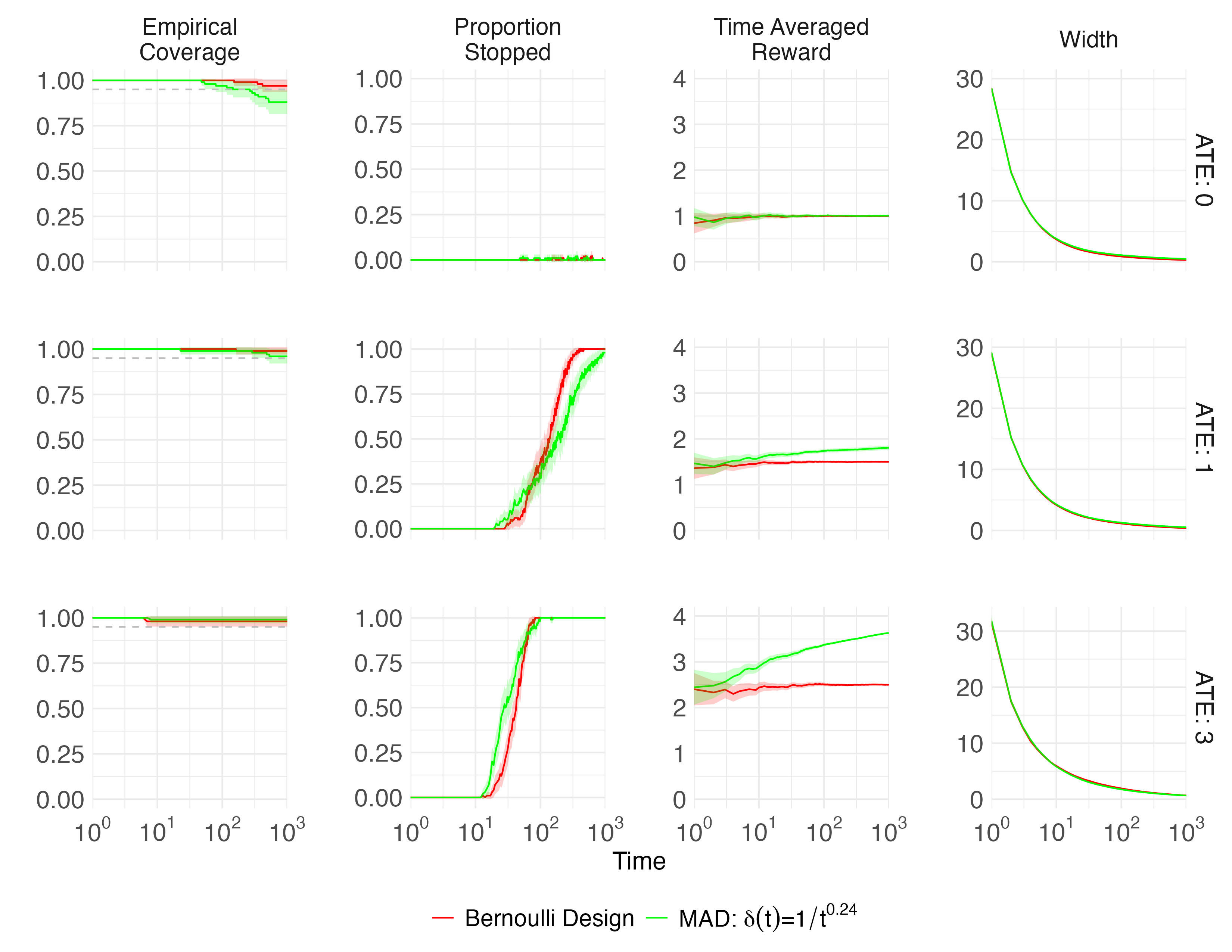}
    \caption{Empirical coverage, proportion stopped, time averaged reward, and width of the CS of Theorem~\ref{thrm1} across $N=100$ random seeds for different experimental designs under a two-armed bandit setting with a Normal outcome model using UCB as the bandit algorithm, as described in Appendix~\ref{simulation_results_appendix}. The dashed grey line represents $1-\alpha$. Error bands depict $\pm 2$ SEs.}
    \label{fig:all_metrics_norm_ucb}
\end{figure}

Figures~\ref{fig:all_metrics_norm}--~\ref{fig:all_metrics_norm_ucb} show the empirical coverage, stopping behavior, time averaged reward, and average width of this experiment using TS and UCB for the adaptive algorithm, respectively. Recall Section~\ref{section:simulations} for descriptions of how each of these metrics are computed. As in the Bernoulli experiments, the Standard Bandit design does not have anytime validity. Though it can have smaller width than the MAD and the Bernoulli designs, the CSs generated via the Standard Bandit design cannot be used to produce valid inference. On the other hand, the MAD achieves the proper anytime validity in finite-samples and performs similarly to the Bernoulli design in terms of both CS width and stopping time while having reward closer to that of the Standard Bandit design.

We also consider misspecified settings where the Thompson sampler still assumes a Normal outcome model with standard normal priors as in the previous experiment, but the outcome model is not normally distributed. Instead, we set the outcome model to be t-distributed or Cauchy, \emph{i.e.}, $Y_i(1) \stackrel{i.i.d.}{\sim} t(\mu_0, \nu)$ and $Y_i(0)  \stackrel{i.i.d.}{\sim} t(\mu_1, \nu)$, where $\nu$ is the degrees-of-freedom ($\nu=1$ corresponds to the Cauchy distribution), and $\mu_i$ is the non-centrality parameter, $i=0, 1$. In these misspecified settings, MAD is implemented with $\delta(t) = \frac{1}{t^{0.2}}$ to better control the variance of the ATE estimator. As in the Normal example, we set $T=1000$ and run all experiments across $N=100$ random seeds.

\begin{figure}
    \centering
\includegraphics[width=\linewidth]{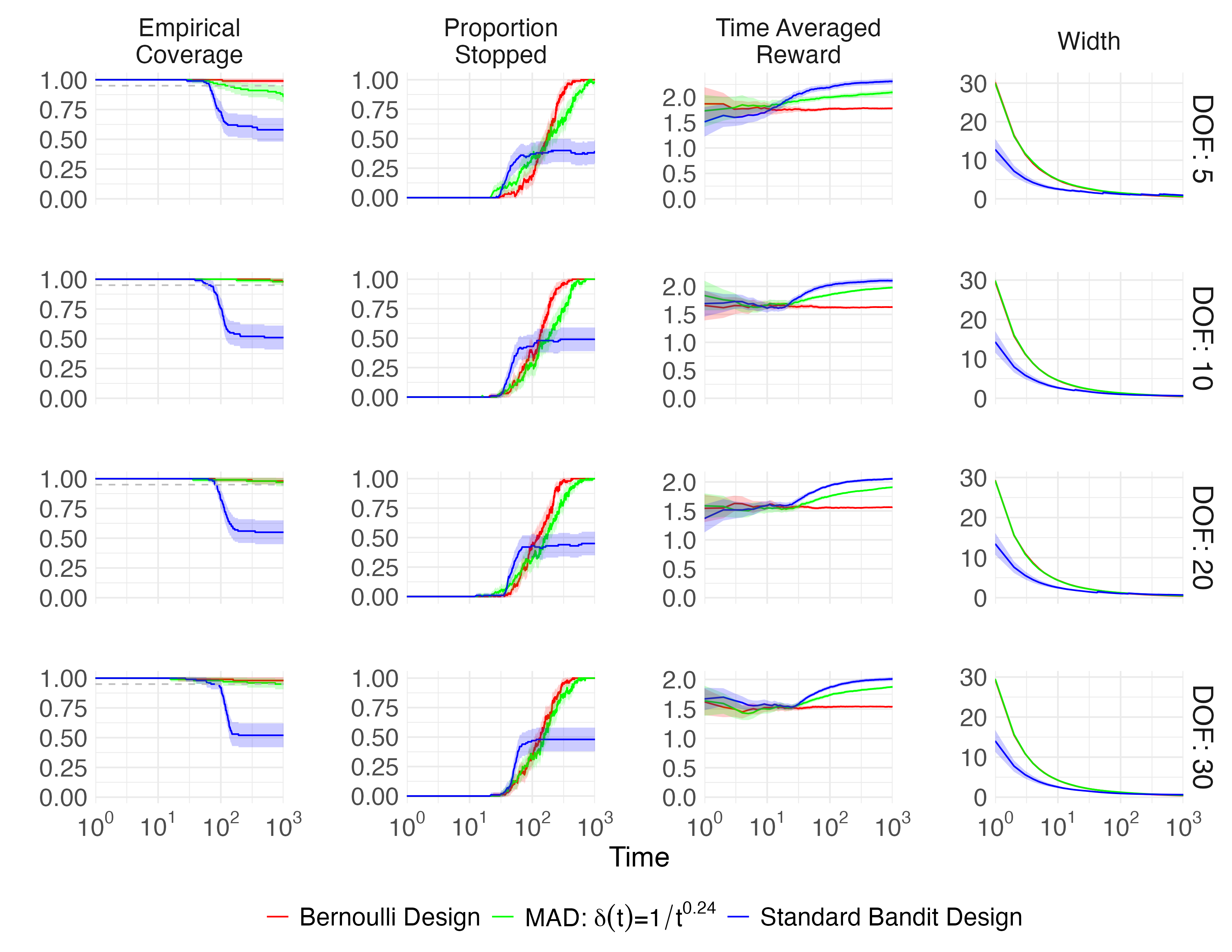}
    \caption{Empirical coverage, proportion stopped, time averaged reward, and width of the CS of Theorem~\ref{thrm1} across $N=100$ random seeds for different experimental designs under a two-armed bandit setting with a t-distributed outcome model using TS as the bandit algorithm, as described in Appendix~\ref{simulation_results_appendix}. The dashed grey line represents $1-\alpha$. Error bands depict $\pm 2$ SEs.}
    \label{fig:all_metrics_t}
\end{figure}

\begin{figure}
    \centering
\includegraphics[width=\linewidth]{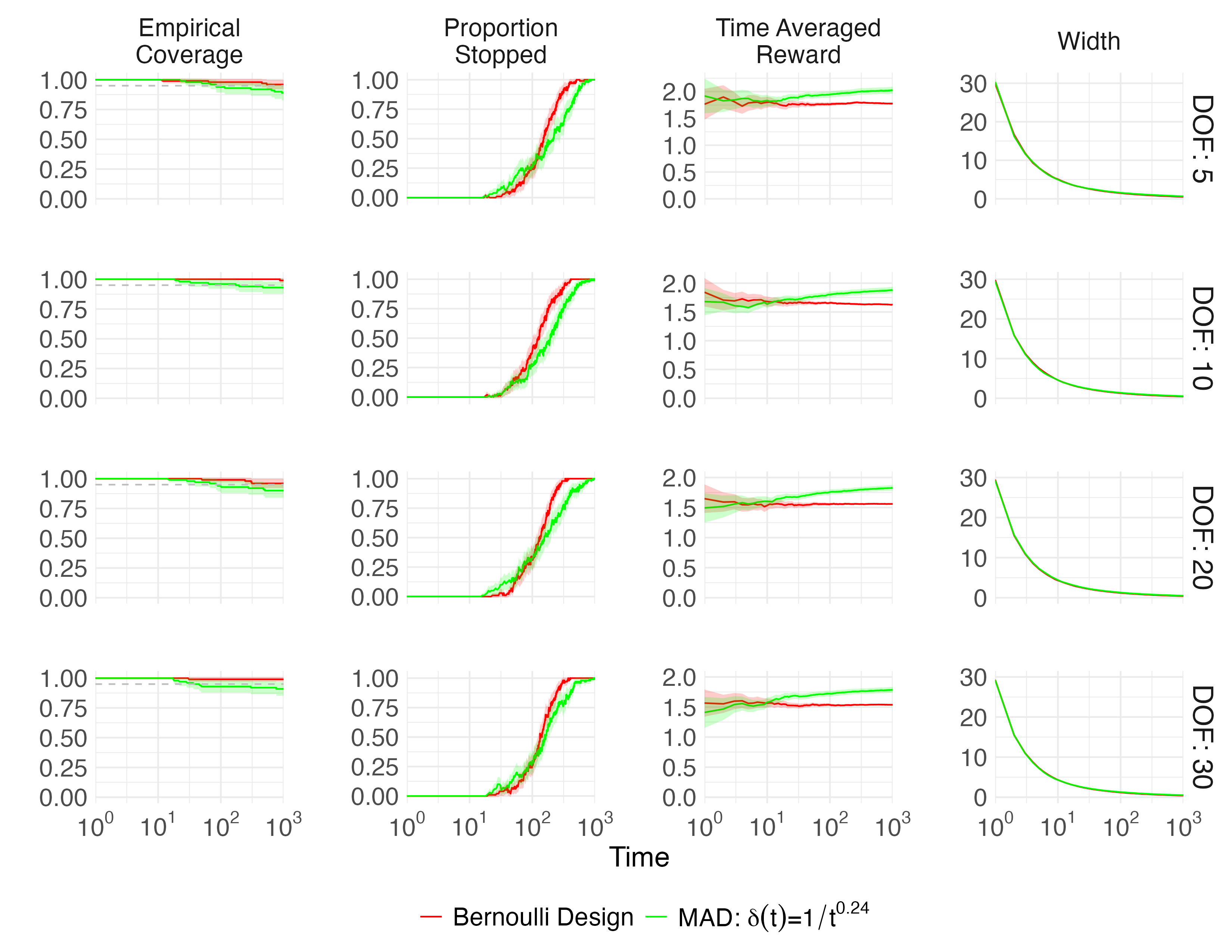}
    \caption{Empirical coverage, proportion stopped, time averaged reward, and width of the CS of Theorem~\ref{thrm1} across $N=100$ random seeds for different experimental designs under a two-armed bandit setting with a t-distributed outcome model using UCB as the bandit algorithm, as described in Appendix~\ref{simulation_results_appendix}. The dashed grey line represents $1-\alpha$. Error bands depict $\pm 2$ SEs.}
    \label{fig:all_metrics_t_ucb}
\end{figure}

\begin{figure}
    \centering
\includegraphics[width=\linewidth]{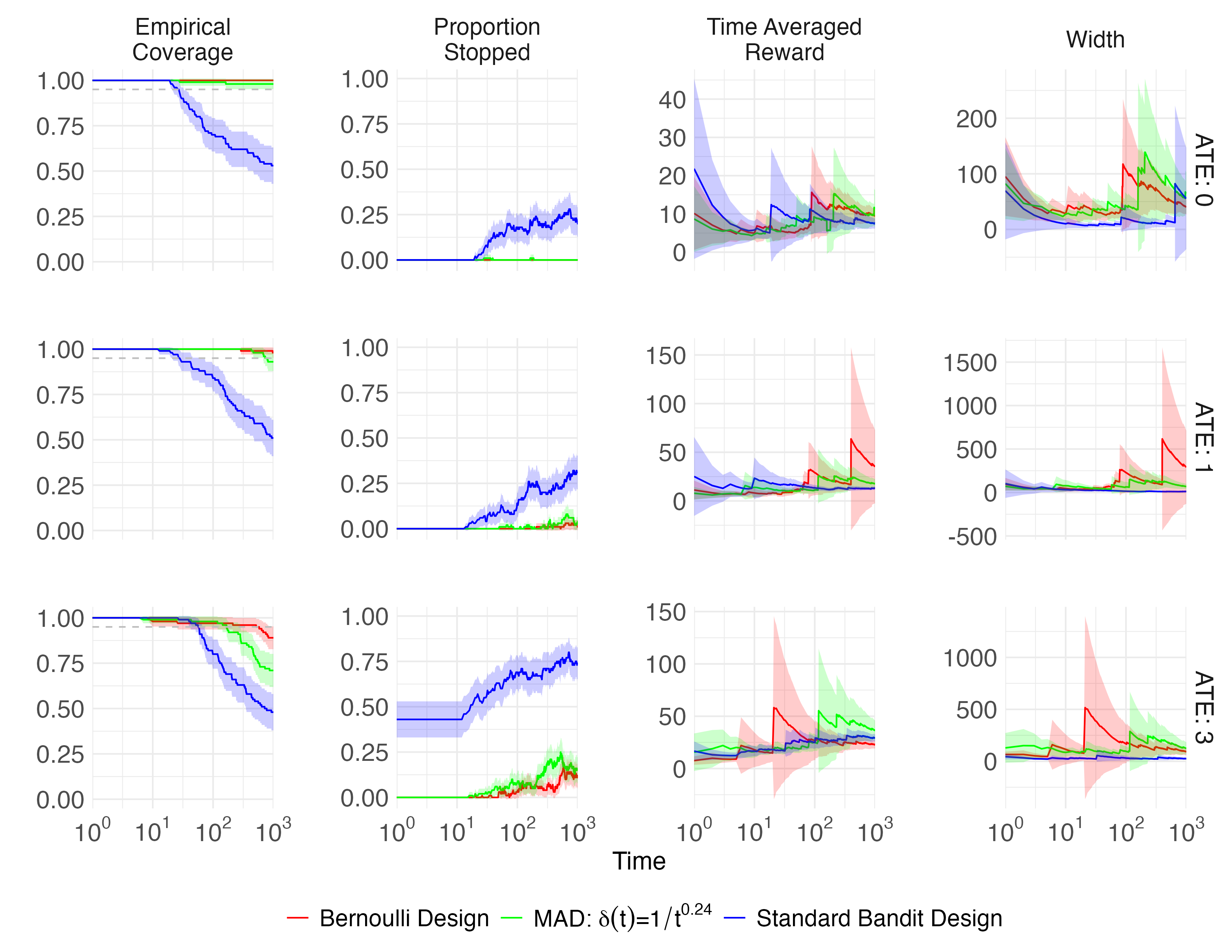}
    \caption{Empirical coverage, proportion stopped, time averaged reward, and width of the CS of Theorem~\ref{thrm1} across $N=100$ random seeds for different experimental designs under a two-armed bandit setting with a t-distributed outcome model using TS as the bandit algorithm, as described in Appendix~\ref{simulation_results_appendix}. The dashed grey line represents $1-\alpha$. Error bands depict $\pm 2$ SEs.}
    \label{fig:all_metrics_cauchy}
\end{figure}

\begin{figure}
    \centering
\includegraphics[width=\linewidth]{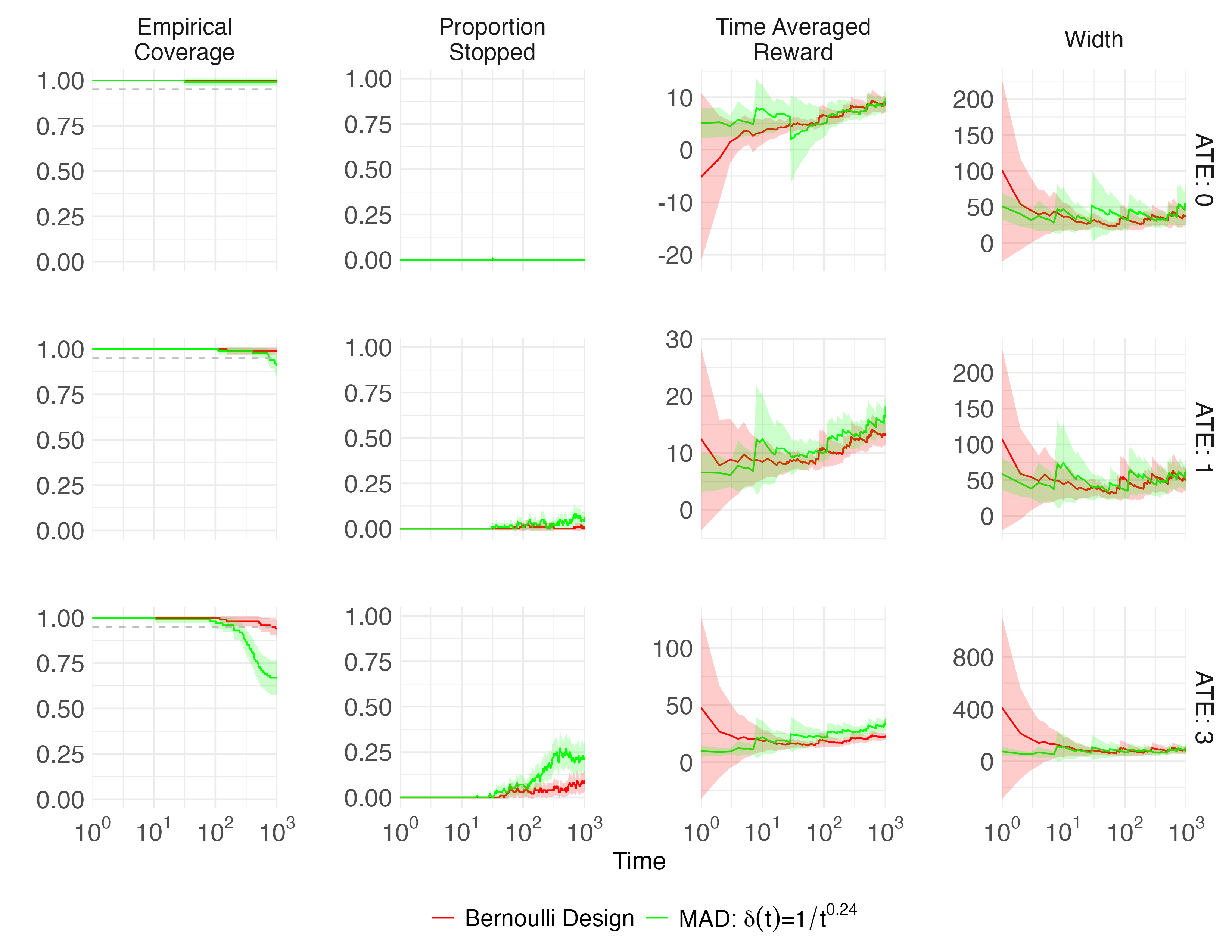}
    \caption{Empirical coverage, proportion stopped, time averaged reward, and width of the CS of Theorem~\ref{thrm1} across $N=100$ random seeds for different experimental designs under a two-armed bandit setting with a t-distributed outcome model using UCB as the bandit algorithm, as described in Appendix~\ref{simulation_results_appendix}. The dashed grey line represents $1-\alpha$. Error bands depict $\pm 2$ SEs.}
    \label{fig:all_metrics_cauchy_ucb}
\end{figure}

Figures~\ref{fig:all_metrics_t}-~\ref{fig:all_metrics_t_ucb} shows the corresponding results across different choices of $\nu$ for t-distributed outcomes with $(\mu_0, \mu_1) = (1, 2)$ using TS and UCB as the bandit algorithms, respectively. ~\ref{fig:all_metrics_cauchy}--~\ref{fig:all_metrics_cauchy_ucb} shows the corresponding results for the Cauchy outcome distribution across  $(\mu_0, \mu_1) = \{(1, 1), (1, 2), (1, 4)\}$. Note, this is the same set of $(\mu_0, \mu_1)$ pairs as used in the Normal outcome distribution examples. 

Notably, even under heavy-tailed t-distributed outcomes (\emph{e.g.}, $\nu = 5$), the MAD still performs similarly to the Bernoulli design in terms of both width, coverage, and stopping times, while still having reward closer to the Standard Bandit design (which once again, does not maintain anytime validity in this setting). However, under a Cauchy outcome model, as shown in Figures~\ref{fig:all_metrics_cauchy}--\ref{fig:all_metrics_cauchy_ucb}, the width and reward of \emph{all} designs, even the Bernoulli, can be unstable due to the heavy tails of the Cauchy distribution, and when the ATE is large, the MAD can sometimes not have the correct coverage. However, we note that the ATE is not well-defined in this setting since the expectation of a Cauchy is undefined, and hence, the IPW estimator is no longer an unbiased estimator for the true ATE and no method is guaranteed to maintain validity in this setting.

\begin{figure}
    \centering
\includegraphics[width=\linewidth]{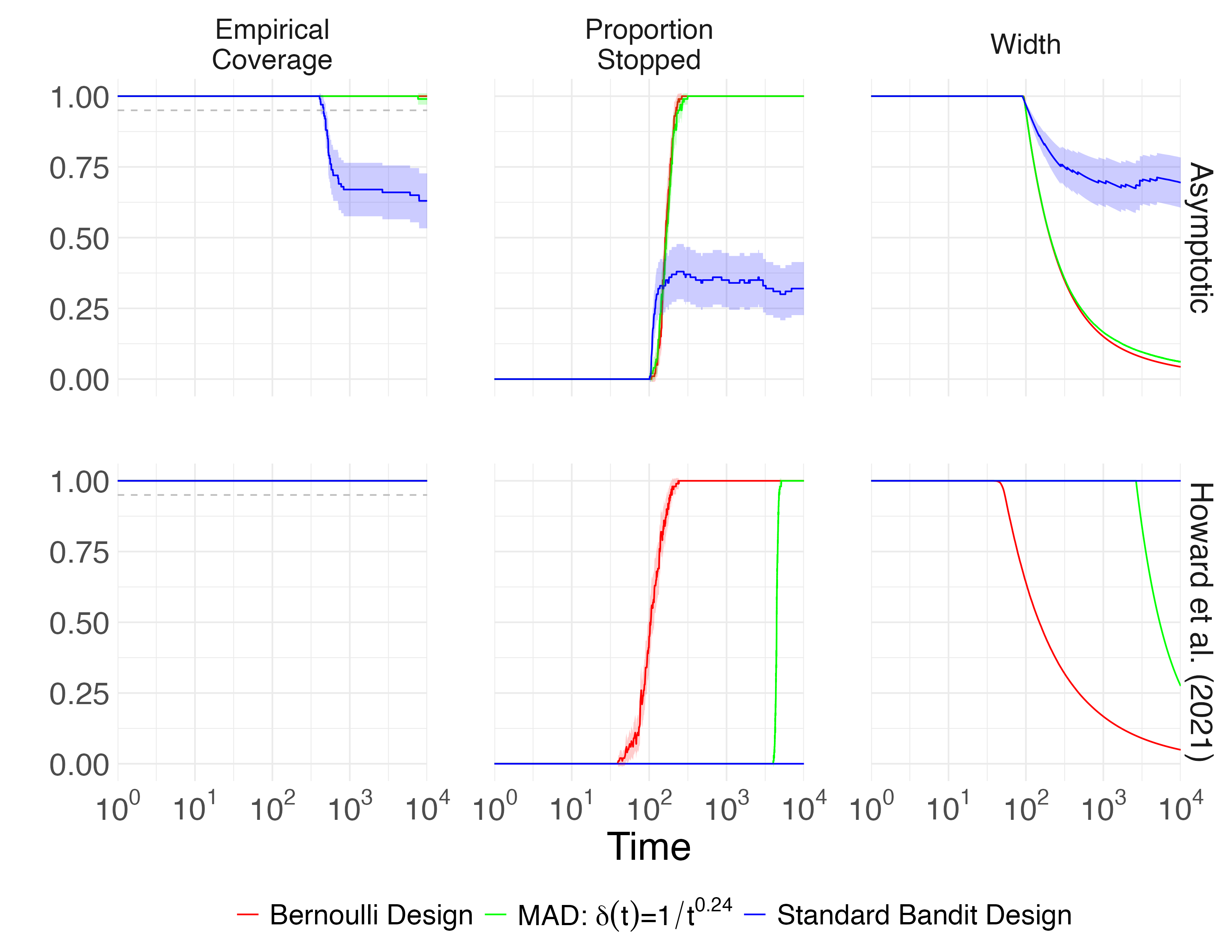}
    \caption{Empirical coverage, proportion stopped, and average width of the CS of Theorem~\ref{thrm1} vs. that of \cite{howard} across $N=100$ random seeds for different experimental designs under a two-armed bandit setting with a Bernoulli outcome model, as described in Section~\ref{section:simulations}. Time averaged reward is not shown since it is the same across both experiments as the same random seed were used. The dashed grey line represents $1-\alpha$. Error bands depict $\pm 2$ SEs.}
    \label{fig:width_howard}
\end{figure}

Finally, we assess all designs using the asymptotic CS of Theorem~\ref{thrm1} and the non-asymptotic CS of \cite{howard} in the Bernoulli outcome setting described in Section~\ref{section:simulations} to exhibit how the MAD with our asymptotic CS generally produces far more powerful inference than the MAD used with the non-asymptotic CS of \cite{howard} while still maintaining finite-sample anytime validity; hence, we recommend the use of our asymptotic CS in practice. In this setting, we use the MAD setting $\delta_t$ like in Example 2 of Section~\ref{section:setting_delta}, as the CS of \cite{howard} requires there exists some $p_{min}$ such that all treatment assignment probabilities are bounded within $[p_{min}, 1-p_{min}]$. Because of this restriction, we can only use Thompson sampling with the CS of \cite{howard}.
We also implement the Standard Bandit and Bernoulli designs and assess the \cite{howard} CS for them. Since $p_{min}$ does not exist for the Standard Bandit design with Thompson sampling, we set $p_{min}$ heuristically to be $1/T$. \footnote{Since we expect Thompson sampling to achieve a regret of $O(\log(T))$ asymptotically \cite[]{agrawal2012analysis}, the share of units that get assigned to the suboptimal treatment is on the order of $\log(T)/T$, so we use $1/T$ as a conservative, heuristic lower bound.} Note, however, that the CS of \cite{howard} is not guaranteed to have error control for the Standard Bandit design because their validity results require the existence of such a $p_{min}$. As shown in Figures~\ref{fig:width_howard}, the CS of \cite{howard} is generally much wider than that of our asymptotic CS, and the stopping times occur much later for the MAD. Since our asymptotic CS still maintains the proper error control in finite samples, we recommend its use in practice.

\end{document}

%% file: cmd.tex
\usepackage{algorithm}
\usepackage{algorithmic}

\newtheorem{innercustomgeneric}{\customgenericname}
\providecommand{\customgenericname}{}

\newcommand{\newcustomtheorem}[2]{%
  \newenvironment{#1}[1]
  {%
  \renewcommand\customgenericname{#2}%
  \renewcommand\theinnercustomgeneric{##1}%
  \innercustomgeneric
  }
  {\endinnercustomgeneric}
}

 \newtheorem{theorem}{Theorem}
 \newtheorem{definition}{Definition}
 \newcustomtheorem{customthm}{Theorem}
\newcustomtheorem{customlemma}{Lemma}
\newcustomtheorem{customassumption}{Assumption}
\newcustomtheorem{customdef}{Definition}

\newtheorem{assump}[theorem]{Assumption}